\documentclass[12pt]{article}
\usepackage{graphicx}
\usepackage{subfigure}
\usepackage{cite}
\usepackage{amsmath}
\usepackage{capt-of}

\begin{document}

\title{\vspace{-2cm} Spontaneous Symmetry Breaking, \\ Group
Decision Making and Beyond \\  1.  Echo Chambers and Random Polarization}

\author{ Serge Galam\thanks{serge.galam@sciencespo.fr} \\
CEVIPOF - Centre for Political Research, Sciences Po and CNRS,\\
1, Place Saint Thomas d'Aquin, Paris 75007, France}

\date{October 2, 2024}

\maketitle

\begin{abstract}

Starting from a symmetrical multiple choice individual I build a sociophysics model of decision making. Reducing the choices to two and interactions to pairs recover the  Ising model from physics at zero temperature. The associated equilibrium state is a spontaneous symmetry breaking with the whole group sharing a unique choice, which is selected at random. However, my focus departs from physics, which aims at identifying the true equilibrium state discarding any possible impact of the initial conditions, the size of the sample and the used update algorithm. Any memory of the past history is erased. In contrast, I claim that dealing with a social system, the history of the system must be taken into account in identifying the relevant social equilibrium state, which is always biased by its history. Accordingly, using Monte Carlo simulations I explore the spectrum of non universal equilibrium states of the Ising model at zero temperature. In particular, I show that different initial conditions with the same value of the order parameter lead to different equilibrium states. The same applies for different sizes and different update algorithms. The results indicate that in the presence of a social network composed of agents sharing different initial opinions, it is their interactions that lead them to share a unique choice and not their mere membership in the network. This finding sheds a new light on the emergence of echo chambers, which appear to be the end of a dynamical process of opinion update and not its beginning with a preferential attachment. Furthermore, polarization is obtained as a side effect of the random selection of the respective unanimous choices of the various echo chambers within a social community. The study points to social media exchange algorithms, which are purely technical levers independent of the issue and opinions at stake, to tackle polarization by either hindering or accelerating the completion of symmetry breaking between agents.

\end{abstract}

\newpage
%%%%%%%%%%%%%%%%%%%%%
\section{Introduction}

The phenomenon of spontaneous symmetry breaking is key in physics to address the origin of ordered states of inert matter in bulk \cite{p1, p2}. In particular, it explains the puzzling observation that large collections of atoms with short range interactions succeed to produce long range order. Well studied in condensed matter and magnetic systems, the issue was solved exactly for the benchmark Ising model at two dimensions as a function of a coupling constant and temperature with zero field \cite{ising, p3}.

In presence of a field and for other dimensions no exact treatment has been performed so far. To fill this gap, the mean field treatment has been and is applied to a large series of related models \cite{mf1, mf2}. While mean field provides a good qualitative description of the equilibrium state the associated quantitative values are wrong. Otherwise, renormalization groups technics yield quite good quantitative results  \cite{w1}, To get ``exact" results requires the use of numerical simulations and Monte Carlo algorithm is mostly used \cite{mc1, mc2}.

The generic character of the Ising model has fostered its wide application to numerous issues both in physics and outside physics. With respect to social sciences we published a paper with Gefen and Shapir in the beginning of the eighties calling for the creation of a new field for which we coined the word sociophysics. To demonstrate our plea for this new domain of research we model in the paper the phenomenon of strike setting an Ising like model in a field \cite{s1}. 

Latter on in the earlier nineties I co-author with Moscovici a paper using for the first time a random field Ising model to explain several unsolved observations in group decision making \cite{s2}. A few year latter I focused on the group decision outcomes of the zero temperature case \cite{s3}.  A large number of papers has been since published in sociophysics using Ising like model \cite{s4, s5, s6, s7, s8, s9, s10}. These works subscribe to the active \cite{a1, a2, a3, a4, a5, a6, a7, a8, a9, a10, a11, a12, ph1, ph2, ph3, ph4, ph5, ph6, ph7, ph8, ph9, ph10, ph11, ph12, g1, g2, g3, g4, g5, g6, g7, g8, g9, g10, g11, g12, g13, g14, g15, g16, g17, g18, g19} and growing \cite{v1, v2, v3, v4, v5, v6} field of sociophysics  \cite{so1, so2, so3, so4, so5, so6}.

Here, I build a sociophysics model of decision making starting from a symmetrical multiple choice individual. Reducing the choices to two and adding pair interactions  recover the  Ising model from physics at zero temperature and an earlier model of decision making \cite{s2, s3}. 

However, standard studies of Ising like models ignore the actual dynamics to reach the equilibrium state, which is aimed to be the absolute one. Even in Monte Carlo simulations for which the implementation of a dynamics of repeated updates is a prerequisite, the dynamics is considered as a purely technical mean, which must be of no effect on the associated equilibrium state  \cite{mc}.

In the present work, contrary to above usual practise, I notice that in any social systems a specific framework is always set up to monitor the individual updates produced by the interactive pairs. This very fact turns the update algorithm and initial conditions to relevant ingredients to identify the equilibrium state, which in turn is expected to depend on its history. Therefore, it is of importance to investigate the non universal aspects of update dynamics in reaching an equilibrium state. 

Accordingly, using Monte Carlo simulations I show that different initial conditions and update scheme produce different equilibrium states. These equilibrium states display either a full spontaneous symmetry breaking or a fragmented spontaneous symmetry breaking with the stable coexistence of two opposed domains of different sizes. 

I also compare four types of update schemes, namely the random, sequential, simultaneous and checkerboard updates. The results show that different algorithms produce different equilibrium states yet starting from the same initial sample of agents.

The results shed a new light on the emergence of echo chambers. The well accepted definition of an echo chamber being the outcome of a preferential attachment of people sharing the same opinion is refuted. At odds, echo chambers appear to be the ending of a dynamics of local updates within an existing social network with agents sharing initial different opinions.

It is the existence of a social network bounding a collection of interacting agents, which leads to their alignment on the same choice. In addition, polarization is found to be a direct side effect of the random selection of the respective unanimous choices made in the various echo chambers present in a given social community.

The findings point to a new purely technical lever to tackle polarization in social media by implementing algorithms designed at hindering or accelerating the completion of the symmetry breaking among agents independently of the issue and opinions at stake.

At this point, it is worth to stress that not all social debates fit to symmetrical multiple choice individuals. Often agents are individually committed to a peculiar choice, which breaks the initial symmetry of some of the agents. External realty may also break the symmetry of initial choices. Accordingly, in a forthcoming paper I will study the effect of external and internal pressures on the agent choices.

The rest of the paper is organized as follows: The symmetrical multiple choice individual is discussed in Section 2  while Section 3 deals with the collective choice of a N-person group. Section 4 studies the N-person interacting group. The Ising model is compared in physics and social systems in Section 5. Section 6 develops an analytical identification of the the equilibrium state. Some comments about the dynamics of update to reach equilibrium are contained in Section 7. Section 8 is about the requirement for a real material framework to monitor social update dynamics. Reports of initial steps of symmetry breaking in Ising systems are listed in Section 9. The requirement of social structures to drive collective symmetry breaking is advocated in Section 10. The robustness of the results is checked in Section 11 by accounting for statistical fluctuations.  The social consequences of social collective symmetry breaking are listed in Section 12. Section 13 contains some concluding remarks about novel algorithmic protocols to either accelerate the symmetry breaking of individual choices or to hamper their completion.

\section{The symmetrical multiple choice individual}

Given a socio-political environment, I consider one single person who has to select a choice to solve or address an issue out of $n$ possibilities, $n$ being a positive integer. All these possibilities are equally feasible with no definite advantage.  All $n$ choices are appropriate to the issue despite being different from one another. The basis for the agent choice is subjectively specific to themselves and is not of indisputable value to other ones. Therefore, all choices are equivalent in terms of usefulness to the individual with the same utility value. There exists no objective proof to order those $n$ choices in term of increasing utility.

Accordingly, at this stage it is not possible to predict the choice to be made by a given person. Their selection is subjective and varies form one person to another. Therefore, for an outside observer the individual choice is random. 

Mathematically, it means that all $n$ choices are equiprobable with probability $1/n$. The potential single person choice is thus invariant under the permutation of choices with respect to their utility value. However, as soon as the person selects one specific choice, their symmetry gets broken. This actual individual choice is no longer invariant under a permutation with other choices for that specific person.

At this stage it is worth to stress that in physics magnetic spins share this symmetrical state for their ordering. But a qualitative difference between a person and a magnetic spin. For a magnetic spin having the freedom to orient itself, the only impact of an orientation is that orientation with no cost at changing orientation. 

In contrast, for a person, each choice has a specific impact on their associated environment. Selecting one choice over the others modifies the environment in a way which can be irreversible. In selecting one choice, the related person excludes all the others creating a social inertia. Once done, there is a cost in shifting their choice. The magnitude of the cost depends on the nature of the associated choice, which can range from the color of a coffee machine to the discipline of academic studies or changing a social norm. That is a significant feature of a human frame, which is absent in inert matter.

\subsection{From one to $N$ non-interacting persons}

Under above conditions, going from one person to a collection of $N$ non-interacting persons, the outcome of aggregating their parallel and independent choices obeys on average an equal repartition among the $n$ choices with a proportion
\begin{equation}
p_i=\frac{1}{N} (\frac{N}{n}) = \frac{1}{n} ,
\label{pn}
\end{equation}
for each choice denoted by $i=1, 2, ..., n$.

However, once every person has selected a choice, the initial choice symmetry is recovered at the collective level of the $N$-person group, despite being broken at each individual level. The symmetry restoration is worth to underline since it adds a new feature at the collective level, which is absent at the level of a single multiple choice individual. 

Indeed, while a permutation symmetry between the $n$ choices does exist for the individual choice, it is never fulfilled. As soon a single person makes a choice, the symmetry is broken. Shifting the selected choice modifies the person choice and thus the associated impact on their environment.

In contrast, having a group of $N$ persons where each one has selected a choice equiprobably, while permuting all individual choices modifies their respective choices, overall at the collective aggregated level, nothing is changed with still an equipartition of choices. 

The existence of a group implements a symmetrical state with respect to the available choices. The group social impact is thus invariant under the permutation of individual choices. Moreover, the impact of each choice is a function of the number of people who have adopted this choice, increasing simultaneously its associated inertia. Here, the respective impacts of choices are equivalent both in terms of amplitude and inertia.

\subsection{External homogeneous symmetry breaking pressure}
All choices being equivalent in term of their utility, above multiple choice individual is vulnerable to any external pressure, even extremely weak, to make a given choice. An infinitesimal pressure exerted on a person in favor of a given choice, is therefore sufficient to have the person select that choice. Accordingly, an external infinitesimal pressure is sufficient to break the symmetry of the $n$ choices along a specific choice.

The same symmetry breaking effect applies identically to each person of a group of non-interacting $N$ persons. An infinitesimal uniform pressure is thus sufficient to align all the $N$ persons along the choice favored  by the external pressure. In such a case, the favored choice is selected by all $N$ persons with zero person for all  the other $n-1$ choices. Associated proportions are
\begin{equation}
p_1=1 \ and \  p_i=0 ,
\label{pn1}
\end{equation}
where index 1 denotes the choice favored by the external pressure and $i=2, 3, ..., n$.

Above symmetry of the aggregated $N$ non-interacting persons is now broken by the external pressure similarly to the breaking at the individual level.  The associated social impact is thus enhanced and amplified as a function of $N$ resulting in a higher social inertia and a larger amplitude of the impact on the related environment. All other $n-1$ choices are simultaneously excluded.

\subsection{Individualized heterogeneous symmetry breaking pressure}

It may also happen that for a given issue a specific person does have an internal bias favoring one choice over the others. Their multi choice symmetry is then broken even by an  infinitesimal internal pressure. However, the related pressure is attached to the person and does not apply to others. Such an internal pressure can be either conscious or unconscious. It can be noticeable or not for an outside observer. In that case, the associated person selects their favored choice while their choice can still be perceived as randomly made.

For a collection of $N$ non-interacting persons, when only a single person has an internal pressure, the aggregated outcome of the collection of agent choices leads on average to one choice being selected by  $1+ (N-1)/n$ persons with a number $(N-1)/n$  persons for all  other $n-1$ choices. Associated proportion are

\begin{equation}
p_1=\frac{1}{N} (1+\frac{N-1}{n}) ,
\label{pn10}
\end{equation}
where index $1$ represent the choice favored by the agent having an internal pressure. Others $n-1$ choices have an equal proportion
\begin{equation}
p_i=\frac{1}{N} \frac{N-1}{n} ,
\label{pni}
\end{equation}
where $i=2, ..., n$.

It is of importance to emphasize that in contrast to above external pressure, which applies equally to all persons, the internal pressure applies only to one person. Accordingly, while the symmetry of choices is broken for one person, it still holds at the collective level of the $n-1$ other persons. Yet, at the full collective level of the $N$ persons, the symmetry is broken along choice 1.

Nevertheless, in case of $n$ internal symmetry breaking pressures, each along one distinct choice of the avalaible $n$ choices, the symmetry at the collective level is recovered including the collective social impact. Nevertheless, in this case above associated social impact and inertia, associated with an external pressure, are now equally divided along each choice as $N/n$.

\subsection{Reducing the number $n$ of choices down to 2}
  
At this stage, to simply the modeling without losing generality I restrict the number $n$ of equivalent choices to 2. In that case,  Eqs. (\ref{pn}, \ref{pn1}, \ref{pn10}, \ref{pni}) write respectively $p_1=p_2=\frac{1}{2}$, $p_1=1$ and $p_2=0$, $p_1=\frac{1}{N}(1+\frac{N-1}{2})$ and $p_2=\frac{1}{N}(\frac{N-1}{2})$. 

Indeed, a large number of issues have two competing choices. In addition, many cases with a larger number of answers can be reduced to two choices at a certain level of approximation  \cite{17}. The two choice case corresponds to a choice between two items A and B with for instance at a referendum, a two candidate election, or many other situations. 

However, it is worth stressing that the two choices could also apply to choosing an activity to organize a community at structural levels. For instance to decide between hunting and gathering in primitive societies. In such a case, the respective related social and cultural impacts were significantly different on the evolution of the related communities with significant associated long term  inertia. Accordingly, the impact of an individual breaking the symmetry of their available choices may vary significantly depending on the issue at stake. However, the underlying symmetry breaking mechanism is identical and universal regardless of the content of the issue.

In addition to the reduction of the number of choices down to 2, from now on I consider individual choices without including neither external nor internal symmetrical breaking pressures. I will study their effect in a future following paper.

\section{The collective choice of a N-person group}  

Binary choices allow to recover the so-called Ising variables from statistical physics \cite{ising}, which are usually represented by $+1$ and $-1$. The multiple choice case could have also been cast in statistical physics using the so called Potts model \cite{pot}.

Along the Ising connexion, I denote the individual's discrete bimodal choices by $\pm1$, which in turn allows to evaluate the level of diversity within a collection of individual choices of a $N$-person group as seen with the quantities
\begin{equation}
C=  \sum_{i=1}^{N}c_i ,
\label{C}
\end{equation}
and \begin{equation}
c= \frac{1}{N} \sum_{i=1}^{N}c_i ,
\end{equation}
where $c_i=\pm 1$ denotes the choice of individual $i$ with $i=1,2,...,N$. $C$ accounts for the amplitude of the actual social impact with  $-N\leq C \leq N$ while $c$ measures the degree of the symmetry breaking with  $-1\leq c \leq +1$. 

In statistical physics $c$, denoted $m$, stands for the order parameter of the system. Dealing with phase transitions $c \neq 0$ marks an ordered phase in opposition to a disordered phase marked by $c=0$.

Here, I introduce the additional quantities $N_+$ and $ N_-$ to count the respective numbers of agents having chosen choice $+1$ and $-1$. Associated social and cultural impacts are proportional to $N_+$ and $ N_-$.  Associated proportions are denoted $p=N_+/N$ and $(1-p)$. The quantity $0\leq p \leq 1$ measures the degree of diversity of the $N$-person group.

The equiprobability of the two choices $\pm 1$ yields 
\begin{equation}
p_{c_i}=\frac{1}{2}\{ \delta (c_i-1)+\delta (c_i+1)\} ,
\end{equation}
for the probability distribution of the choice of individual $i$ where $\delta$ is the Kronecker function and $i=1, 2$. The associated probability distribution for the collection of $N$ persons is thus
\begin{equation}
p_c=\prod_{i=1}^{N}p_{c_i} .
\end{equation}

Eq.  (\ref{C}) shows that adding all individual choices from a collection of $N$ persons extends drastically the number of possible outcomes with $2^N$ different possible choices instead of two for one single person. In physics this number is divided by 2 to account for the reversal symmetry between $+1$ and $-1$ referred as time reversal. However, here this reduction does not hold due to the different social impacts associated to the choices $-1$ and $+1$.

In any case, while the choices $\pm 1$ have each a well defined meaning, it is not the case for $C$, which can vary from $-N$ to $+N$ by increment of $+2$. Yet, a simple majority rule like $C>0 \rightarrow +1$ and $C<0 \rightarrow -1$ restores the individual $\pm 1$ choices besides a stronger impact, which increases with $N$.

But  majority rule does not apply to the outcome $C=0$. One easy way to tackle the case without changing the symmetry is to assign the case $C=0$ to either $-1$ or $+1$ with equal probability $1/2$. But such a treatment would erase a major feature of the dynamics of choices since the outcome $C=0$ is the expected average outcome for the collection of symmetrical N-person group. More precisely $C=\pm \sqrt N$ and $c=\pm \frac{1}{\sqrt N}$ to account for statistical fluctuations.

On this basis, the case $C=0$ must be preserved and requires additional ingredients to be grounded with both social meaning and impact.  $C=0$ is obtained when $N_+=N_-=N/2$. Both activities at stake are then implemented with equal amplitudes on the related environment. At the same time, individuals can trade their initial choices with no visible effect at the collective level. It means that neither activity can feature the group identity, which can be labelled ``neutral".

The state determined by $C=0$ allow to implement the initial symmetry available to each person prior to their choice. In that state the symmetry is broken at the individual level but not at the collective level. Nevertheless, although the distribution of independent choices tends on average towards $C=0$, it is rarely reached due to random fluctuations of order $\pm N$.

\section{The N-person interacting group}

Above N-person non interacting group has allowed the setting of the basic ingredients of the model. Given a pair of agents $i, j$, their respective possible choices yield four different configurations:
\begin{enumerate}
\item $c_i=c_j=+1$ ,
\item $c_i=c_j=-1$ ,
\item $c_i=-c_j=+1$ ,
\item $c_i=-c_j=-1$ ,
\end{enumerate}
where both agents cooperate in the first two configurations and oppose in the other two. The product $c_i c_j =\pm 1$ allows labelling each case with $c_i c_j =1$ for cooperation and $c_i c_j = - 1$ for opposition. At this stage all configurations are equiprobable.

But in most situations where people face a choice within a group, either small or large, individuals interact between themselves before reaching their respective final decisions since that decision will impact them through the social impact of the choice on their environment.

To account for these interactions within the group I introduce pair interactions among people adding a utility $J$ associated with the implementation of the issue at stake with a benefit $+J$ for cooperation between the agents of the interacting pair and a cost $-J$ for opposite choices, i. e., a conflict. This setting implies that $J>0$ favors cooperation while $J<0$ favors conflict. In the following of the paper I restrict the study to $J>0$.

The product $J c_i c_j =\pm J$ incorporates the utility of a given pair configuration and thus breaks the symmetry between the four configurations. The cooperating ones being favored over the conflicting ones. For a N-person group with agents interacting by pairs within a configuration $\{c_1, c_2, \dots, c_N \}$, the total magnitude of the group utility writes
\begin{equation}
U \equiv J \sum_{<i,j>}c_ic_j ,
\label{U}
\end{equation}
where I have assumed a constant amplitude $J$ for all pairs of connected agents and $<i,j>$ represents all interacting pairs in the group.  Eq. (\ref{U}) recovers an earlier model of group decision making \cite{s2, s3}.

I emphasize that not all pairs of agents have their agents interacting with each other.  In physics, this feature is illustrated for instance with so-called short range interactions in contrast to long range interactions on a lattice. 

Here the situation is similar although agents are not located on a geometrical lattice and do not need to be neighbors in the real world. They need to be socially connected independently of their geometrical distance. But latter to keep simulations simpler to handle and visualize, agents socially connected are placed as nearest neighbors on a two-dimensional grid. Then, the geometrical neighboring means a social connection and not a geographical connection.

\section{The Ising model in physics versus social systems}

In statistical physics Eq. (\ref{U}) is formally identical to the Hamiltonian of the nearest neighbor ferromagnetic Ising model in zero field ($H=0$) with the orientation $S_i=\pm 1$ for the individual spin instead of the individual choice $c_i=\pm 1$. The associated criterium to reach equilibrium is the minimization of energy instead of maximization of the utility. Yet, this criterium applies only at zero temperature ($T=0$), otherwise it is the so-called free-energy which has to be minimized. In the present study, I maximize $U$ implying a $T=0$ like situation.

Having stressed that not all pairs of agents interacts, I must also emphasize that a given agent may be interacting simultaneously by pairs with several other agents, who themselves are interacting by pair with other agents. I then define a social network as the collection of all agents who are connected to at least one other agent in the group. 

A society facing a collective choice is thus fragmented into many social networks, which are disconnected on from the other in term of interacting pairs. Members of each social network exchange between themselves about the issue at stake but none discusses with agents from the other networks.

On above basis, given $c_i= \pm 1$ and the maximization of the pair utilities, it is straightforward to figure out that all N-person of a social network will end up  sharing the same choice to maximize their utility. But, from what was seen earlier, this collective uniform choice can be either $+1$ or $-1$ with equal probability. Those two results hold true for both the social and physical cases. 

However, at this point major qualitative differences arise between the two cases:
\begin{description}
\item[In physics] what matters is finding the absolute equilibrium state, which must be independent of the initial conditions. Here, it depends solely on $J$. The actual dynamics to reach the equilibrium state is a technical issue with no physical meaning. The outcome must be independent of the algorithm used to reach it.
\item[In social systems] I claim that  the algorithm used to reach the equilibrium state is instrumental and characterizes the related social system. Here the equilibrium state is a function of both the initial conditions and the concrete procedure implementing the dynamics of choices among the agents of the social network.
\item[Moreover,] while in inert matter the number of spins is astronomical of the order of the Avogadro number $10^{23}$, social systems may involve small and large number of agents. This fact has solid consequence on the actual solving of the problem. In physics the thermodynamic limit $N \rightarrow + \infty $ is used to perform calculations. And in simulations, large samples are used with periodic boundary conditions to get rid of  finite size effects and mimic an ``infinite system". 
\end{description}

\section{Analytical identification of the the equilibrium state}

Physics being concerned by the equilibrium state of a system, a powerful technics has been developed to solve analytically the problem. The technics is called a mean field treatment of the Hamiltonain. It basically ignores the fluctuations of individual spins by considering identical spins with the same average value of the magnetization. 

Indeed, I can justify a mean field treatment of Eq. (\ref{U}) by suggesting that in a social environment agents involve anticipation of the outcome in the making of their own choice.  To implement this anticipation feature I rewrite Eq. (\ref{U}) as
\begin{equation}
U=\frac{J}{2}\sum_{i=1}^{N}  \left\{\sum_{j=1}^{k}c_{j(i)}\right\} c_i ,
\label{U2}
\end{equation}
where $k$ is the number of persons individual $i$ interacts with. The choices of these $k$ persons are denoted by $c_{j(i)}$. To keep the presentation simple
this number is assumed equal for everyone. When  everyone interacts with everyone $k=N$. This case applies to small groups.

More generally I assume that every person is aware of the on going symmetry breaking process and thus anticipates the outcome as a collective choice of the network. Then, individual $i$ extrapolates the $k$ choices $c_j$ to be all equal to the expected collective choice of the without its own choice, which leads to
\begin{equation}
c_j = \frac{1}{N-1}(C-c_i) ,
\end{equation}
where $C$ is the collective choice defined as before. In turn, Eq. (\ref{U2}) writes
\begin{equation}
U_a=\frac{J}{2}\sum_{i=1}^{N}\left\{\sum_{j=1}^{k}\frac{1}{N-1}(C-c_i)\right\}c_i ,
\end{equation}
and,
\begin{equation}
U_a=\frac{k J}{2(N-1)}\left\{C\sum_{i=1}^{N}c_i-\sum_{i=1}^{N}c_i^2\right\} ,
\end{equation}
where superscript $a$ signals the anticipating process. Using $C=\sum_{i=1}^{N}c_i$ and $c_i^2=1$ leads to
\begin{equation}
U_a=\delta \frac{C}{N}\sum_{i=1}^{N}c_i-\delta ,
\end{equation}
where $\delta  \equiv\frac{n J N}{2(N-1)}$ is a constant independent of the group choice and thus does not affect  the expected collective choice $C$.

Defining a group pressure
\begin{equation}
P_g \equiv\delta  \frac{C}{N} ,
\end{equation}
I get
\begin{equation}
U_a^g=P_g\sum_{i=1}^{N}c_i-\delta ,
\end{equation}
where the  product $P_g c_i$ measures the conflict between the choice of the person $i$ and the group collective expected choice. A positive pressure  $P_g$ favors a positive choice $+1$, while $-1$ is associated to a negative pressure The conflict 
amplitude is given by  $P_g$. Rewriting $\sum_{i=1}^{N}c_i= C$ leads to 
\begin{equation}
U_a^g=\gamma  \frac{C^2}{N}-\delta ,
\end{equation}
which is maximum for $C^2=N ^2\rightarrow C=\pm N$. The anticipation process leads to a symmetry breaking along either +1 or -1 with equal probability. 

The result of spontaneous random symmetry breaking obtained from the simulations is thus recovered. Applying the anticipating effect separately to a series of network yield a polarization among the different networks. 

However, the technics gives no light about the dynamics reaching the equilibrium state. The related path and time are absent from the calculation. No information is provided on the protocol implementing individual updates, which drives the dynamics. 

Moreover, the exact solving of the Ising model at two dimensions and in zero field \cite{p3} has shown that mean treatment \cite{p2} provides a good qualitative description of the equilibrium state but the associated quantitative values are wrong. The same discrepancy holds true in higher dimensions and with a field as seen when applying renormalization groups technics \cite{}, which yield quite good quantitative results. But yet, all these analytical handlings of the Hamiltonian ignore the dynamics of repeated updates, which are required and taking place to reach the equilibrium.

\section{The dynamics of update to reach equilibrium}

To get ``exact" results about the equilibrium state of the Ising model at any dimensions for any temperature and possibly with a field requires the use of numerical simulations. And to reach that goal, an actual update dynamics must be defined and implemented. Monte Carlo algorithm is most used and efficient. 

However, it is worth to emphasize that the dynamics part is viewed as a mere necessary technical protocol, which must have no effect on the final outcome expected to be the equilibrium state of the model. To comply with this constraint a series of technical ``tricks" are incorporated to the algorithm to avoid artefacts.

For instance, when simulating a two-dimensional Ising model at zero temperature,  reaching the expected full symmetry breaking is found to be hindered by the formation of walls separating domains with opposite spin orientations. These domain walls appear randomly and stabilze at some size of the domains when the cost in energy to reverse a full domain to have it coalesce with a neighboring domain, becomes higher than the cost in energy of the domain walls.

To overcome this obstacle, an efficient tool is applying a very small external field during the first steps of the simulation before cancelling it when the equilibrium gets near. Annealed cooling is also used with turning the temperature to a non zero value allowing the sample to pass by the domain wall barriers and then putting back the temperature slowly to zero. However, the associated process requires long time to be implemented to avoid trapping in metastable states.

In addition, in physics the equilibrium state must be independent of the initial conditions, i.e., the actual value of the magnetization as well as the distribution of the initial random distribution of spins. The update scheme used in the Monte Carlo simulations must be also of no effect on the final equilibrium state. To reach that goal some simulations discard a large number of Monte Carlo steps before taking the following ones into account. Other simulations average thousands of different runs to wipe out the impact of initial conditions. At zero temperature the true equilibrium state is found to have all spins aligned in the same direction, irrespective of that direction. Some simulations end up with all spins at $+1$ and others at $-1$.The associated outcome is called a spontaneous symmetry breaking of the system.

\section{The requirement for a real material framework to monitor social update dynamics}

In a social network two interacting agents need to coordinate their choice to select the same choice and benefit a utility $J$ instead of a $-J$ cost  utility. However, that goal does not identify the common choice since both choices yield the same utility. The individual symmetry is thus recovered for the pair of agents and likewise, the choice of the pair is perceived as being selected randomly by an observer. As for the one person choice, once a choice is selected by the pair its symmetry is broken.

At this stage, it is reasonable to assume that self alignment of the pair is not too demanding to be achieved in terms of associated frame to monitor the selection of the choice to be made by both agents. However, monitoring the update dynamics of a N-person group with many interacting pairs becomes more complicated in terms of associated infrastructure. To elaborate further on how precisely individuals coordinate to reach the maximum utility of the full group is an open issue, which is out the scope of the present paper. 

At this stage of the modeling, pointing to the fact that an external frame is required to drive the dynamics, is both instrumental and sufficient to proceed. indeed, dealing with humans instead of atoms creates differences in the respective mechanisms driving the symmetry breaking although some overlaps hold. 

I thus claim that unlike the situation in physics, the update scheme used to implement the local dynamics driven by the interactions between connected agents, is not a purely technical issue whose effects must be discarded. On the contrary, I claim that the update procedure implementing the coordination of local individual choices is a major component of each specific human community unlike physics where the goal of a simulation is designing the appropriate and most efficient protocol to reach the equilibrium state.

In human systems reaching the true equilibrium state of a given community is rarely feasible since the implementation of the update dynamics has to be achieved via an established social structures. That is a complex and challenging task, which requires the building of a framework to support the update protocol between connected agents.

In this respect, the update procedure is not neutral for a social system. On this opposite, the actual procedure is instrumental and a signature of a specific social structure and often varies from one social network to another. These operating structures determine the time laps needed to achieved a symmetry breaking among all agents of a social community.  It also impacts the extent of the size at which a given symmetry breaking is achieved in a social community.

\section{Initial steps of symmetry breaking in Ising systems}

To illustrate my assumption of centrality of the update procedure in social systems, I revisit the main update schemes used in Monte Carlo simulations of the two-dimensional Ising model.  These schemes amounts to four, respectively, random, sequential, simultaneous and checkerboard updates.

Contrary to usual studies in statistical physics, here I explore finite size samples and the related impact of the details of the initial conditions. More precisely, I show that given the same collective aggregated choice $N_d=0$ ($N_+=N_-$ ), the actual distribution of initial individual choices impacts the final equilibrium state of the collective choice. I thus highlight the effect of the randomness of the choice distribution at  individual level, which is absent at the macroscopic level, on the associated final equilibrium state. 

I also focus on the initial steps of the Monte Carlo simulations, which are the relevant ones for a social application. That is the opposite of statistical physics where the initial steps Monte Carlo simulations are always discarded being deemed irrelevant. 

The same holds for the size of the sample which is an actual parameter of a social system but dismissed in physics. In other words, all the features which are qualified as artifact in statistical physics, are the relevant ones with respect of a social system.

Accordingly, I ran a series of simulations of samples with a fixed number of  $900$ spins localized at node of a two-dimensional grid of size $30 \times 30$. All initial conditions have an equal number $N_+=N_-$ of spins sharing respectively choices $+1$ and $-1$. The difference between the various samples lies in the actual random distribution of initial spin values. A peculiar distribution of spins is labelled with a seed value whose actual value is arbitrary and irrelevant. It only guarantee to recover exactly the same initial conditions in case of re-running the simulation.

All simulations include spins with nearest neighbor interactions at $T=0, J=1, H=0$. Almost all simulations include Periodic Boundary Conditions (PBC) unless otherwise specified with no PBC. In addition to the path used for the update of spins, the actual scheme to perform each spin update has to be chosen. Metropolis and the Glauber are the two major algorithms used in Monte Carlo simulations. 

Metropolis algorithm always flips the value of a spin when the move decreases the energy. That makes Metropolis algorithm
more efficient in reaching the equilibrium state most of the time. On the opposite, Glauber algorithm uses probabilities to shift a spin value when the shift increase the energy. However, here I am restricting the simulation to $T=0$, which makes the two algorithms identical and deterministic.

\subsection{Figure (\ref{r1})}

Fig. (\ref{r1}) shows the results of three simulations using a random update. Sub-cases (a, b, c) represent three different distributions (Seed = 10, 50, 70) of spins $\pm1$ (450 $+1$ in red, 450 $-1$ in blue), with the same initial value zero for their respective order parameters. Sub-case (a) shows a full symmetry breaking along $-1$, which is achieved after about 150 Monte Carlo steps (Seed = 10). Sub-case (b) shows a full symmetry breaking along $+1$ after less than 100 Monte Carlo steps (Seed = 70). Sub-case (c) shows a full symmetry breaking along $-1$ after about 750 Monte Carlo steps (Seed = 50). However, in this case, the order parameter has been positive during almost 500 Monte Carlo first steps before starting to turn negative to eventually reach a full negative symmetry breaking. Sub-cases (d, e, f) show the respective initial distribution of the three samples with zero order parameter associated to (a, b, c). Sub-cases (g, j), (h, k), (i, l) show related intermediate snapshots  toward full symmetry breaking for the three samples (d, e, f).

\subsection{Figure (\ref{r2})}

Fig. (\ref{r2}) shows the results of two simulations using a random update with initial distributions of spins (Seed = 40, 90) different than in  Fig. (\ref{r1}) (Seed = 10, 50, 70). However, contrary to Fig. (\ref{r1}), these two distributions lead to final states with no full symmetry breaking as  exhibited in Sub-cases (a, c). Indeed two domains of opposite distributions are found in the final equilibrium state as seen in Sub-cases (b, d). In both Sub-cases, the domain are of different sizes (magnetization -0.0667 versus 0.267).

\subsection{Figure (\ref{r3})}

Fig. (\ref{r3}) shows the results of two simulations using a random update with initial distributions of spins (Seed = 10, 40) as in Sub-case a  in Fig. (\ref{r1}) (Seed = 10) and Sub-case  a  in Fig. (\ref{r2}) (Seed = 40). However, contrary to Figs. (\ref{r1}, \ref{r2})), these two simulations do not include Periodic Boundary Conditions (PBC). The related results are very different with a full symmetry breaking along $+1$ instead of $-1$ after about 400 Monte Carlo steps instead of 180 and two coexisting domains of different sizes (magnetization -0.533) instead of (magnetization -0.0667) after about 300 Monte Carlo steps instead of 150.

\subsection{Figure (\ref{s1})}

Fig. (\ref{s1}) shows the results of three simulations in sub-cases a, b, c with identical initial conditions (Seed = 10, 50, 70) than in Fig. (\ref{r1}) but using sequential update instead of random update. The sequential update leads to very different results from Fig. (\ref{r1}) with respectively a full symmetry breaking along $+1$ instead of $-1$ after about only 15 Monte Carlo steps instead of 180, a full symmetry breaking along $-1$ instead of $+1$ after about only 10 Monte Carlo steps instead of 90, and two coexisting domains of different sizes (magnetization 0.0933) instead of a full symmetry breaking along $-1$ (magnetization -1) after about 20 Monte Carlo steps instead of about 700.

\begin{figure}[!htb]
\centering
\subfigure[]{\includegraphics[width=0.40\textwidth]{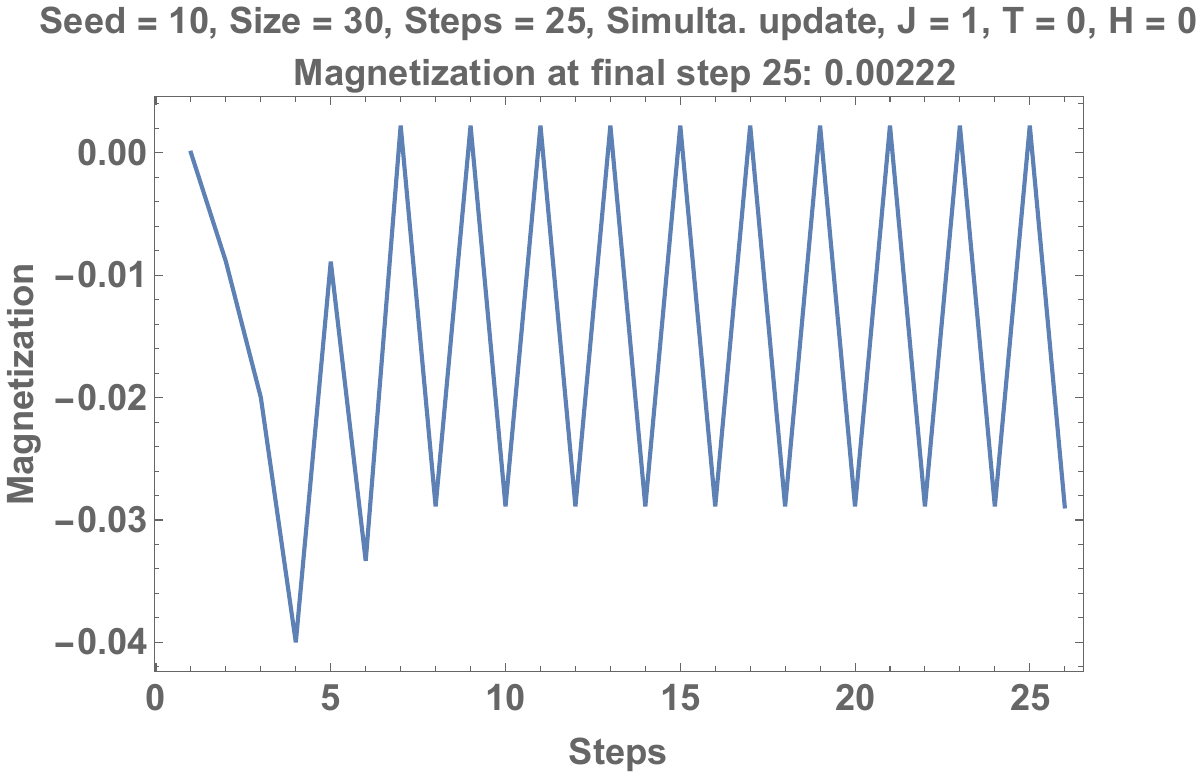}} \hspace{0.002\textwidth}
\subfigure[]{\includegraphics[width=0.26\textwidth]{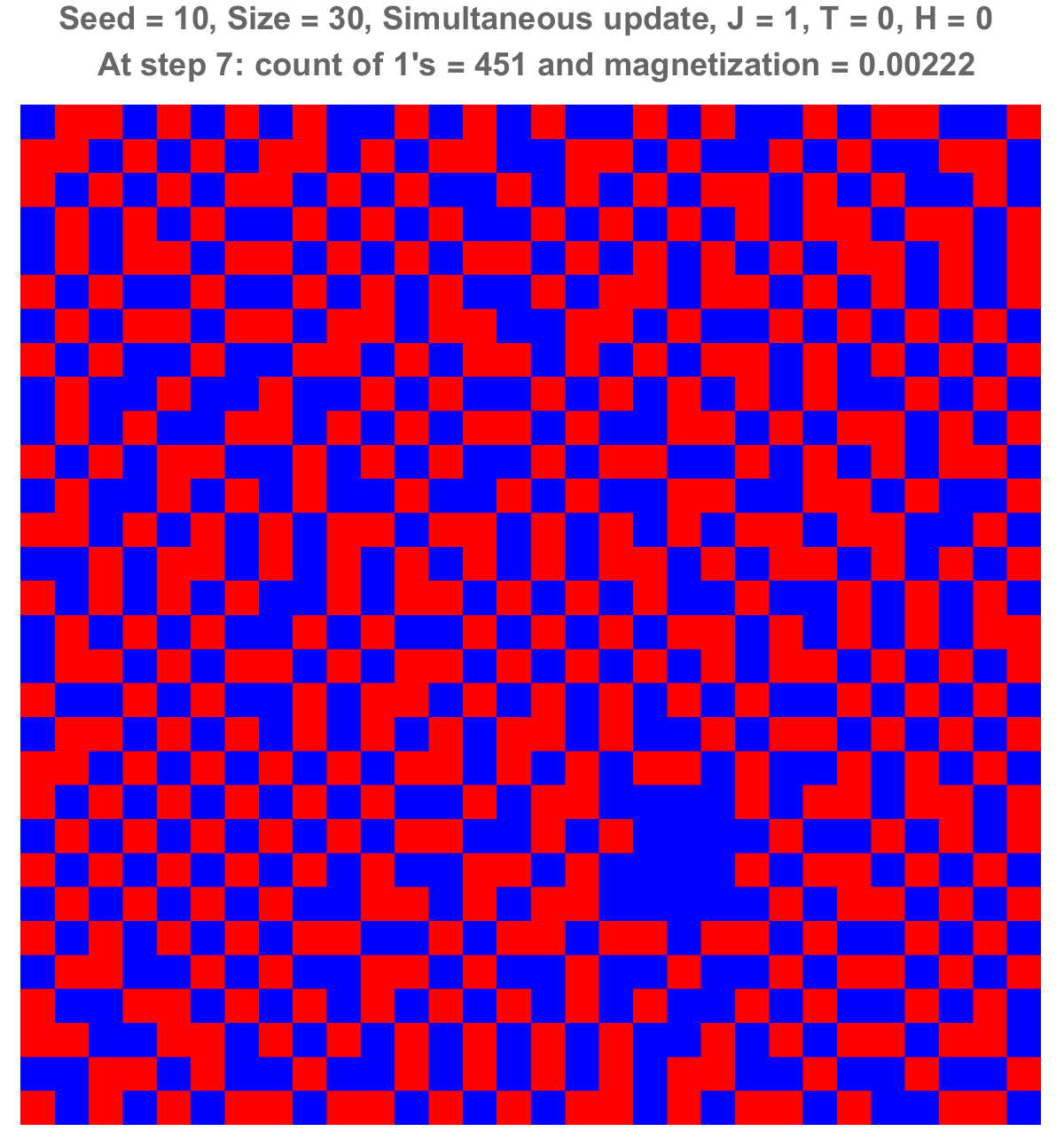}} \hspace{0.002\textwidth}
\subfigure[]{\includegraphics[width=0.26\textwidth]{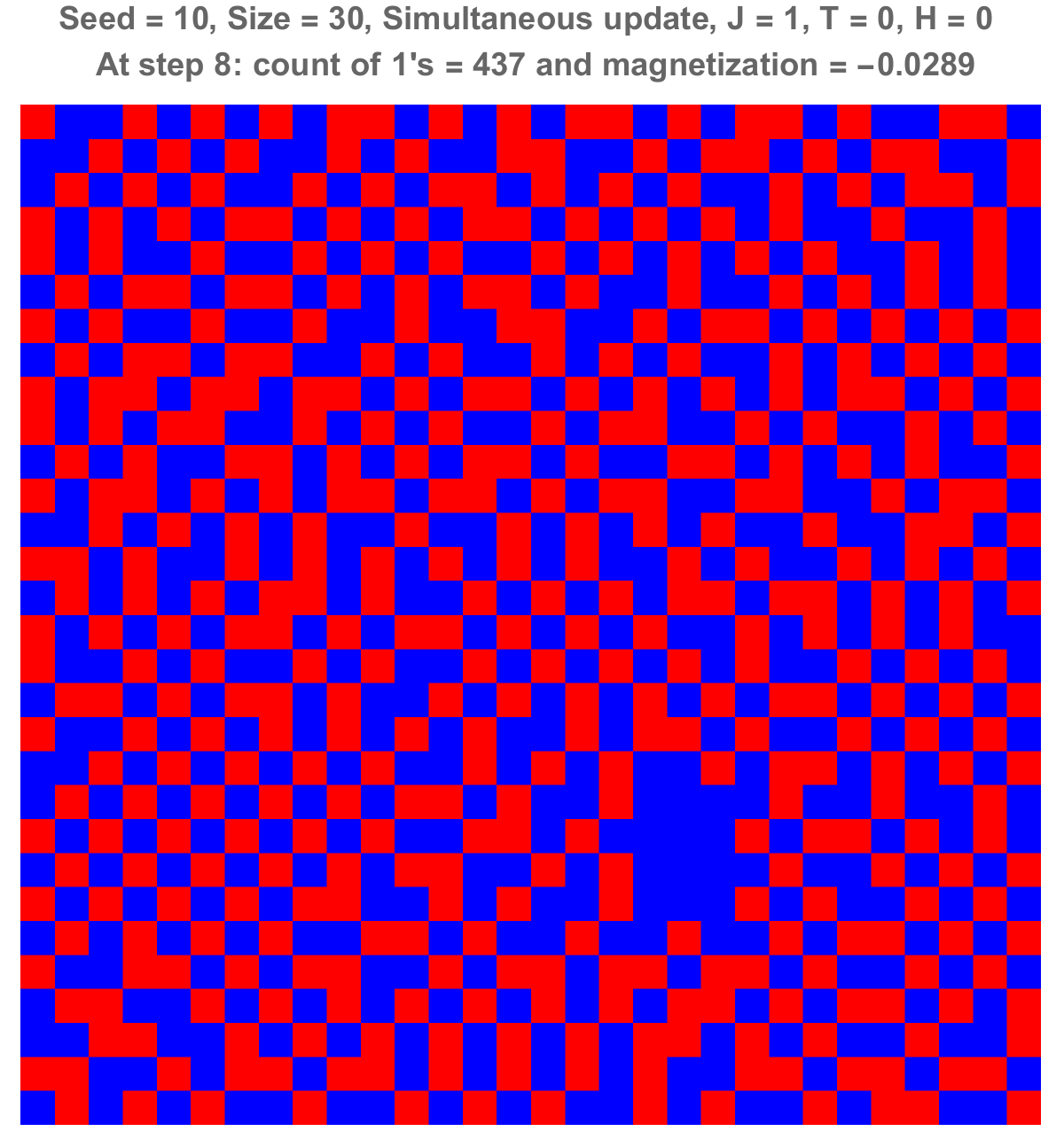}}
\\[0.03\textwidth]\
\subfigure[]{\includegraphics[width=0.40\textwidth]{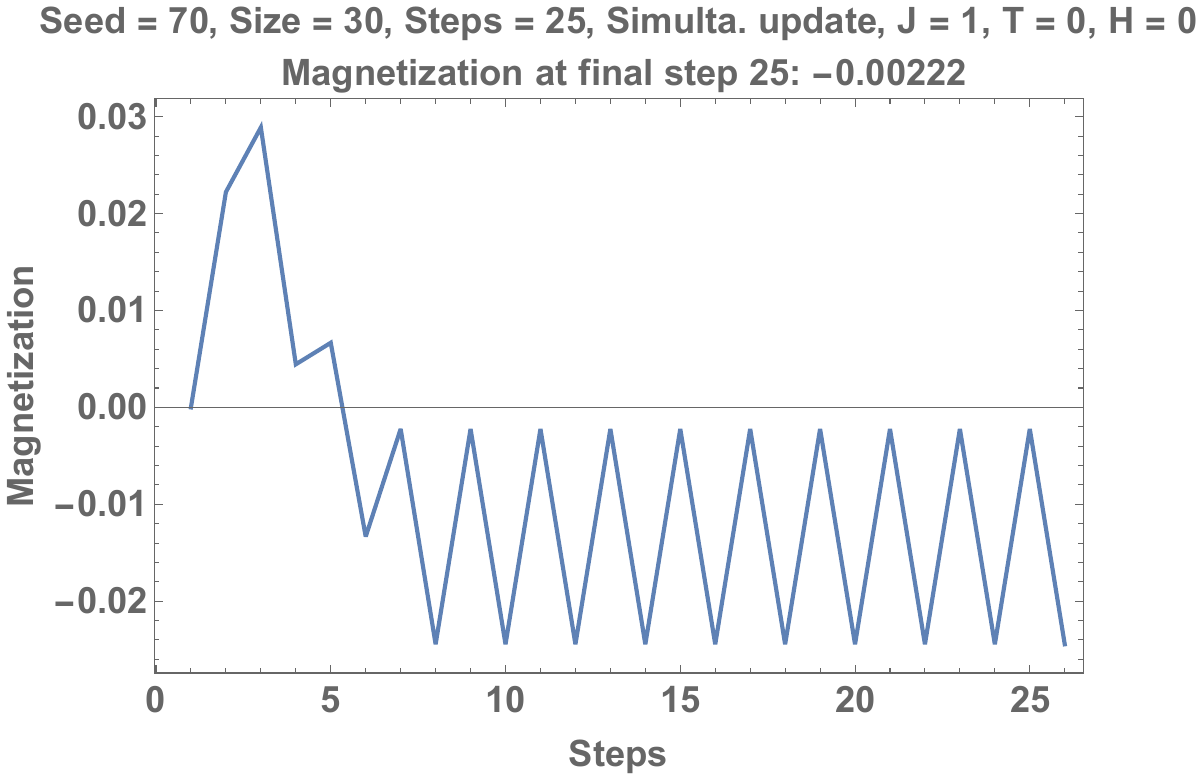}} \hspace{0.002\textwidth}
\subfigure[]{\includegraphics[width=0.26\textwidth]{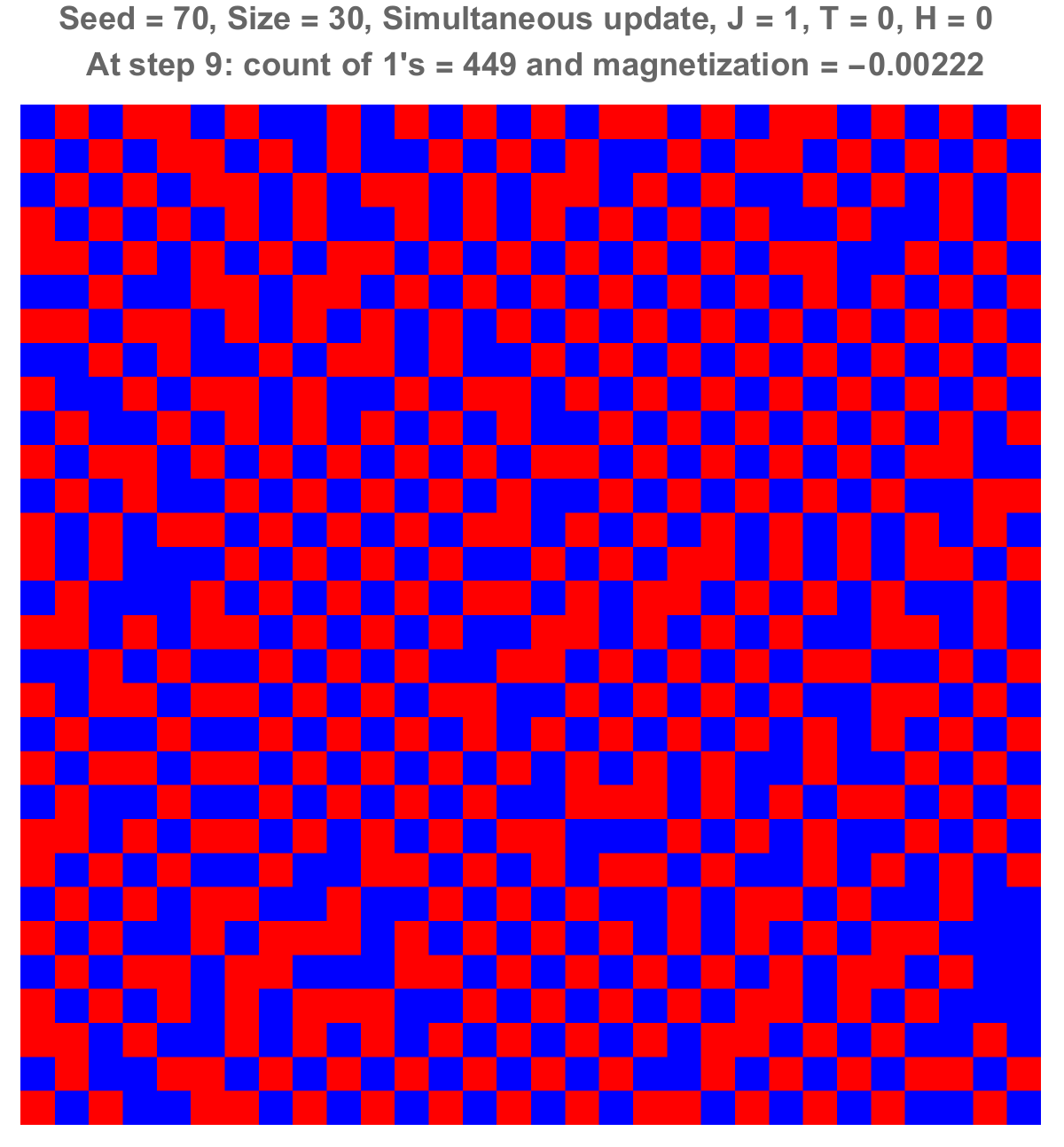}} \hspace{0.002\textwidth}
\subfigure[]{\includegraphics[width=0.26\textwidth]{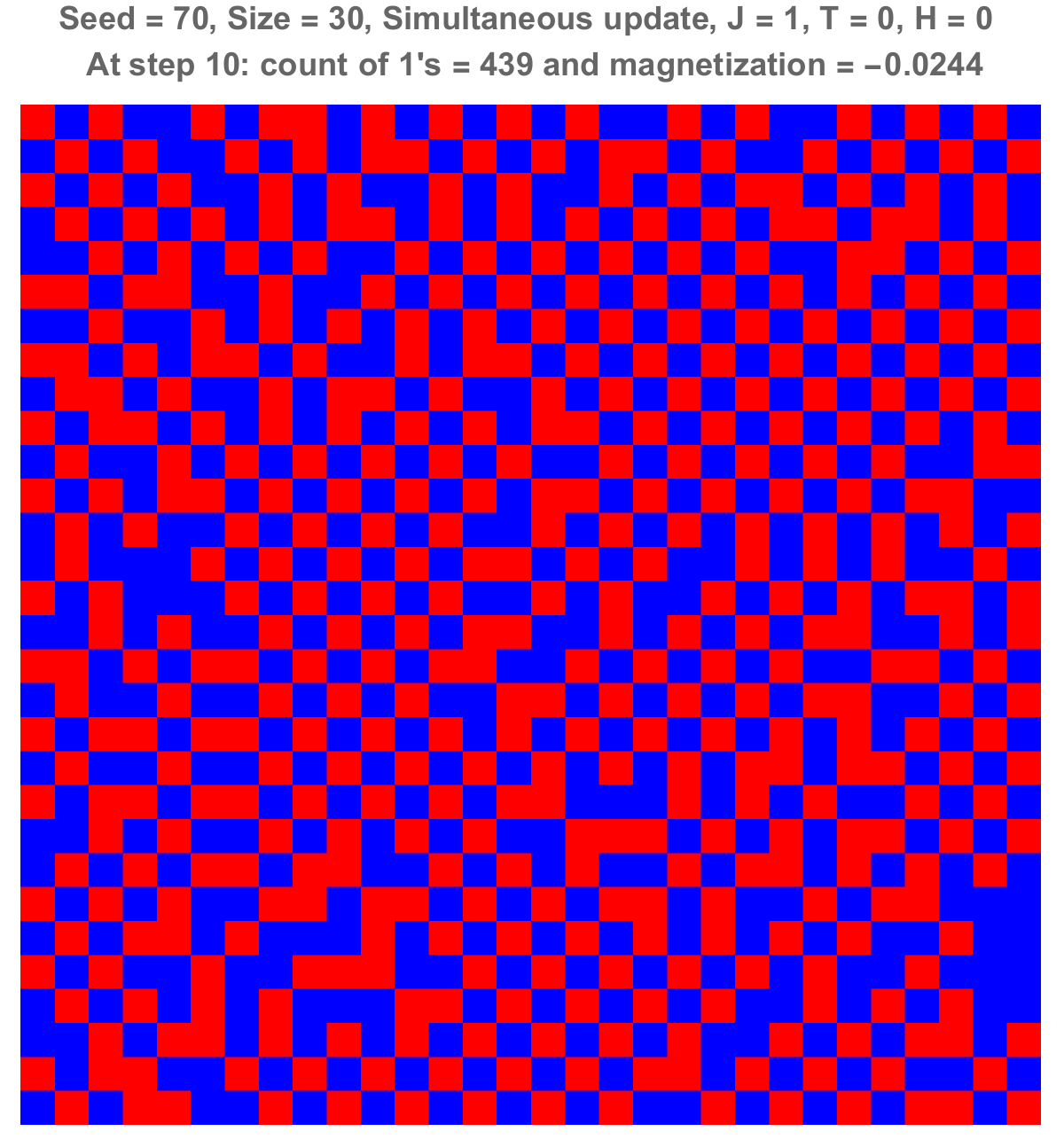}}
\caption{Results of two simulations in sub-cases a, b with identical initial conditions (Seed = 10, 70) using simultaneous update. The system gets trapped very quickly after only a few Monte Carlo steps as seen in the Figure. Both cases lead to periodic shift between two fixed configurations.}
\label{si1}
\end{figure}

\subsection{Figure (\ref{si1})}

Fig. (\ref{si1}) shows the results of two simulations in sub-cases a, b with identical initial conditions (Seed = 10, 70) using simultaneous update. The system gets trapped very quickly after only a few Monte Carlo steps as seen in the Figure. Both cases lead to periodic shift between two fixed configurations.

 \subsection{Figure (\ref{c1})}

Fig. (\ref{c1}) shows the results of a two step simultaneous update denoted checkerboard update. All sites of each sub-lattice are updated simultaneously one after the other sequentially. Three simulations (sub-cases a, b, c ) are performed with identical initial conditions (Seed = 10, 50, 70) as in Fig. (\ref{r1}) but using checkerboard update instead of random update. The checkerboard update leads to very different results from Fig. (\ref{r1}) with respectively a full symmetry breaking unchanged along $-1$ but now after about only 15 Monte Carlo steps instead of 180, two coexisting domains of different sizes (magnetization 0.253) instead of a full symmetry breaking along $+1$ after about only 15 Monte Carlo steps instead of 90, and two coexisting domains of different sizes (magnetization 0.142) instead of a full symmetry breaking along $-1$ (magnetization -1) after about 20 Monte Carlo steps instead of about 700.

Sub-cases (d, e, f) exhibits the same simulations as in sub-cases (a, b, c) but without Periodic Boundary Conditions (PBC). The associated results are slightly different with respectively still a full symmetry breaking along $-1$ but with about 20 Monte Carlo steps instead of 15, a full symmetry breaking along $+1$ instead of two coexisting domains with similar numbers of Monte Carlo steps, still two coexisting domains of different sizes with magnetization 0.133 instead of magnetization 0.142.

\begin{figure}
\centering
\subfigure[]{\includegraphics[width=0.32\textwidth]{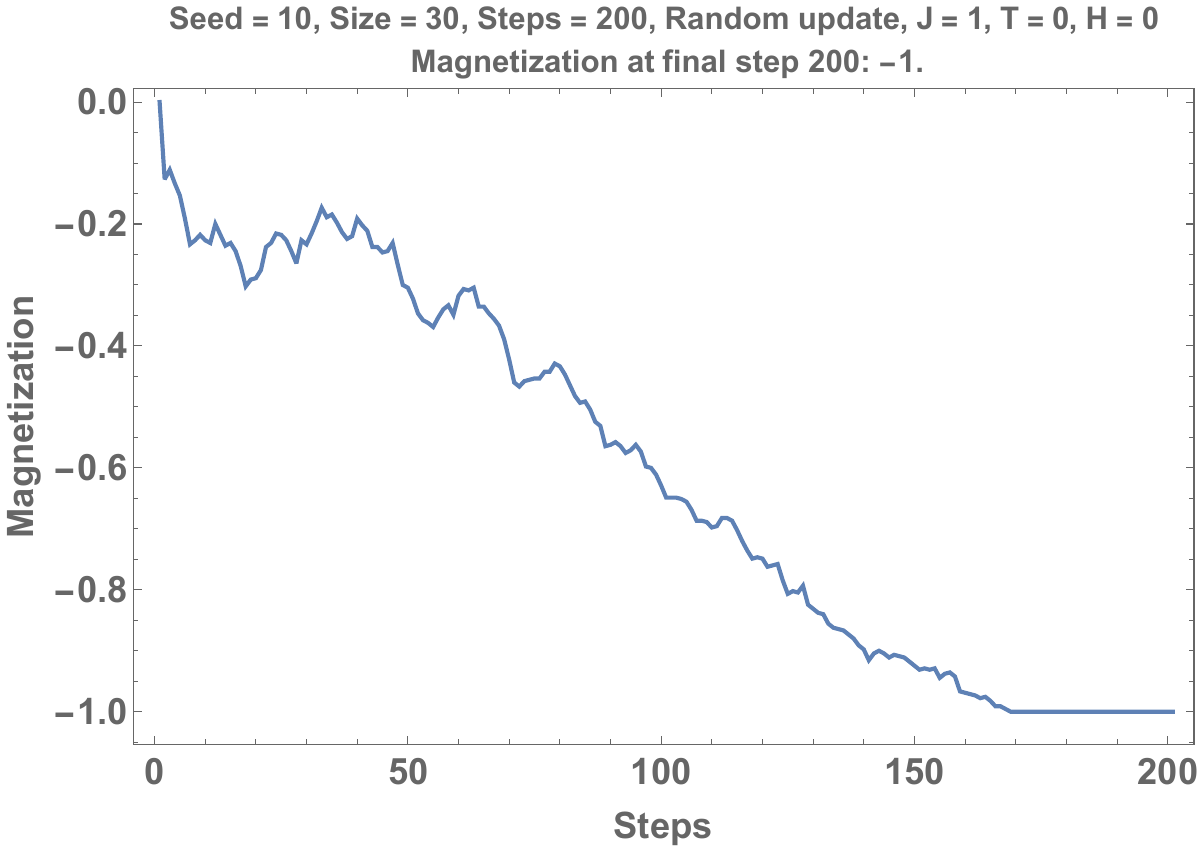}} \hspace{0.002\textwidth}
\subfigure[]{\includegraphics[width=0.32\textwidth]{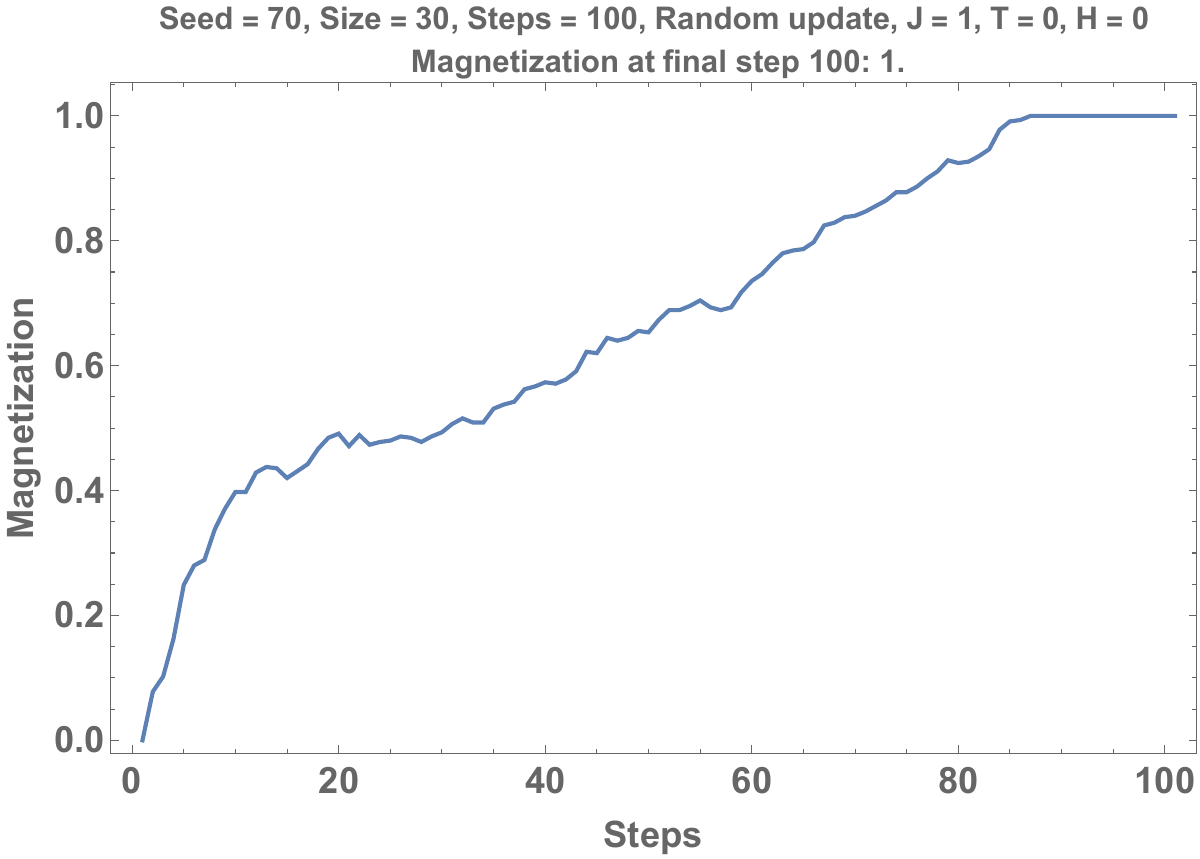}} \hspace{0.002\textwidth}
\subfigure[]{\includegraphics[width=0.32\textwidth]{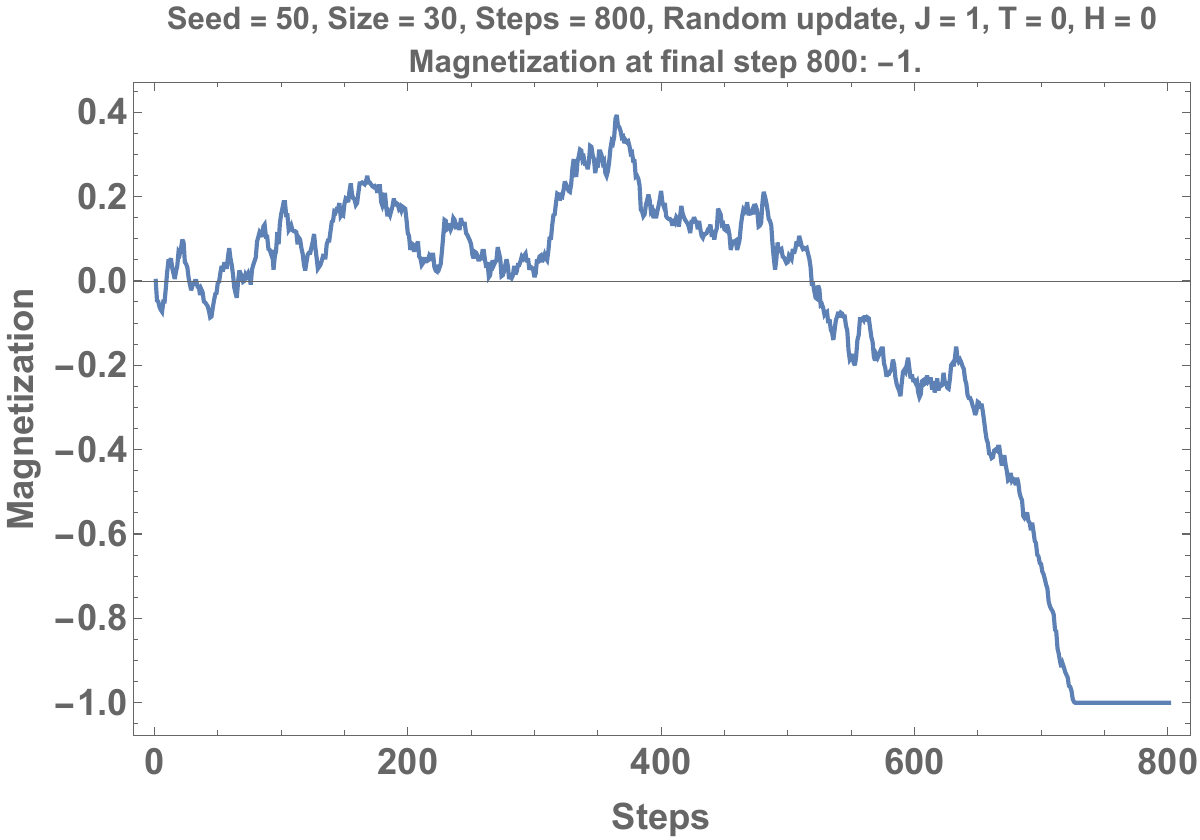}}
\\[0.03\textwidth]
\subfigure[]{\includegraphics[width=0.32\textwidth]{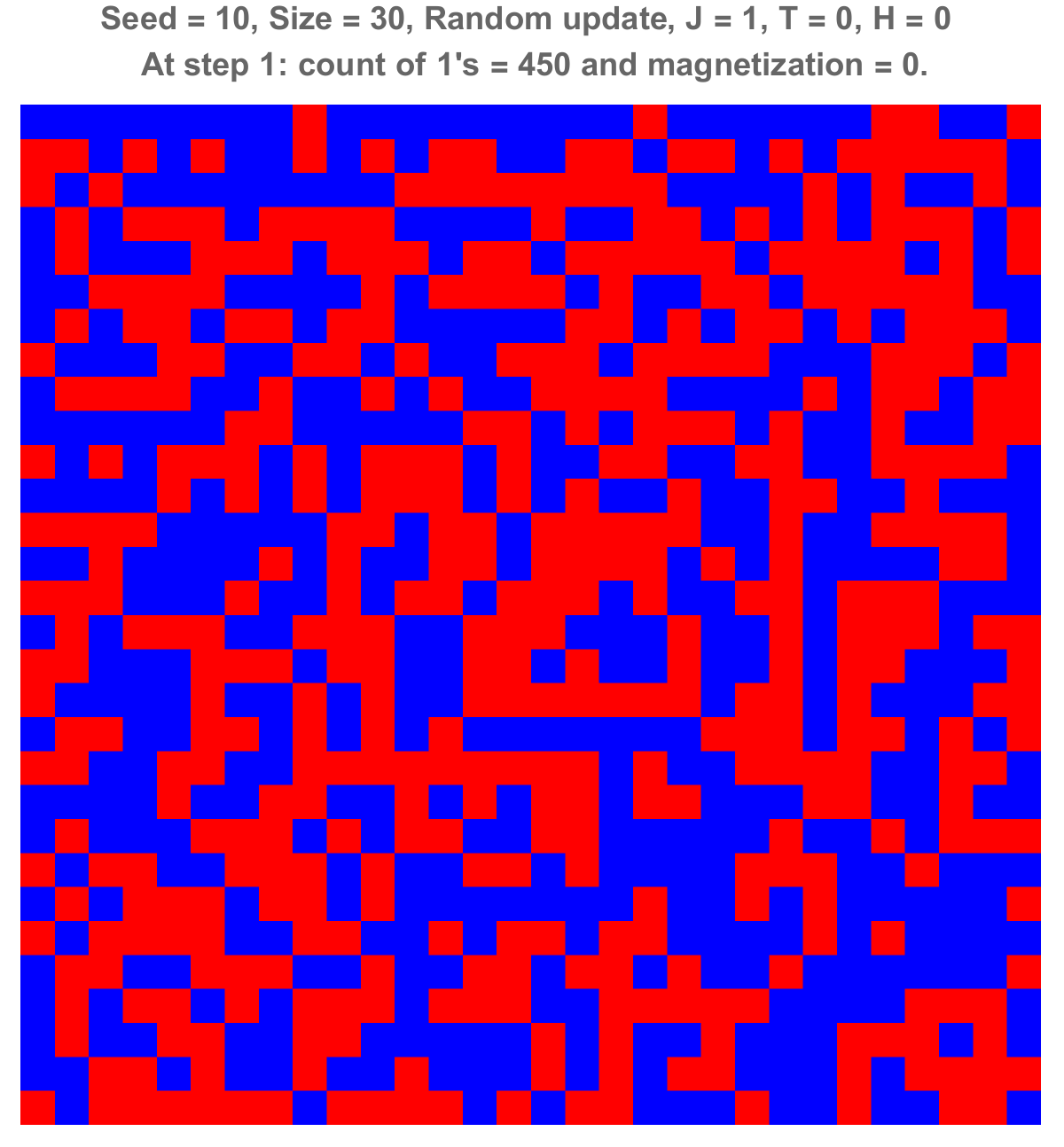}} \hspace{0.002\textwidth}
\subfigure[]{\includegraphics[width=0.32\textwidth]{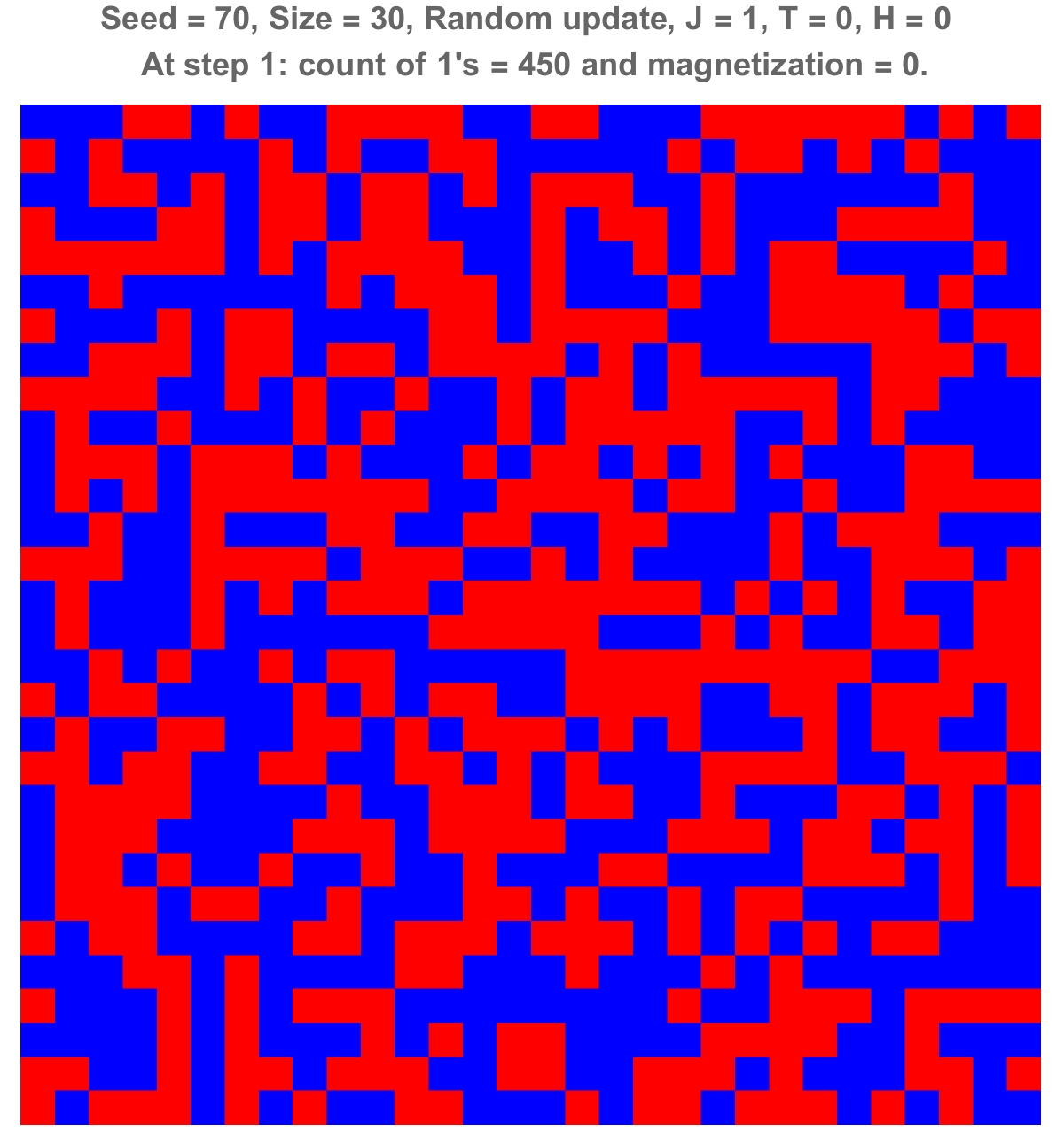}} \hspace{0.002\textwidth}
\subfigure[]{\includegraphics[width=0.32\textwidth]{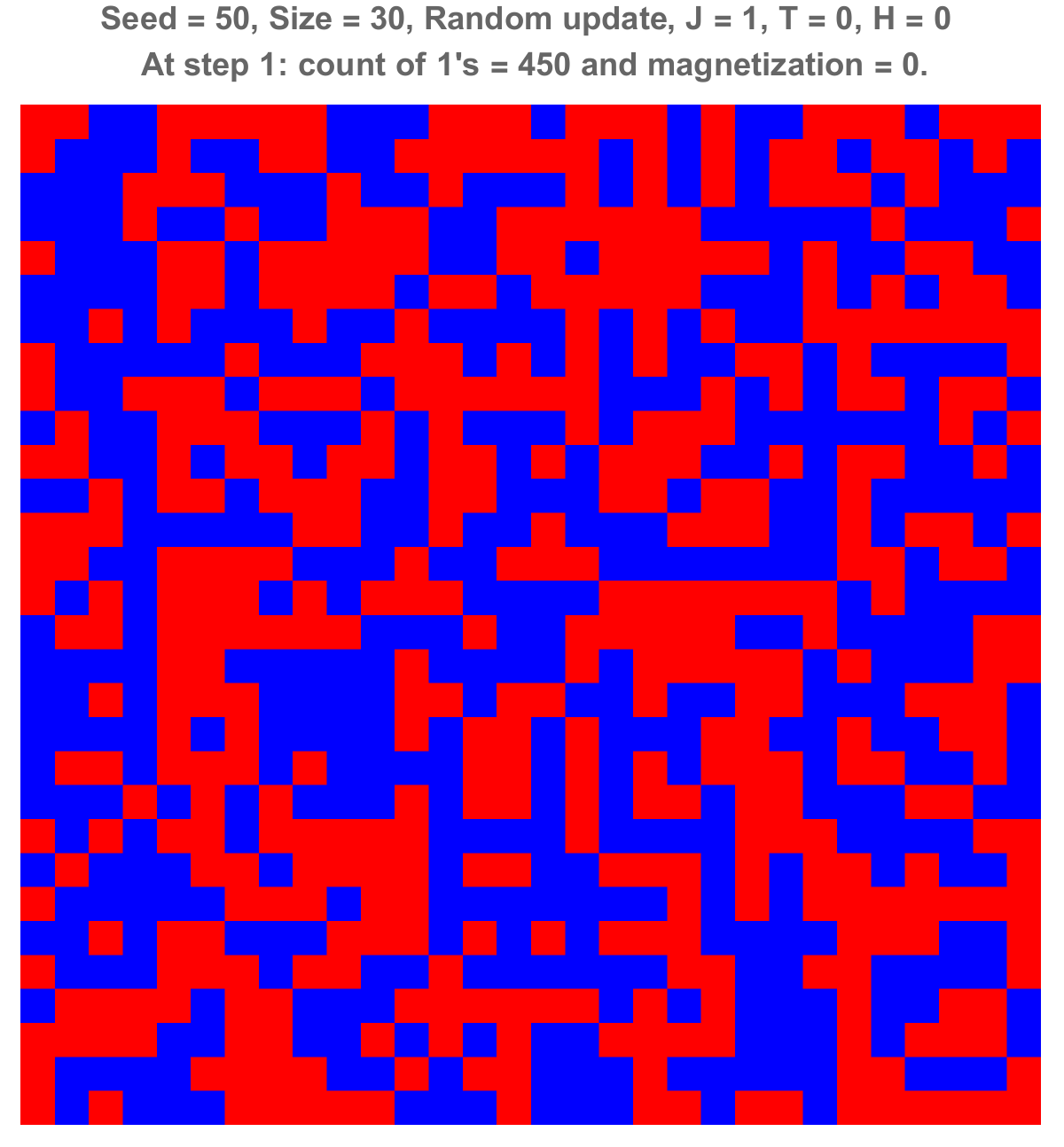}}
\\[0.03\textwidth]
\subfigure[]{\includegraphics[width=0.32\textwidth]{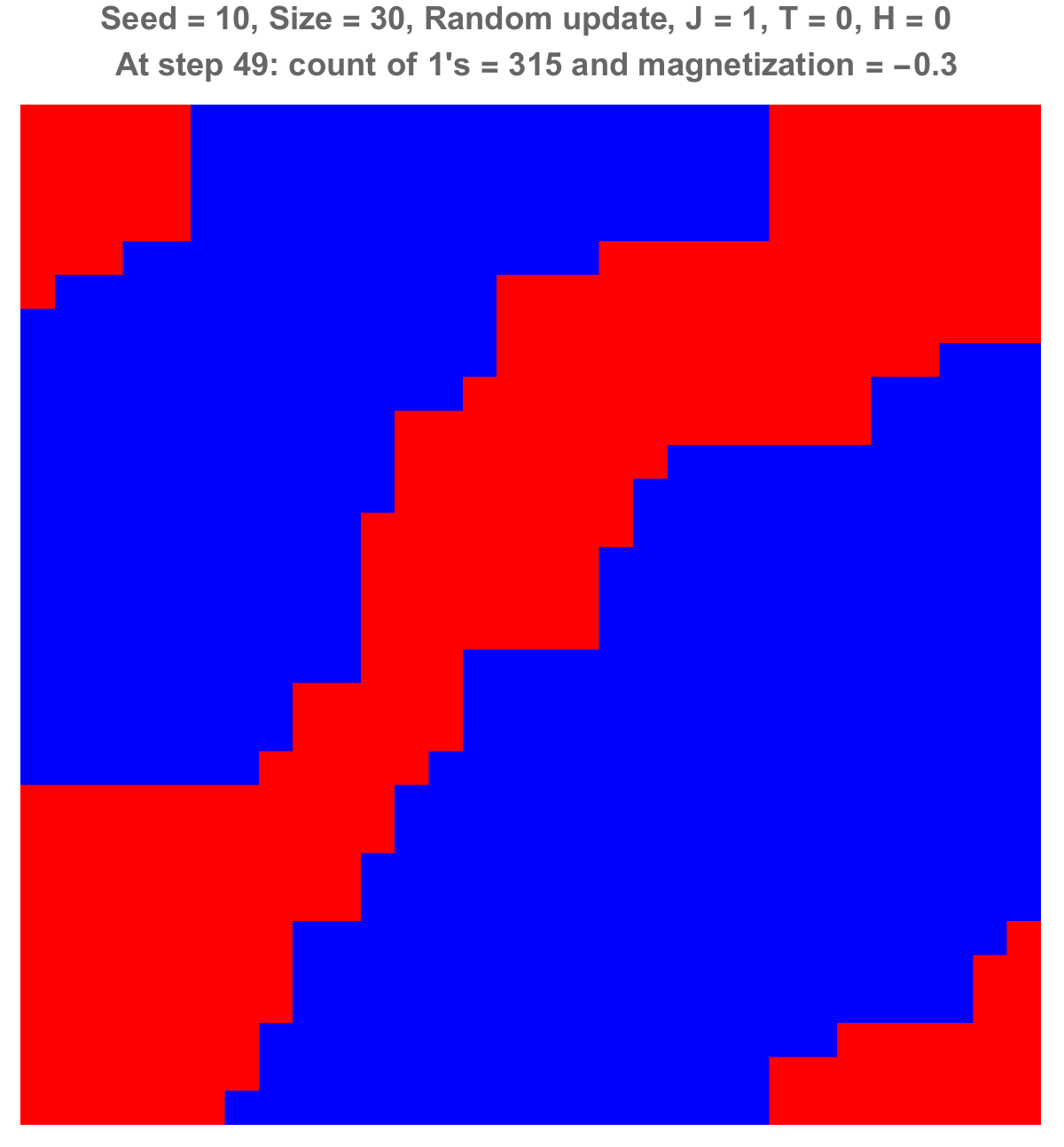}} \hspace{0.002\textwidth}
\subfigure[]{\includegraphics[width=0.32\textwidth]{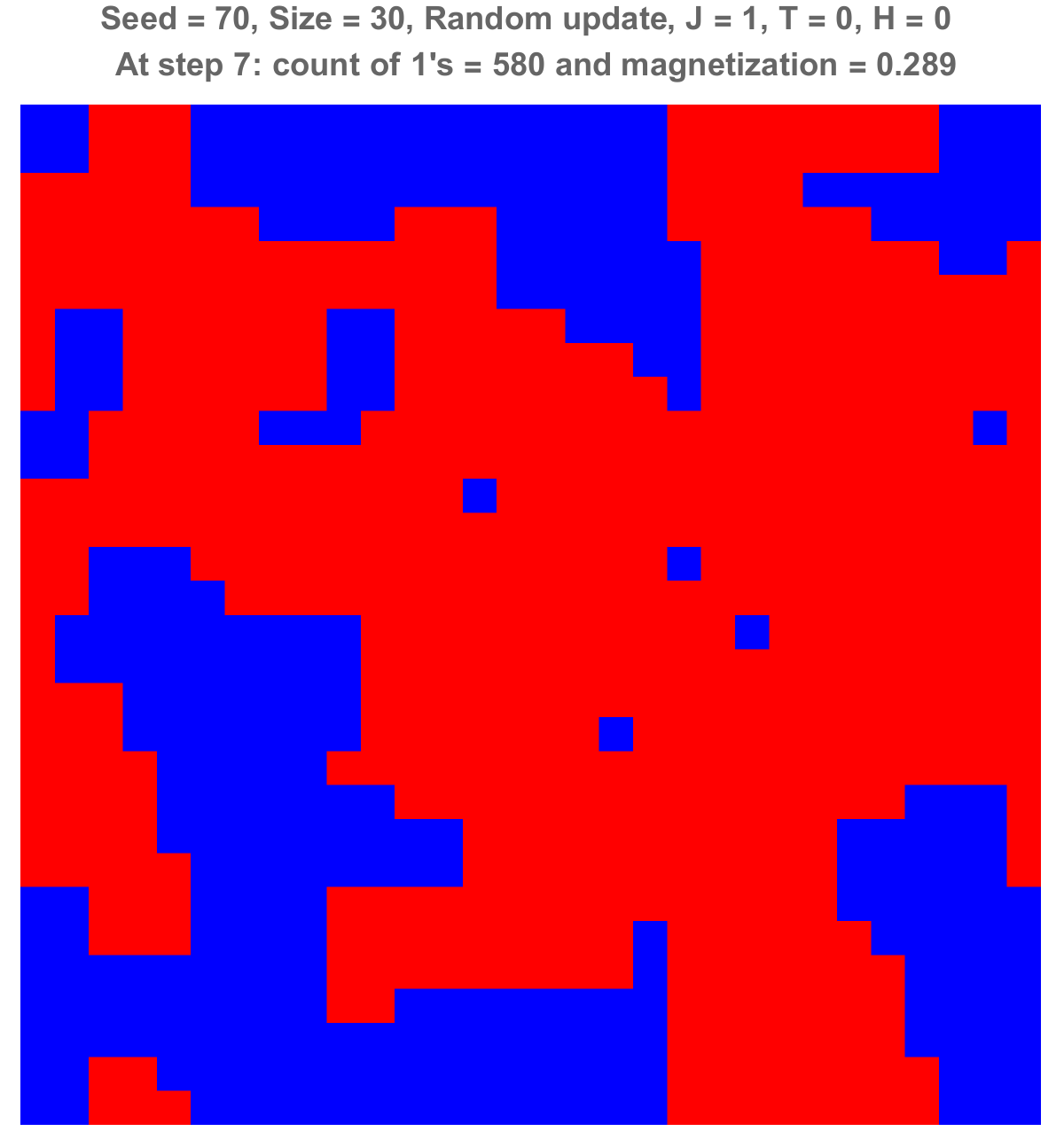}} \hspace{0.002\textwidth}
\subfigure[]{\includegraphics[width=0.32\textwidth]{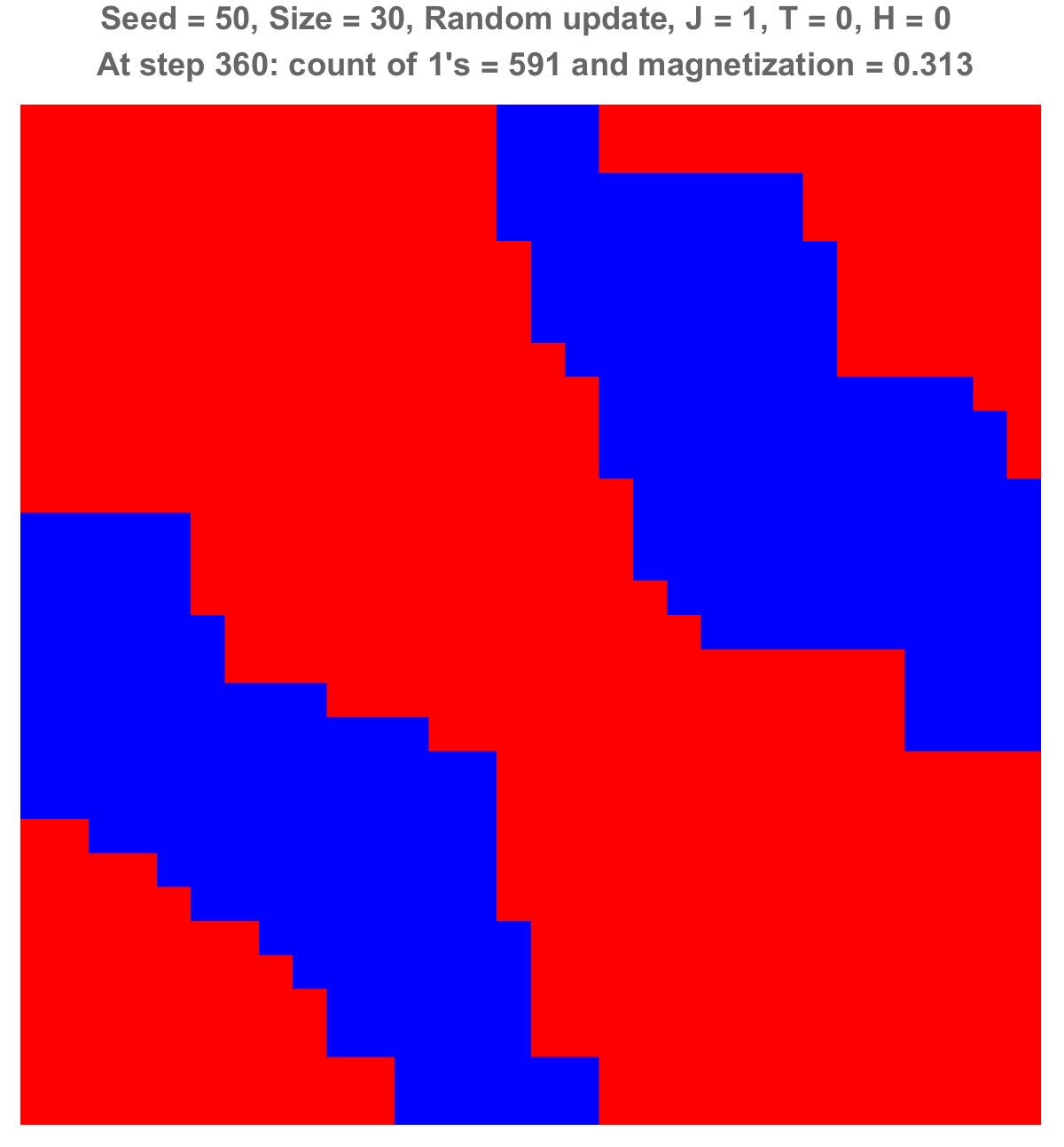}}
\\[0.03\textwidth] 
\subfigure[]{\includegraphics[width=0.32\textwidth]{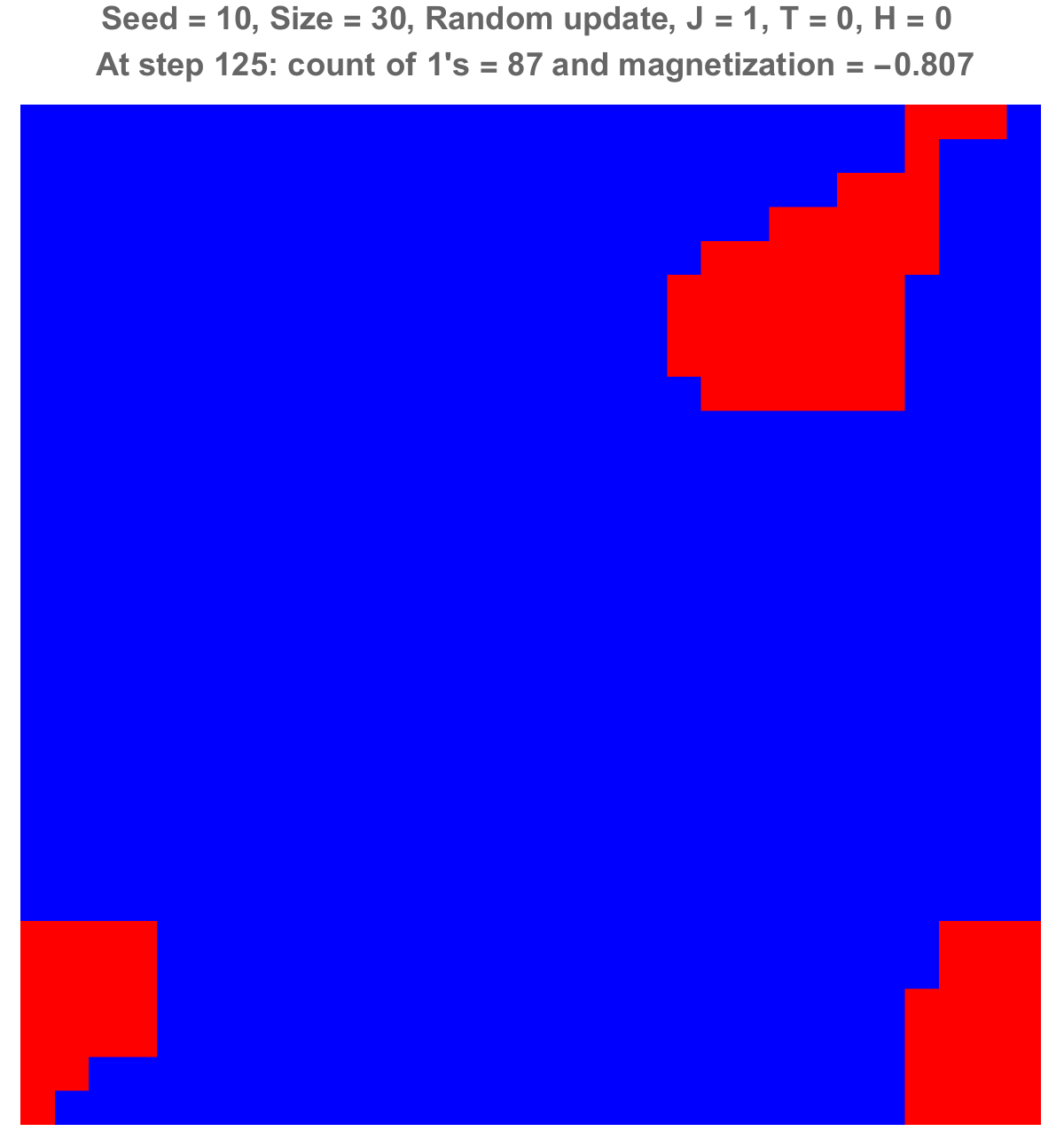}} \hspace{0.002\textwidth}
\subfigure[]{\includegraphics[width=0.32\textwidth]{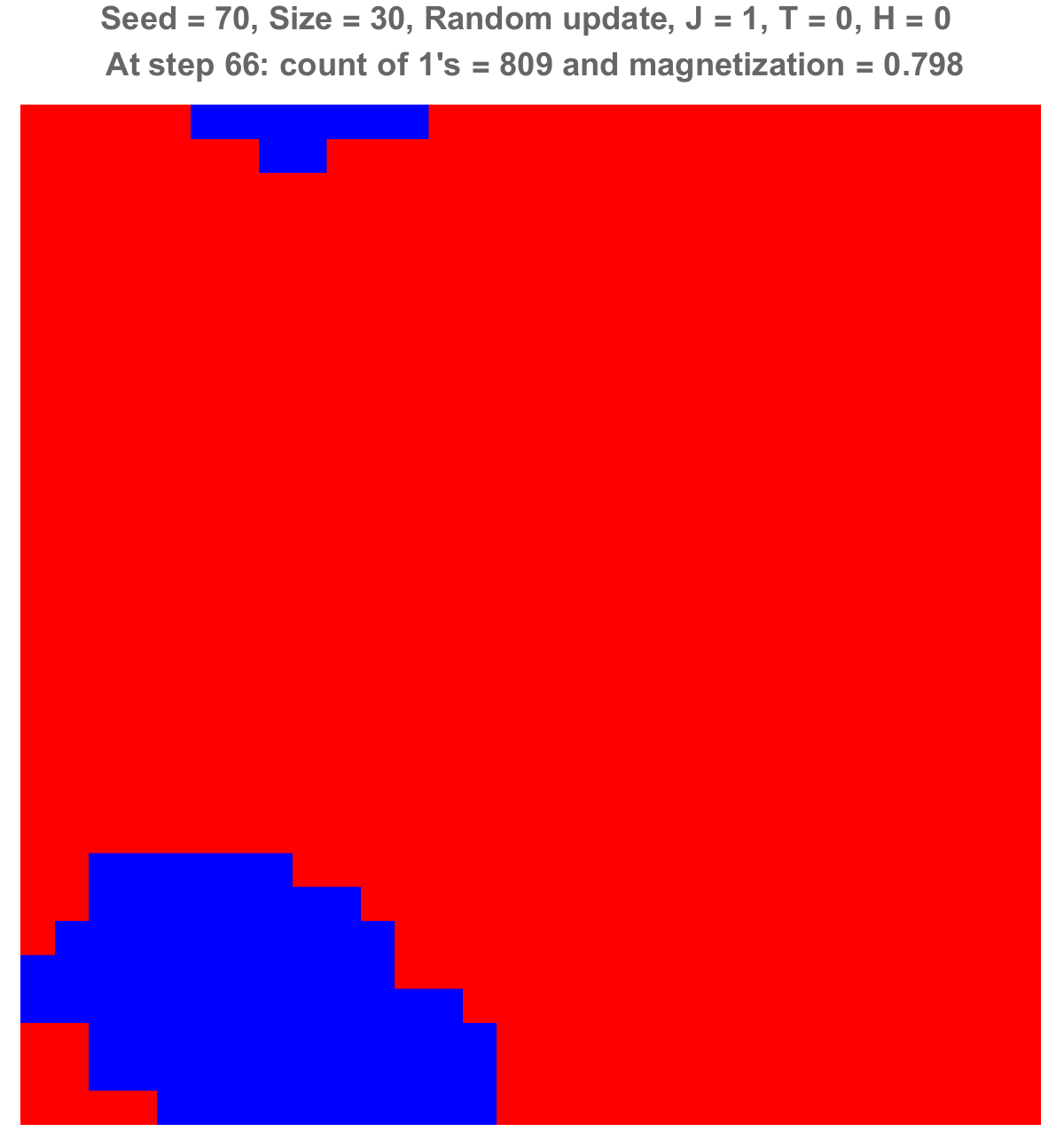}} \hspace{0.002\textwidth}
\subfigure[]{\includegraphics[width=0.32\textwidth]{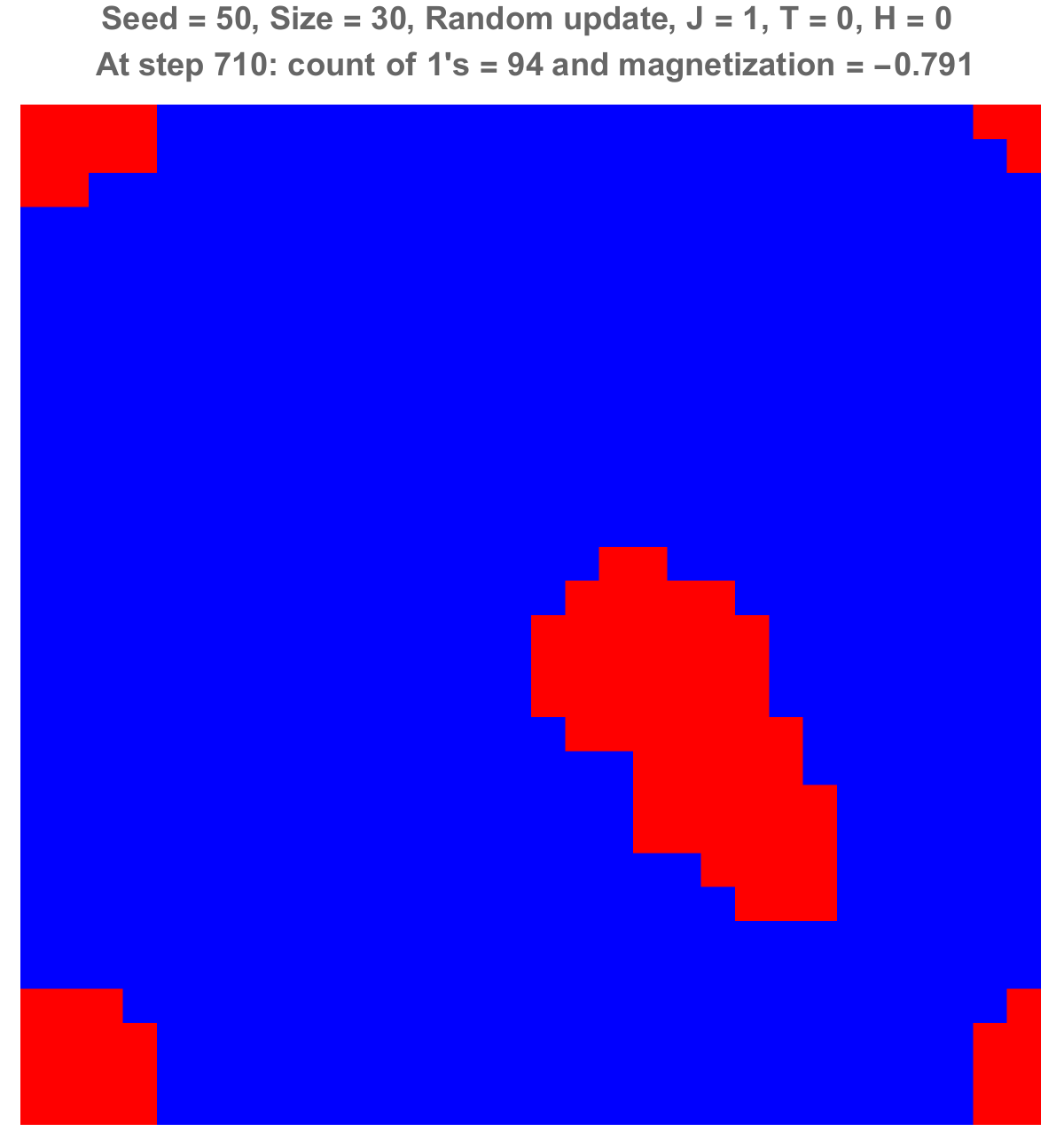}}
\\[0.03\textwidth] 
\end{figure}

\newpage 

\noindent\captionof{figure}{Results of three simulations using a random update. Sub-cases (a, b, c) represent three different distributions (Seed = 10, 50, 70) of spins $\pm1$ (450 $+1$ in red, 450 $-1$ in blue), with the same initial value zero for their respective order parameters. Sub-case (a) shows a full symmetry breaking along $-1$, which is achieved after about 150 Monte Carlo steps (Seed = 10). Sub-case (b) shows a full symmetry breaking along $+1$ after less than 100 Monte Carlo steps (Seed = 70). Sub-case (c) shows a full symmetry breaking along $-1$ after about 750 Monte Carlo steps (Seed = 50). However, in this case, the order parameter has been positive during almost 500 Monte Carlo first steps before starting to turn negative to eventually reach a full negative symmetry breaking. Sub-cases (d, e, f) show the respective initial distribution of the three samples with zero order parameter associated to (a, b, c). Sub-cases (g, j), (h, k), (i, l) show related intermediate snapshots  toward full symmetry breaking for the three samples (d, e, f).
}
\label{r1}

\begin{figure}
\centering
\subfigure[]{\includegraphics[width=0.45\textwidth]{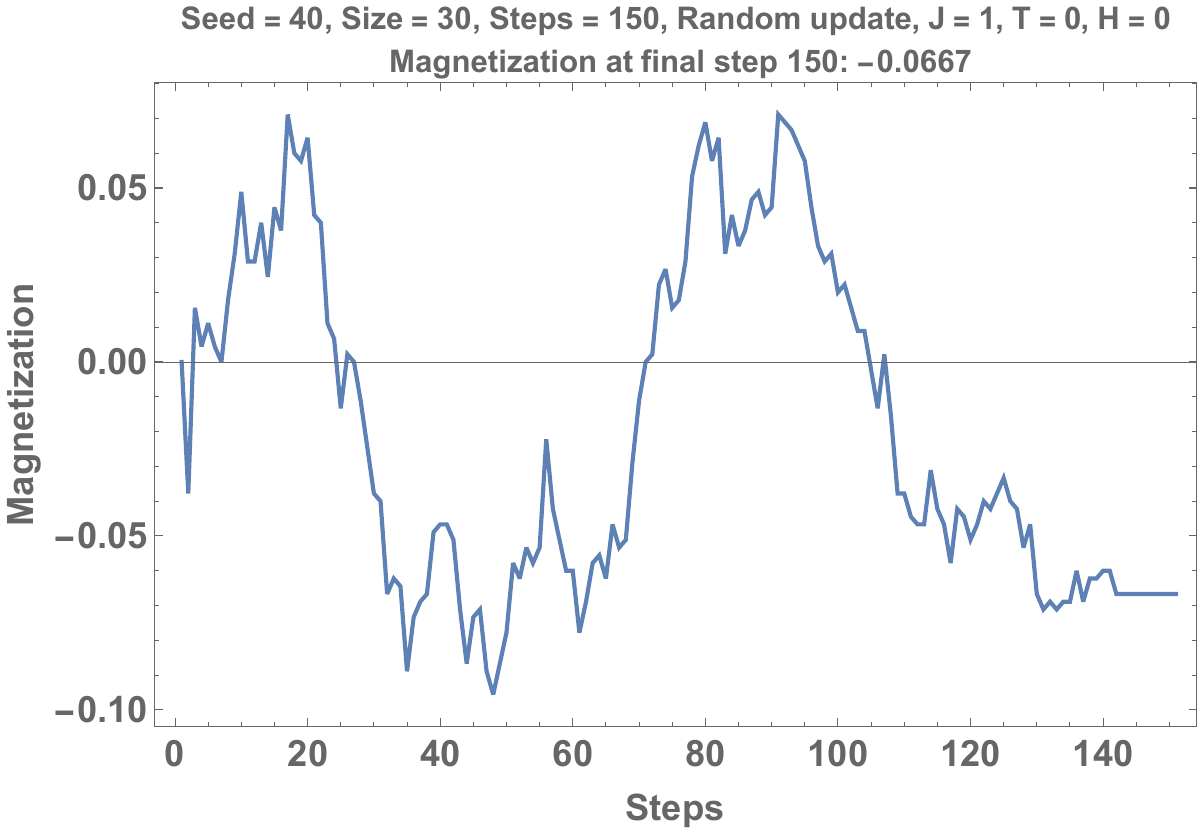}} \hspace{0.002\textwidth}
\subfigure[]{\includegraphics[width=0.32\textwidth]{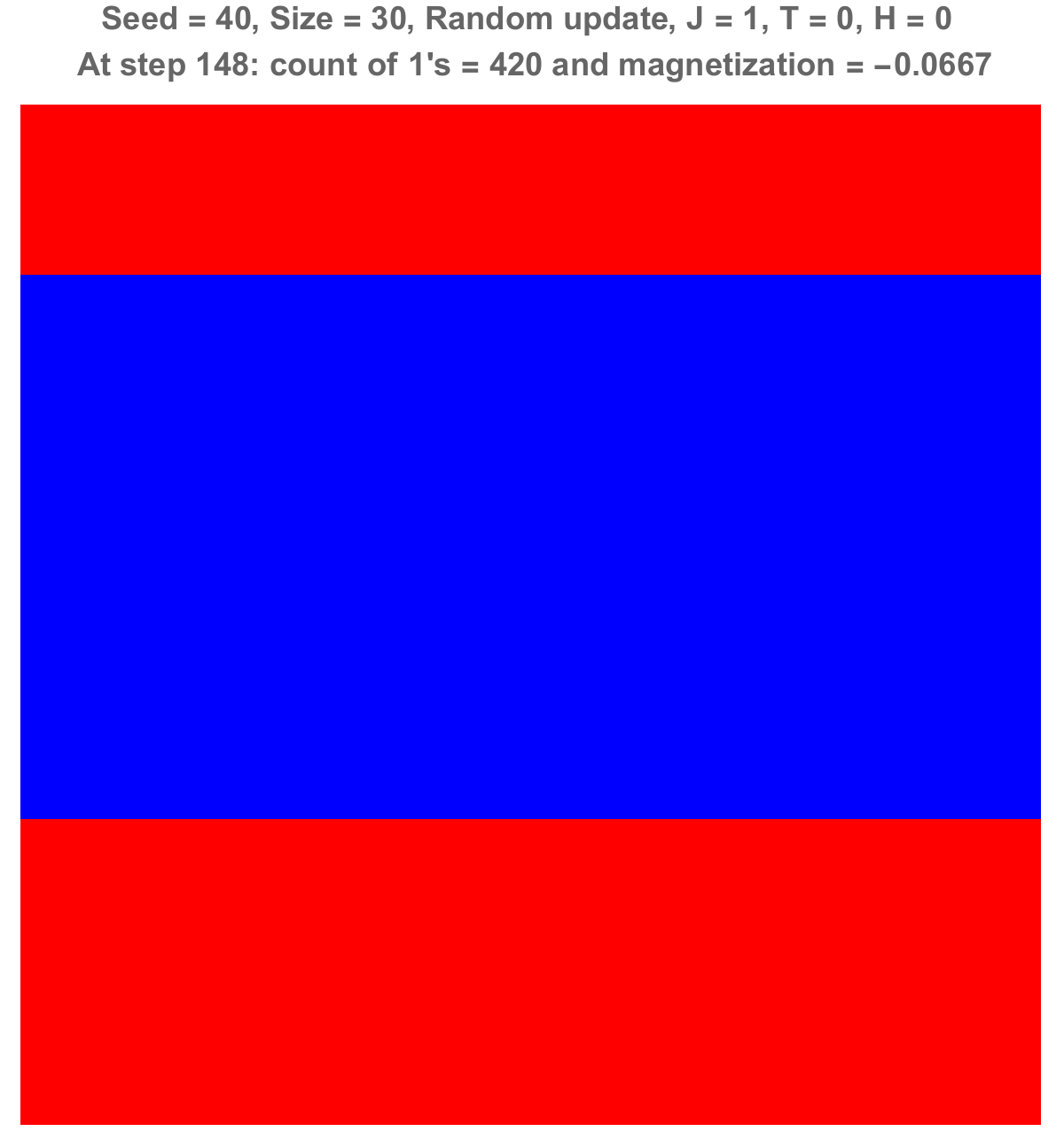}} \hspace{0.002\textwidth}
\\[0.03\textwidth]
\subfigure[]{\includegraphics[width=0.45\textwidth]{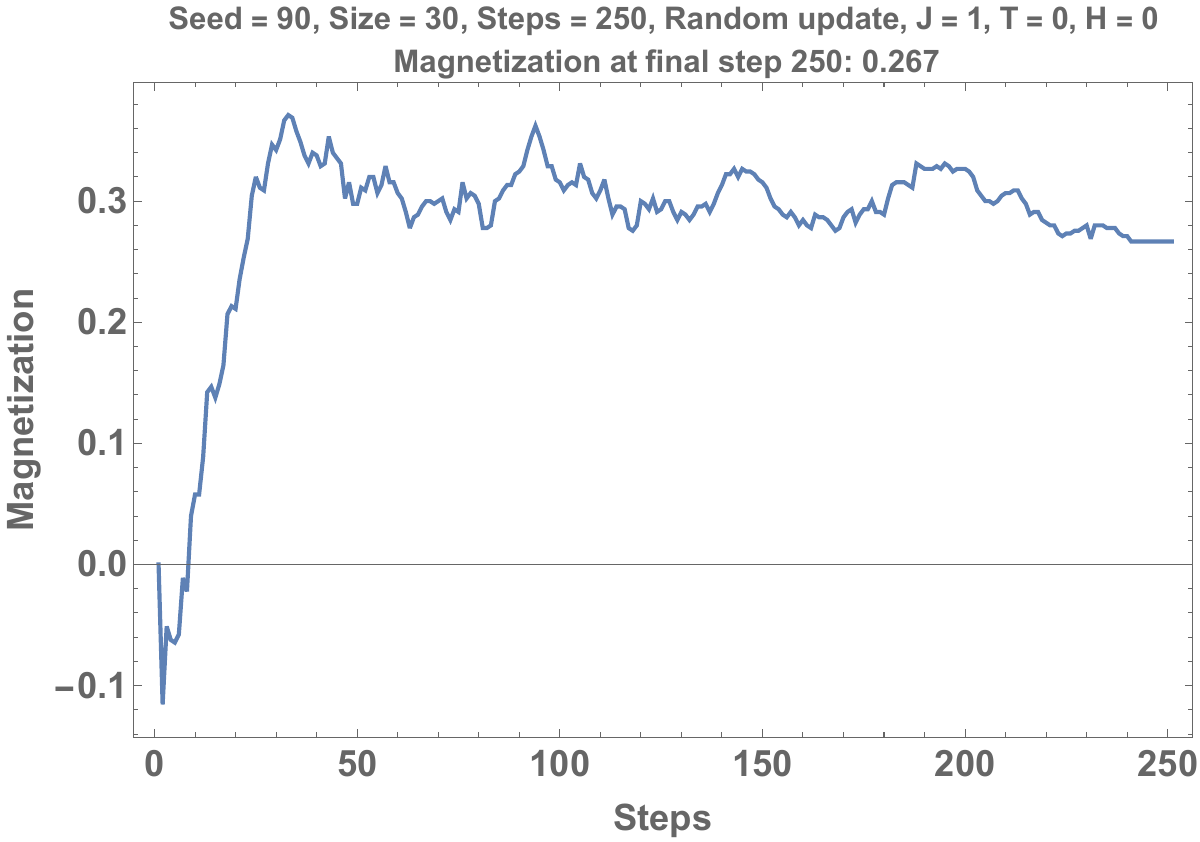}} \hspace{0.002\textwidth}
\subfigure[]{\includegraphics[width=0.32\textwidth]{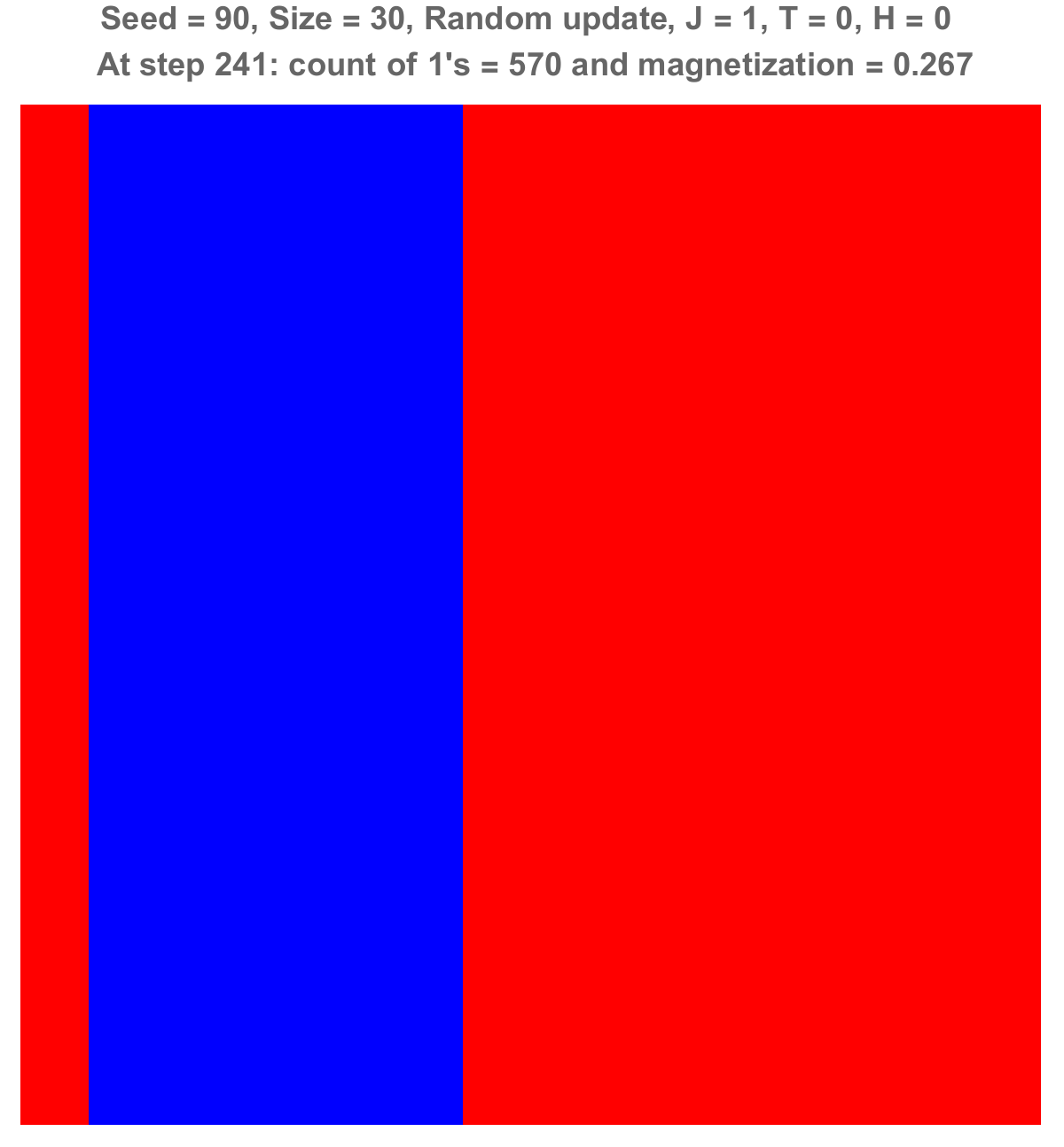}} \hspace{0.002\textwidth}
\caption{Results of two simulations using a random update with initial distributions of spins (Seed = 40, 90) different than in  Fig. (\ref{r1}) (Seed = 10, 50, 70). However, contrary to Fig. (\ref{r1}), these two distributions lead to final states with no full symmetry breaking as  exhibited in Sub-cases (a, c). Indeed two domains of opposite distributions are found in the final equilibrium state as seen in Sub-cases (b, d). In both Sub-cases, the domain are of different sizes (magnetization -0.0667 versus 0.267).
}
\label{r2}
\end{figure}

\begin{figure}
\centering
\subfigure[]{\includegraphics[width=0.45\textwidth]{ar1.pdf}} \hspace{0.002\textwidth}
\subfigure[]{\includegraphics[width=0.45\textwidth]{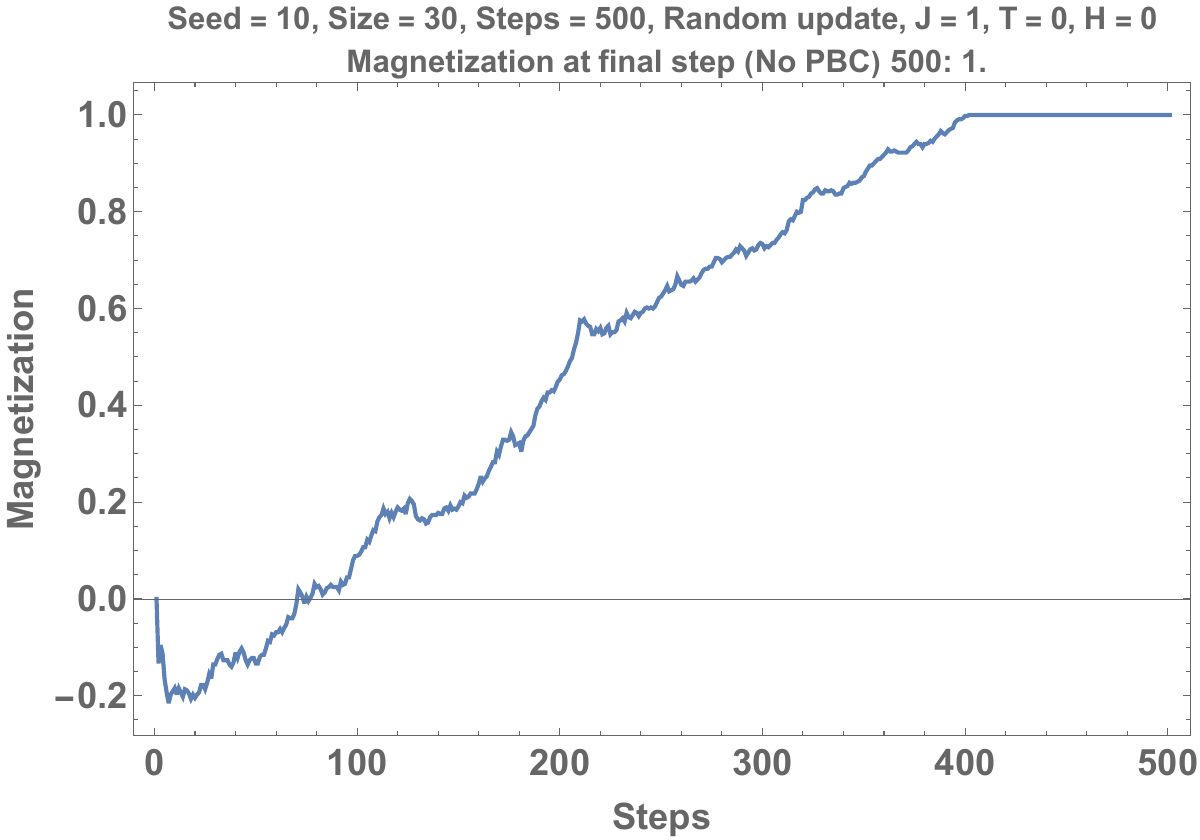}} \hspace{0.002\textwidth}
\\[0.03\textwidth]
\subfigure[]{\includegraphics[width=0.45\textwidth]{dr1.pdf}} \hspace{0.002\textwidth}
\subfigure[]{\includegraphics[width=0.45\textwidth]{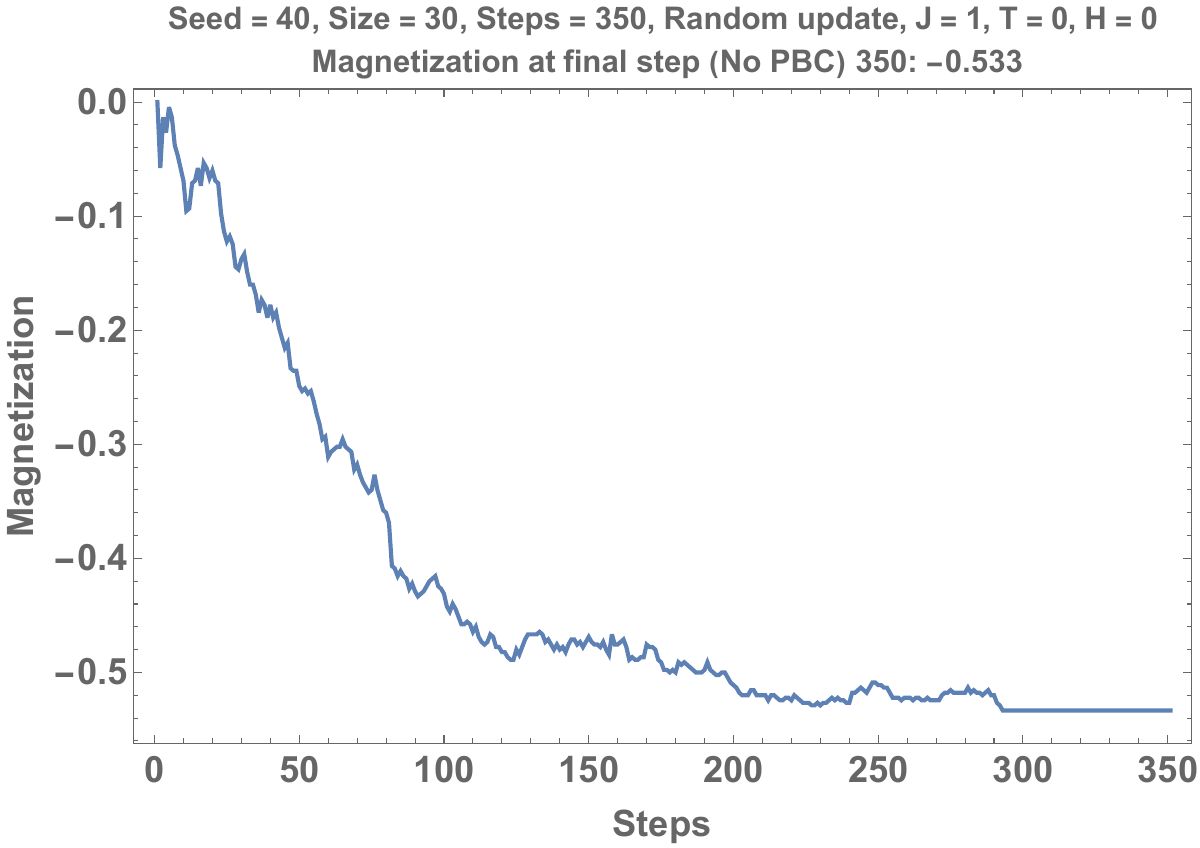}} \hspace{0.002\textwidth}
\caption{Results of two simulations using a random update with initial distributions of spins (Seed = 10, 40) as in Sub-case a  in Fig. (\ref{r1}) (Seed = 10) and Sub-case  a  in Fig. (\ref{r2}) (Seed = 40). However, contrary to Figs. (\ref{r1}, \ref{r2})), these two simulations do not include Periodic Boundary Conditions (PBC). The related results are very different with a full symmetry breaking along $+1$ instead of $-1$ after about 400 Monte Carlo steps instead of 180 and two coexisting domains of different sizes (magnetization -0.533) instead of (magnetization -0.0667) after about 300 Monte Carlo steps instead of 150.}
\label{r3}
\end{figure}

\begin{figure}
\centering
\subfigure[]{\includegraphics[width=0.32\textwidth]{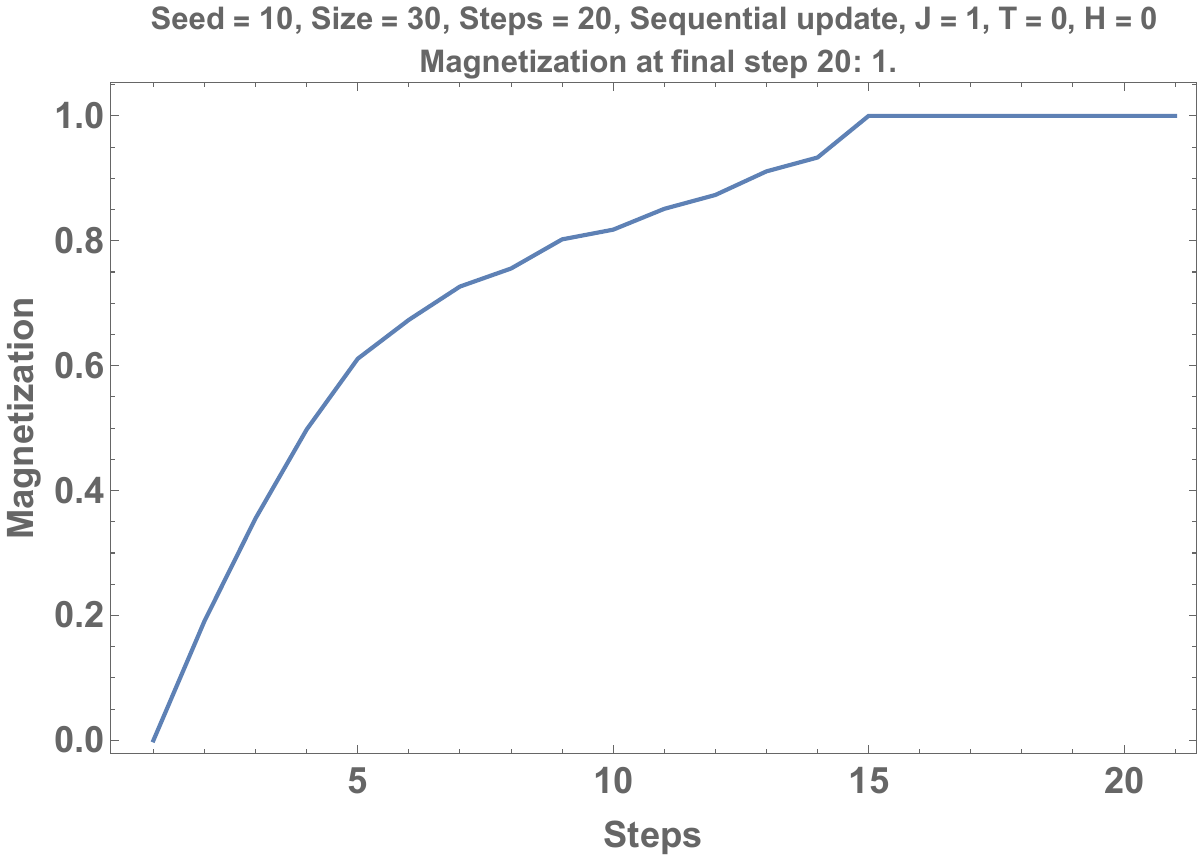}} \hspace{0.002\textwidth}
\subfigure[]{\includegraphics[width=0.32\textwidth]{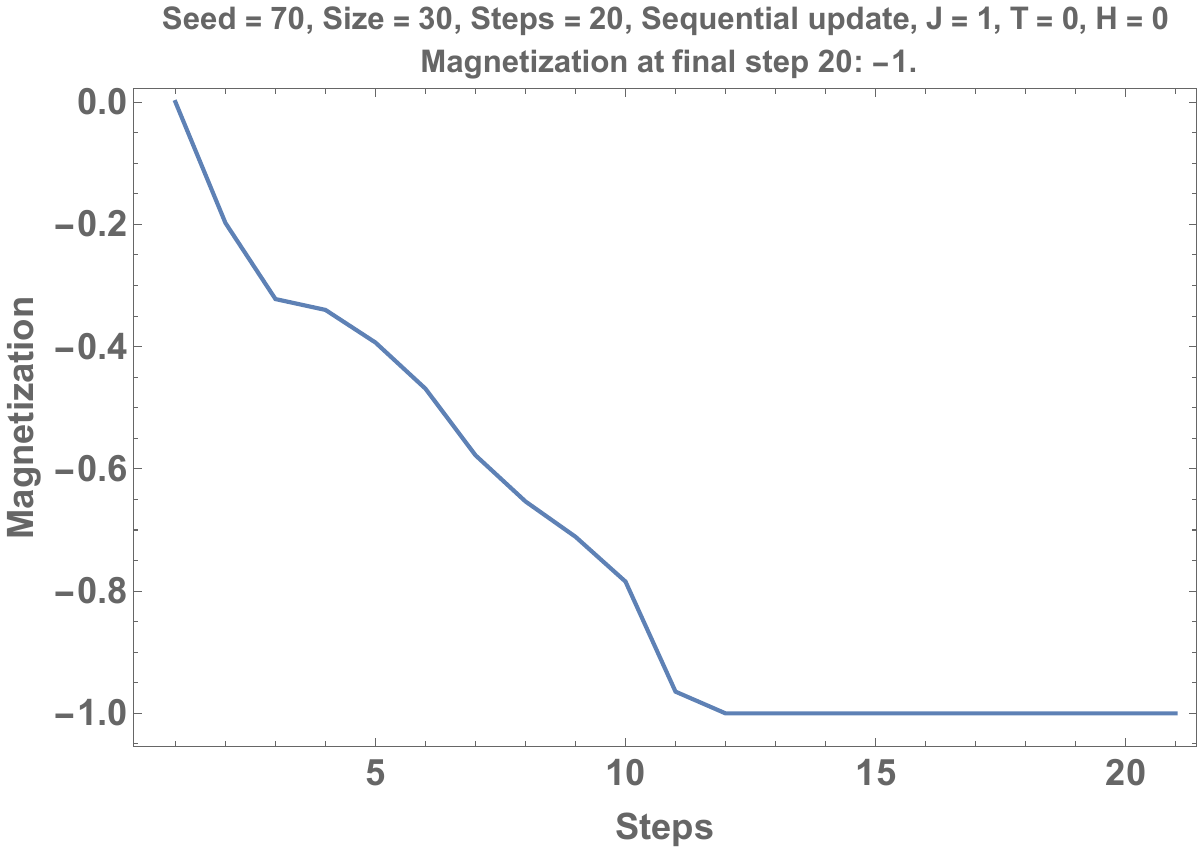}} \hspace{0.002\textwidth}
\subfigure[]{\includegraphics[width=0.32\textwidth]{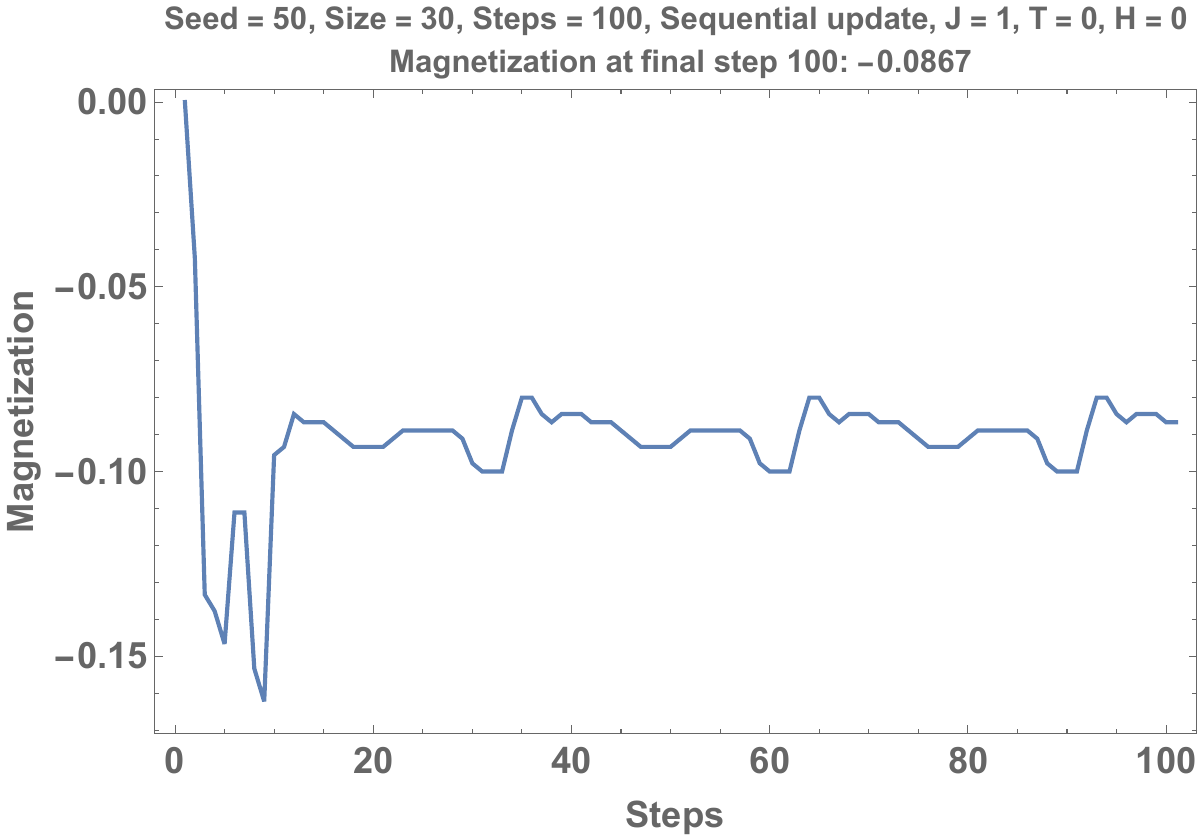}}
\\[0.03\textwidth]
\subfigure[]{\includegraphics[width=0.32\textwidth]{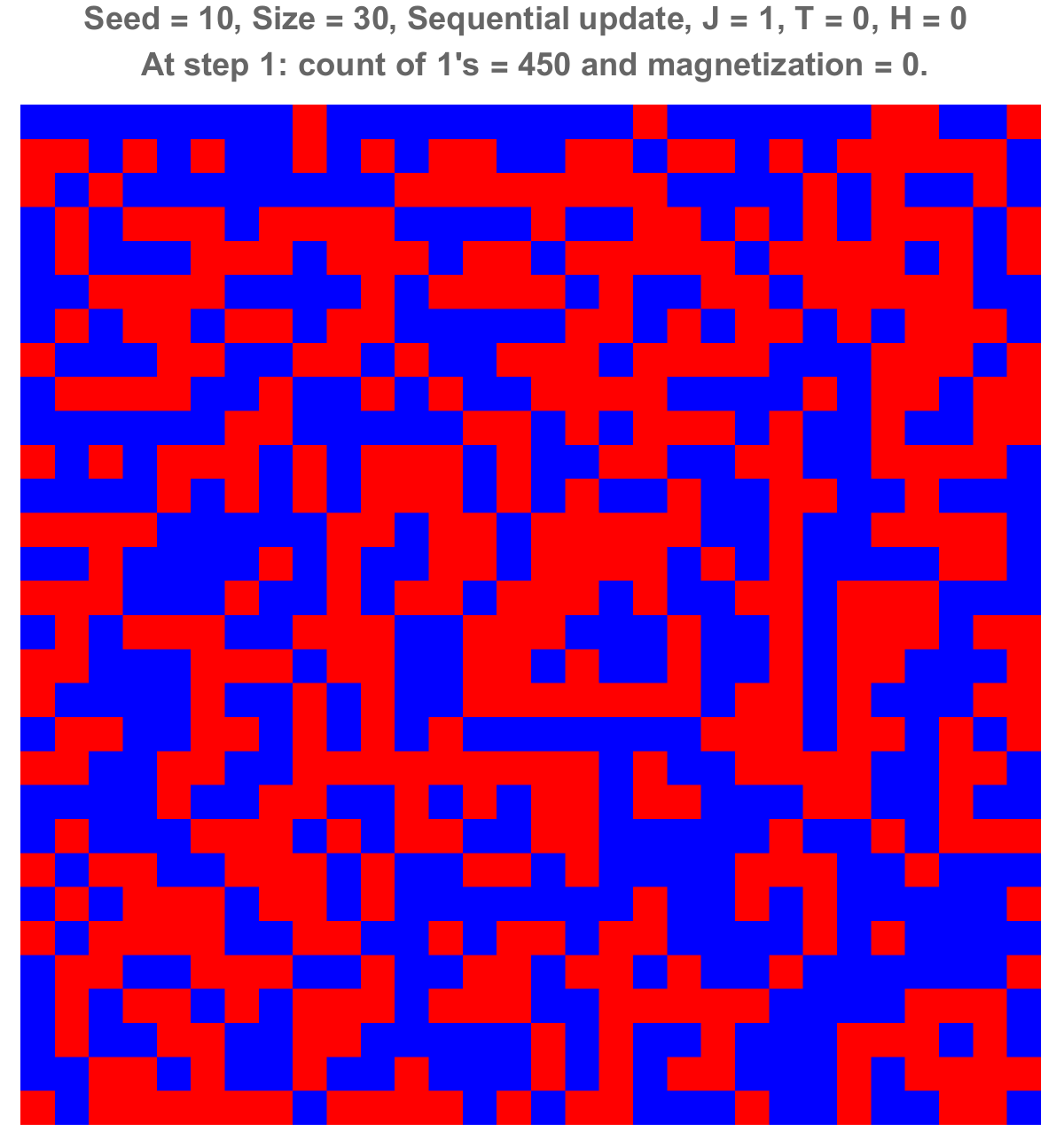}} \hspace{0.002\textwidth}
\subfigure[]{\includegraphics[width=0.32\textwidth]{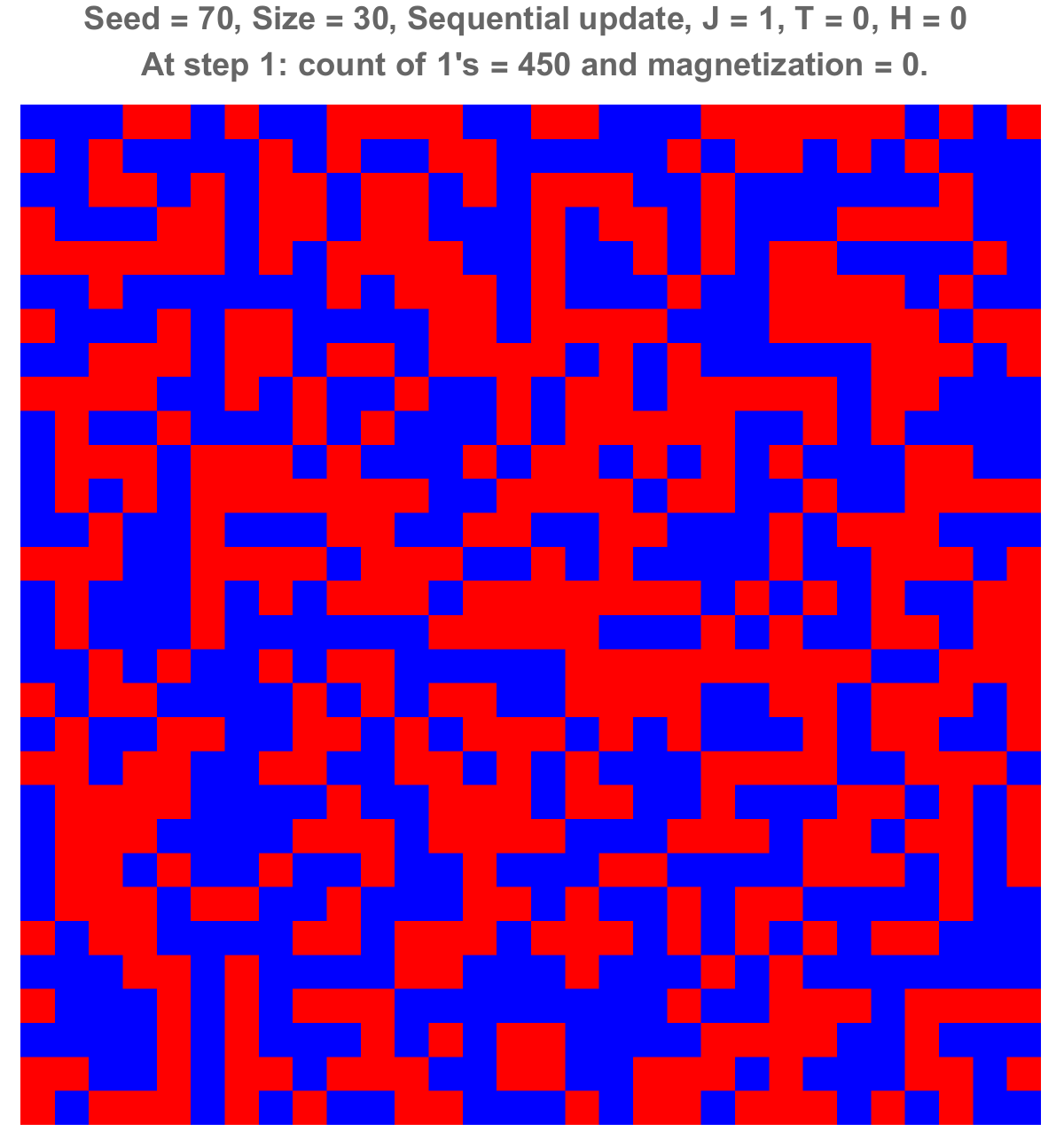}} \hspace{0.002\textwidth}
\subfigure[]{\includegraphics[width=0.32\textwidth]{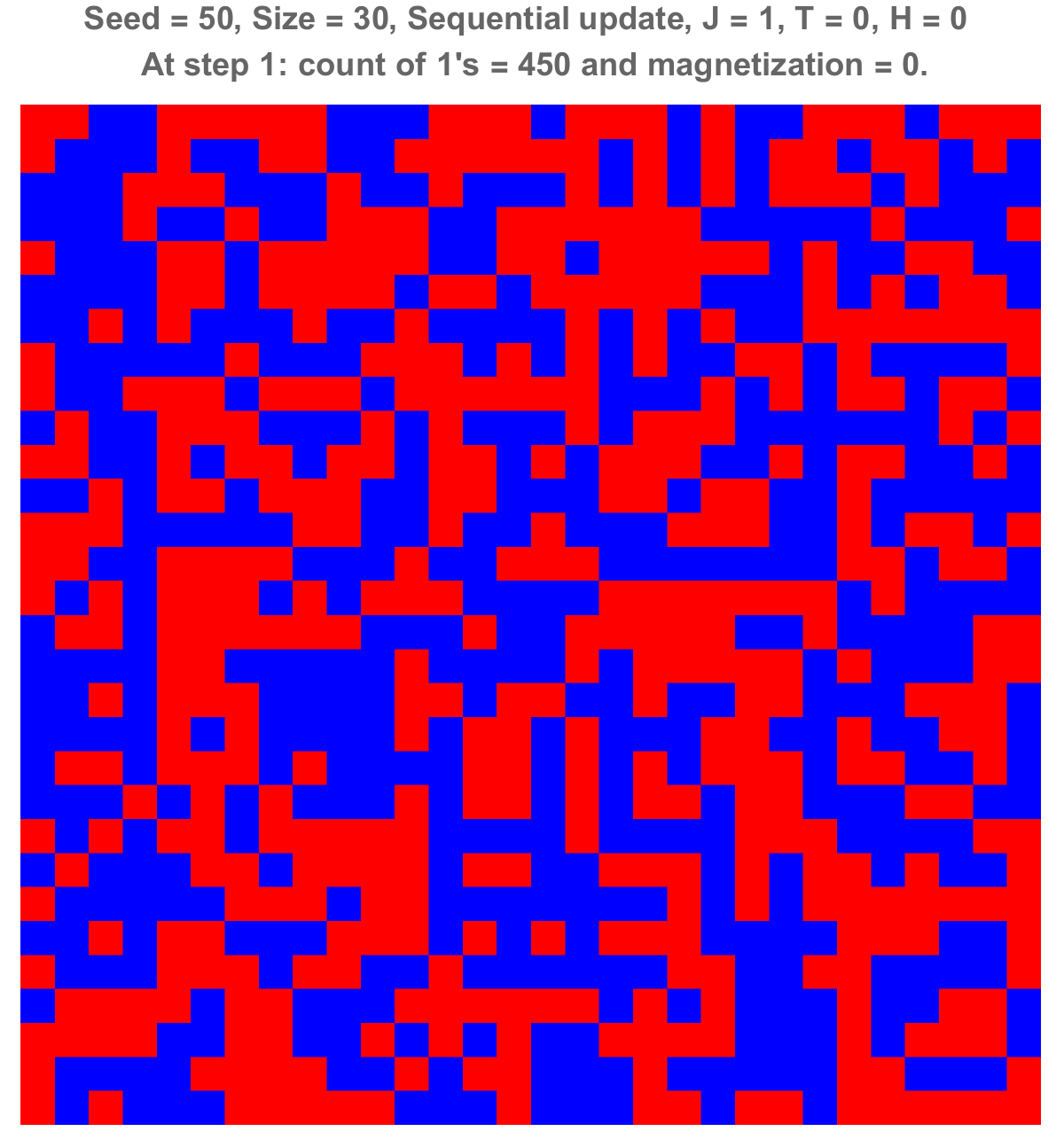}}
\\[0.03\textwidth]
\subfigure[]{\includegraphics[width=0.32\textwidth]{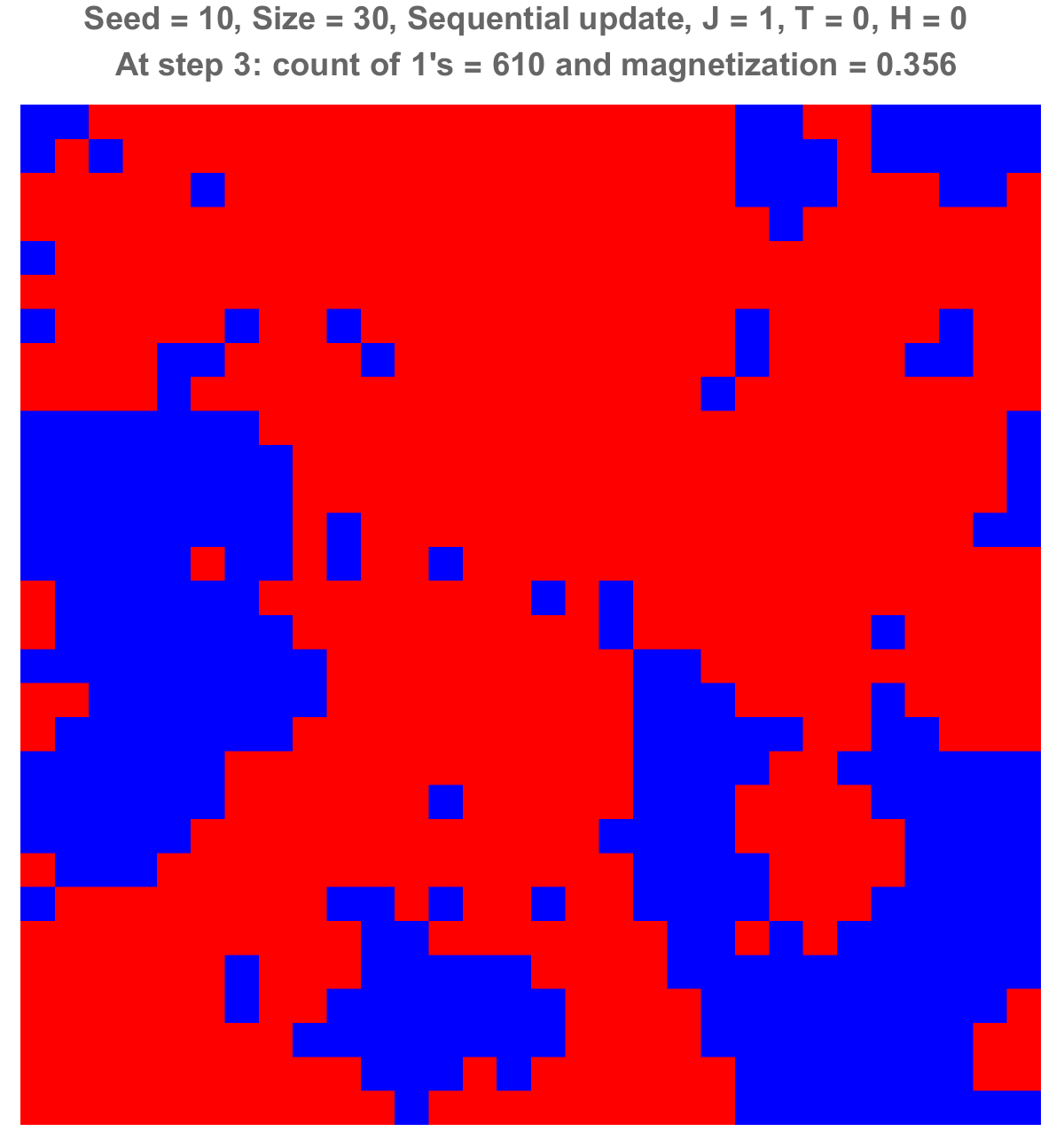}} \hspace{0.002\textwidth}
\subfigure[]{\includegraphics[width=0.32\textwidth]{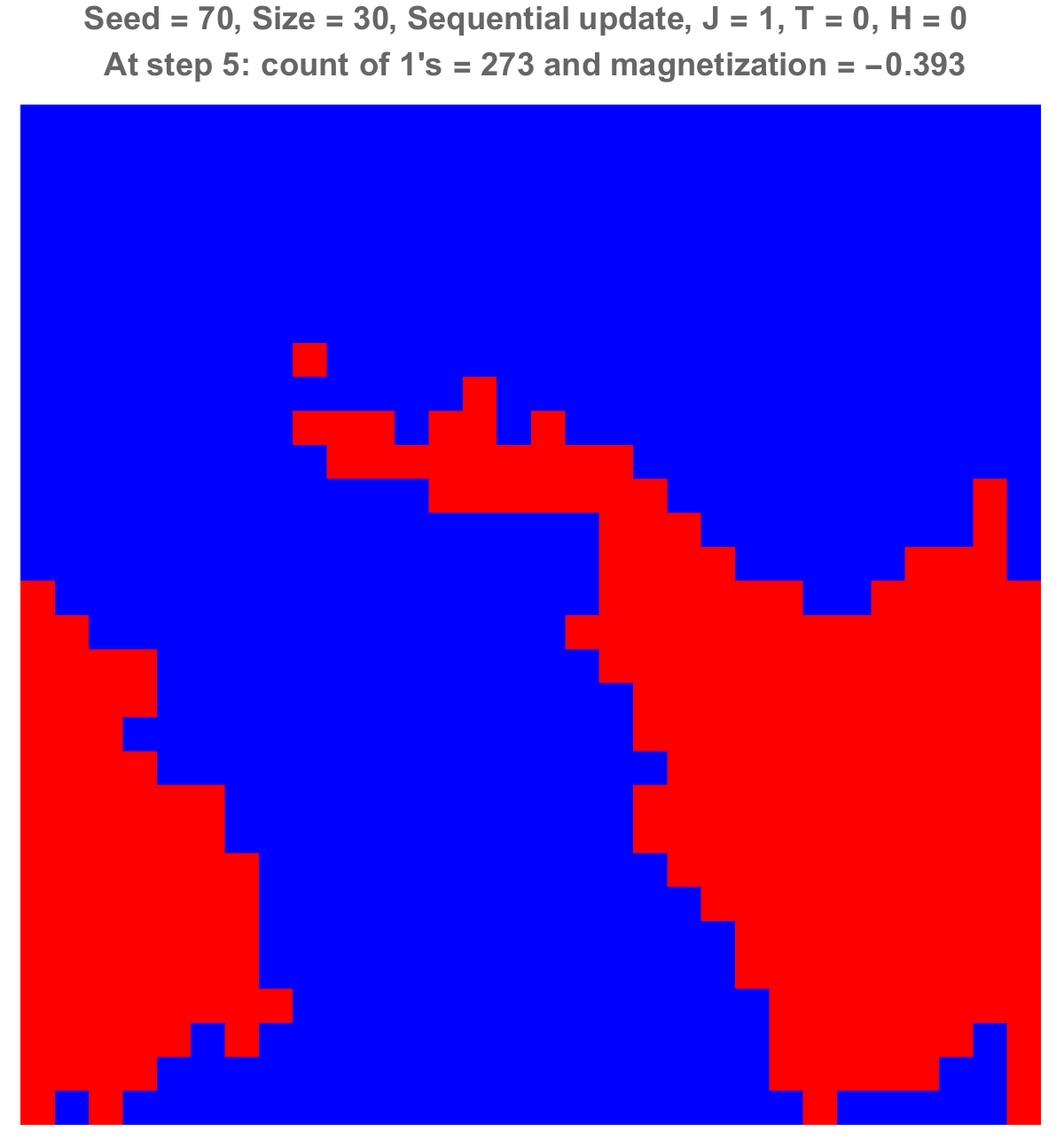}} \hspace{0.002\textwidth}
\subfigure[]{\includegraphics[width=0.32\textwidth]{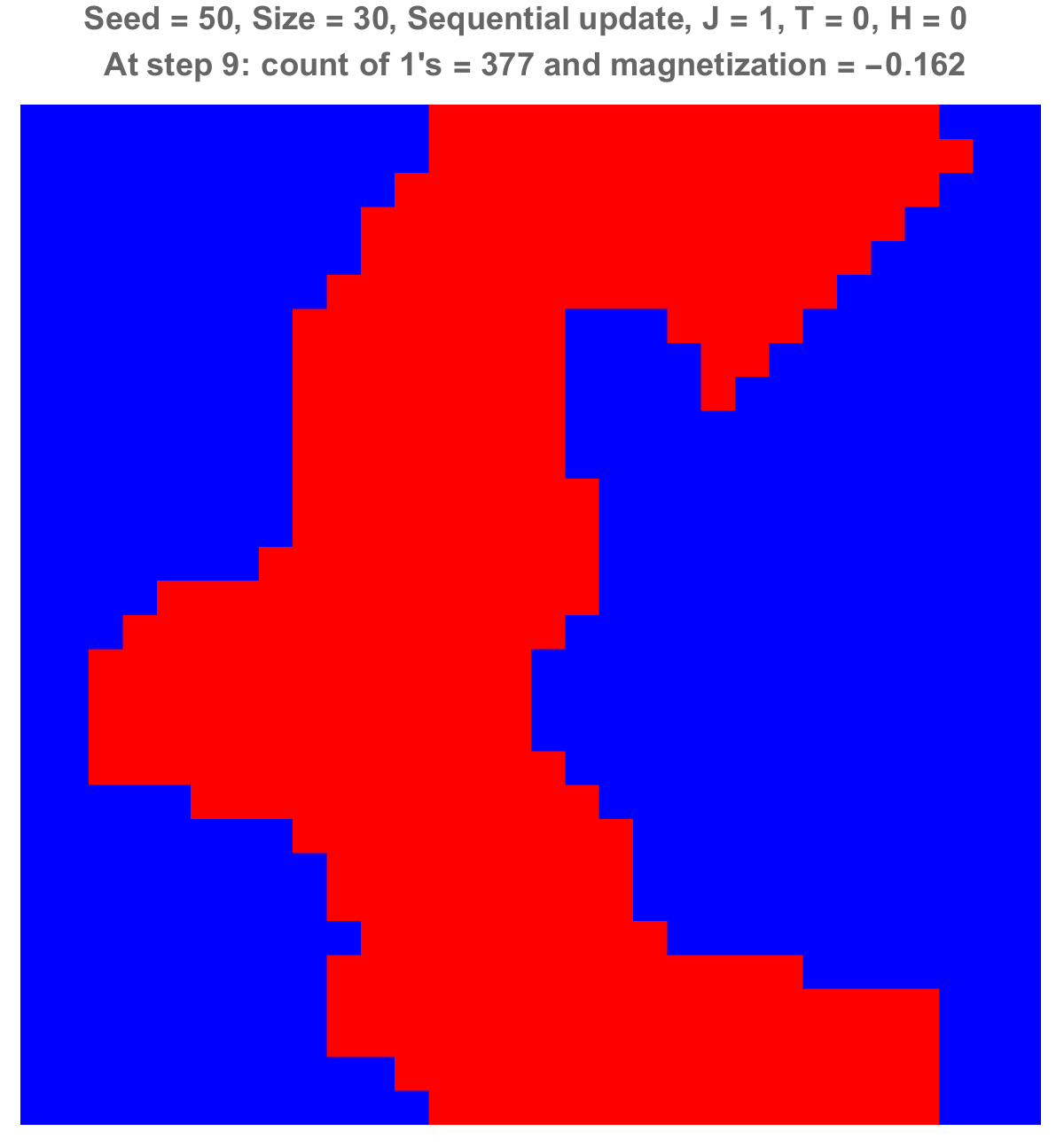}}
\\[0.03\textwidth]
\subfigure[]{\includegraphics[width=0.32\textwidth]{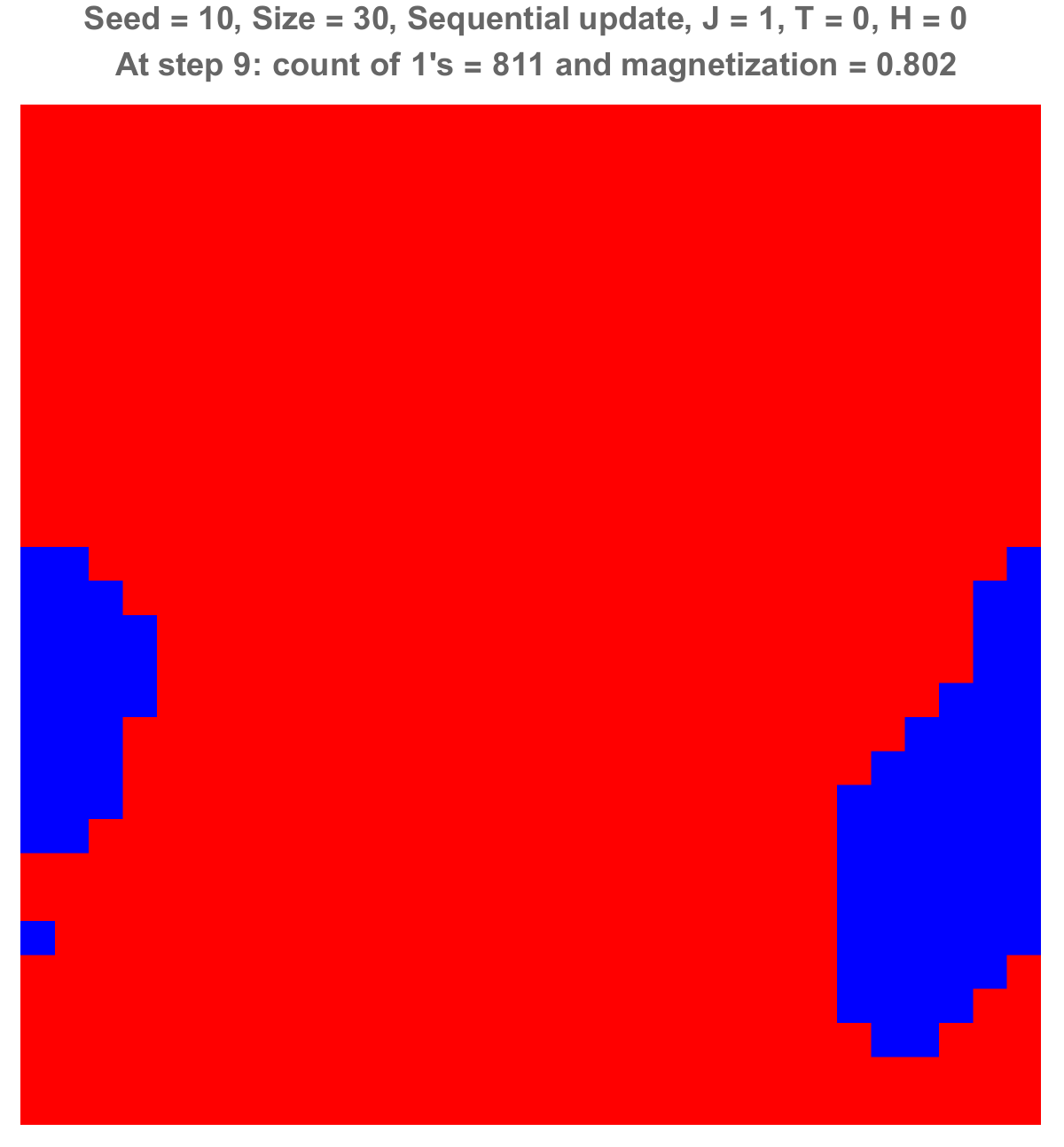}} \hspace{0.002\textwidth}
\subfigure[]{\includegraphics[width=0.32\textwidth]{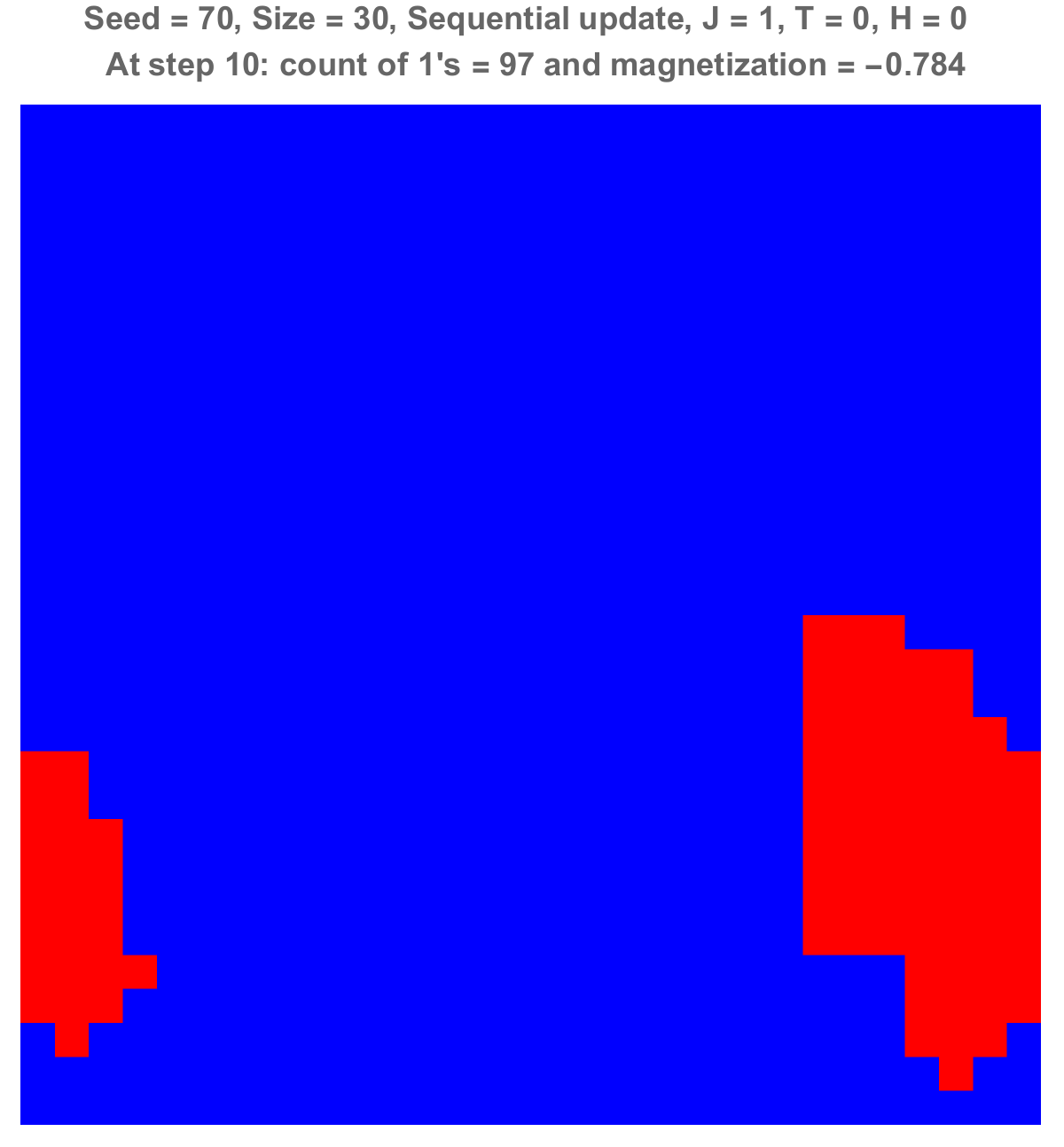}} \hspace{0.002\textwidth}
\subfigure[]{\includegraphics[width=0.32\textwidth]{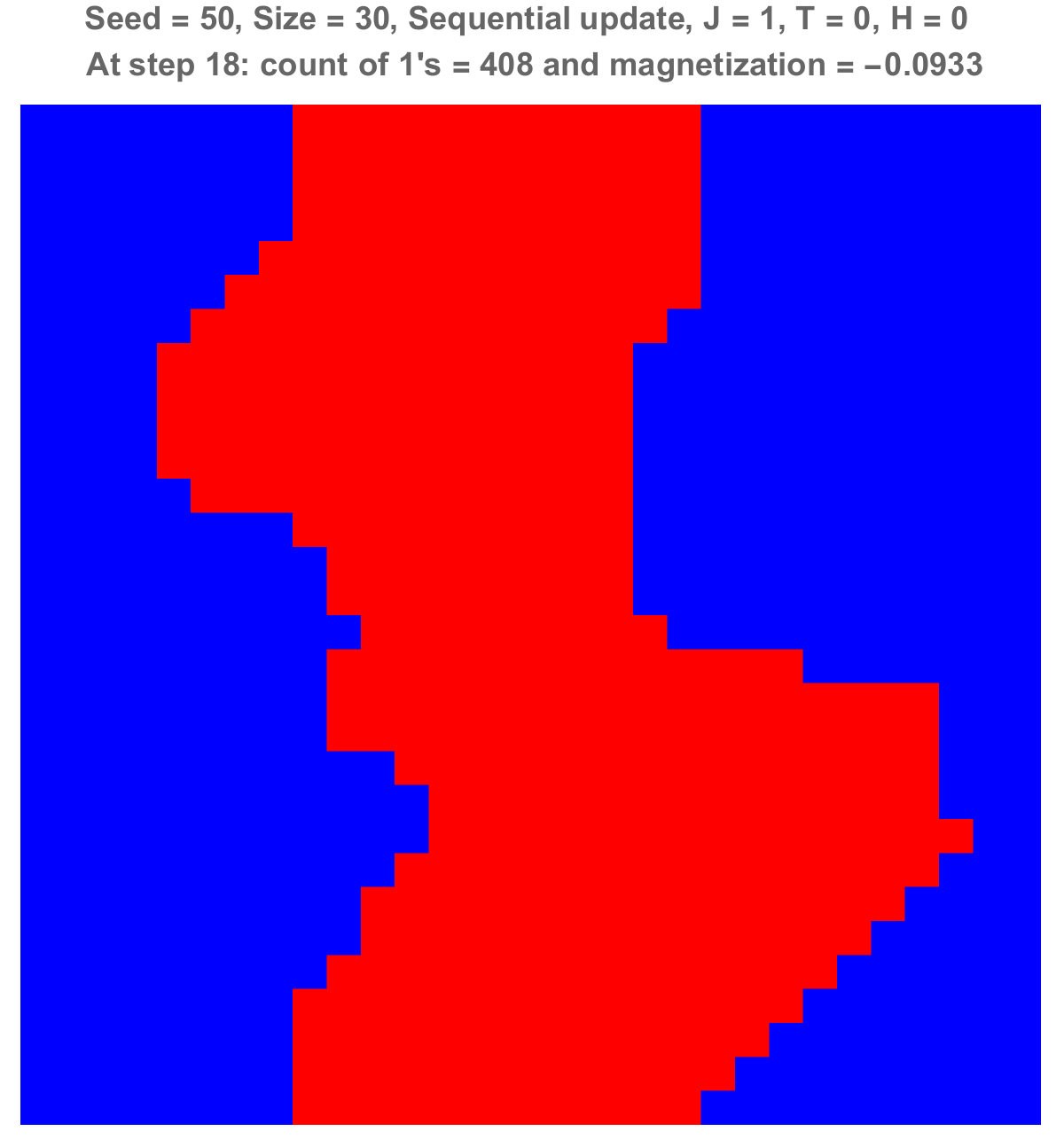}}
\\[0.03\textwidth]
\end{figure}

\newpage 

\noindent\captionof{figure}{Results of three simulations in sub-cases a, b, c with identical initial conditions (Seed = 10, 50, 70) than in Fig. (\ref{r1}) but using sequential update instead of random update. The sequential update leads to very different results from Fig. (\ref{r1}) with respectively a full symmetry breaking along $+1$ instead of $-1$ after about only 15 Monte Carlo steps instead of 180, a full symmetry breaking along $-1$ instead of $+1$ after about only 10 Monte Carlo steps instead of 90, and two coexisting domains of different sizes (magnetization 0.0933) instead of a full symmetry breaking along $-1$ (magnetization -1) after about 20 Monte Carlo steps instead of about 700.
}
\label{s1}

\begin{figure}
\centering
\subfigure[]{\includegraphics[width=0.32\textwidth]{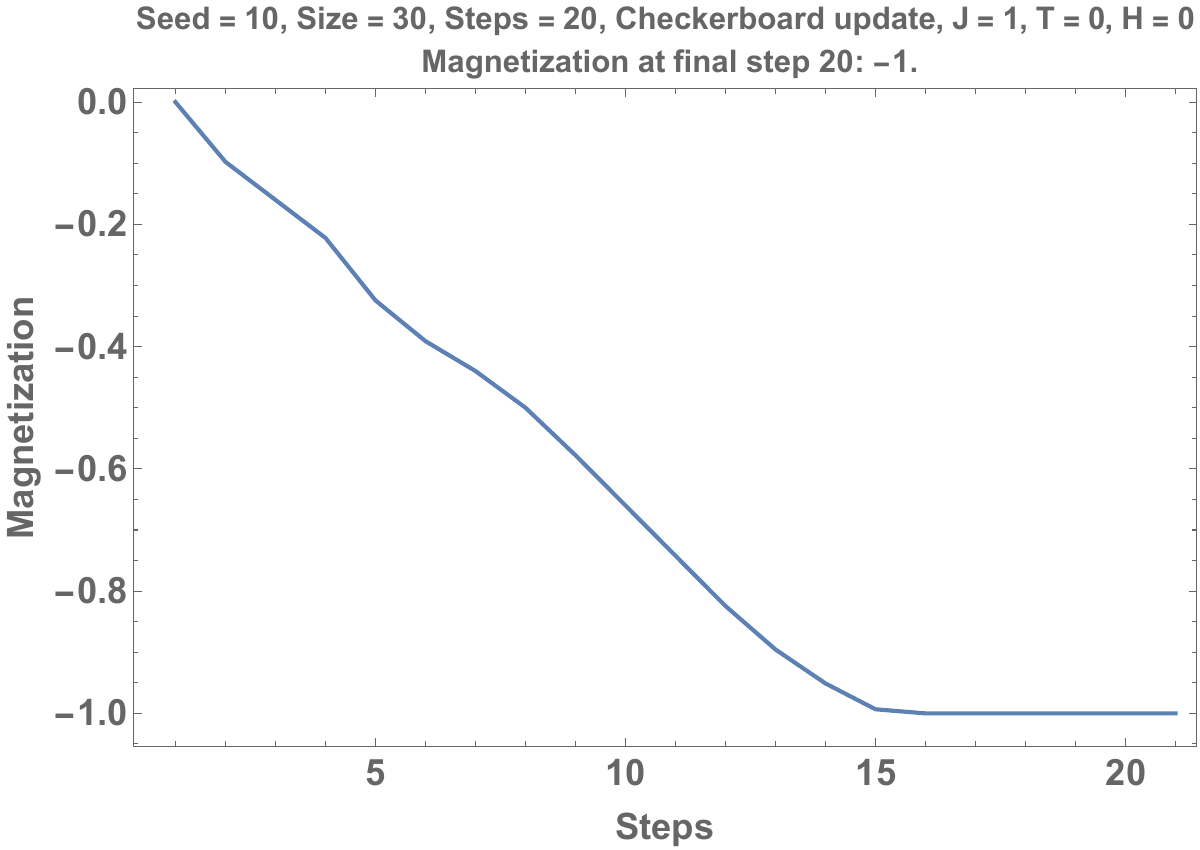}} \hspace{0.002\textwidth}
\subfigure[]{\includegraphics[width=0.32\textwidth]{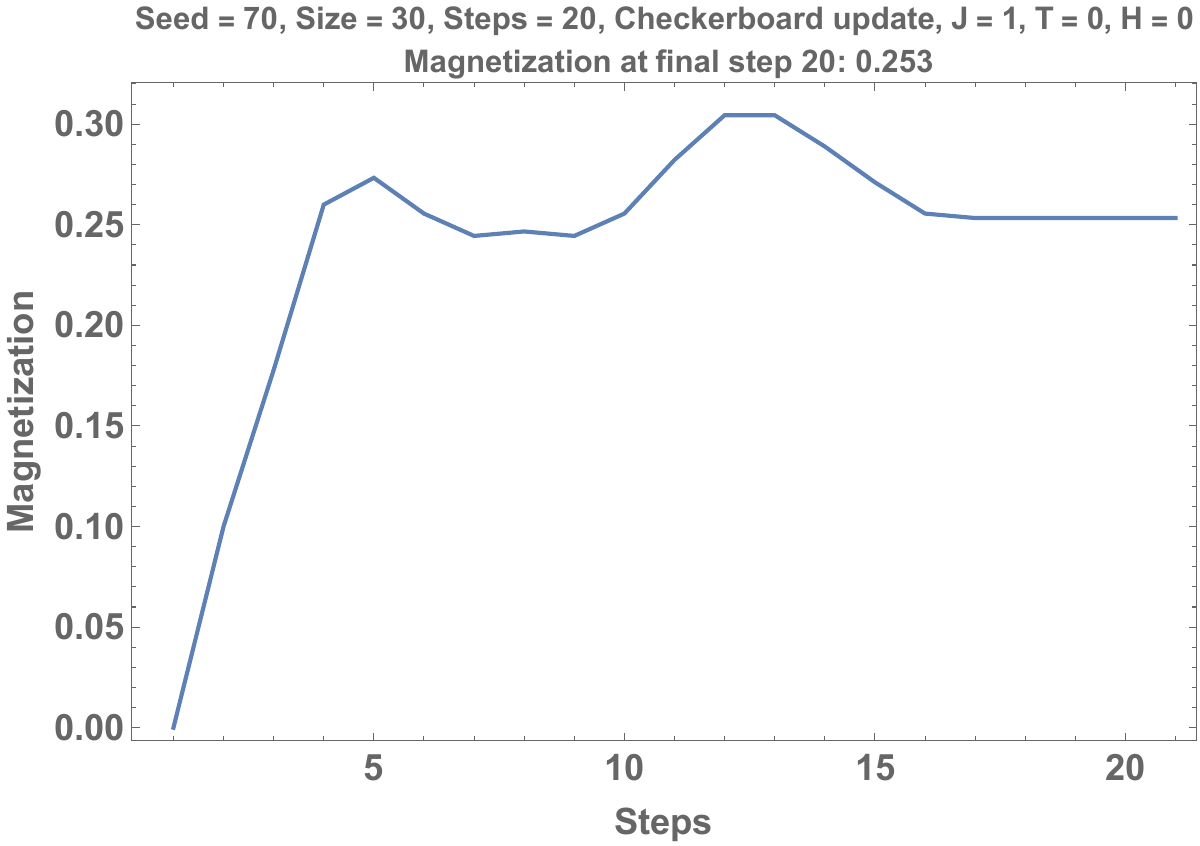}} \hspace{0.002\textwidth}
\subfigure[]{\includegraphics[width=0.32\textwidth]{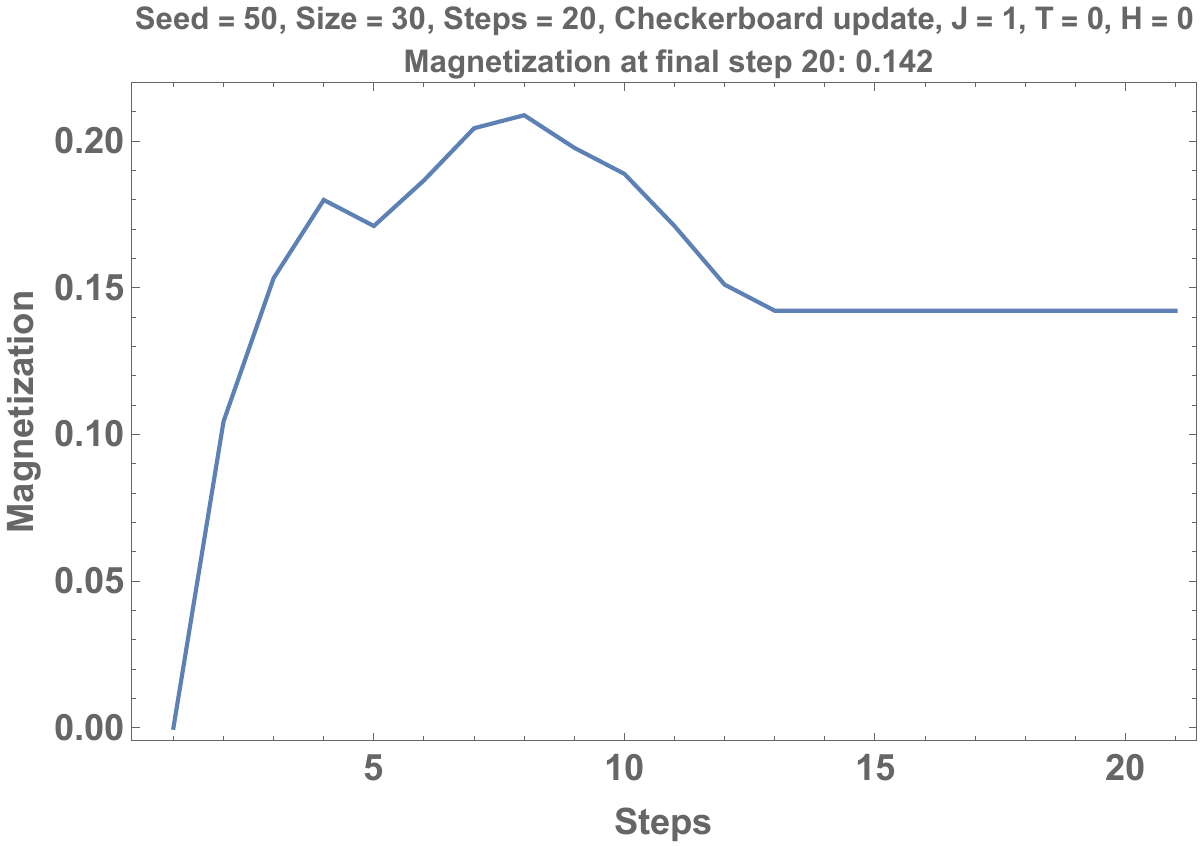}}
\\[0.03\textwidth]
\subfigure[]{\includegraphics[width=0.32\textwidth]{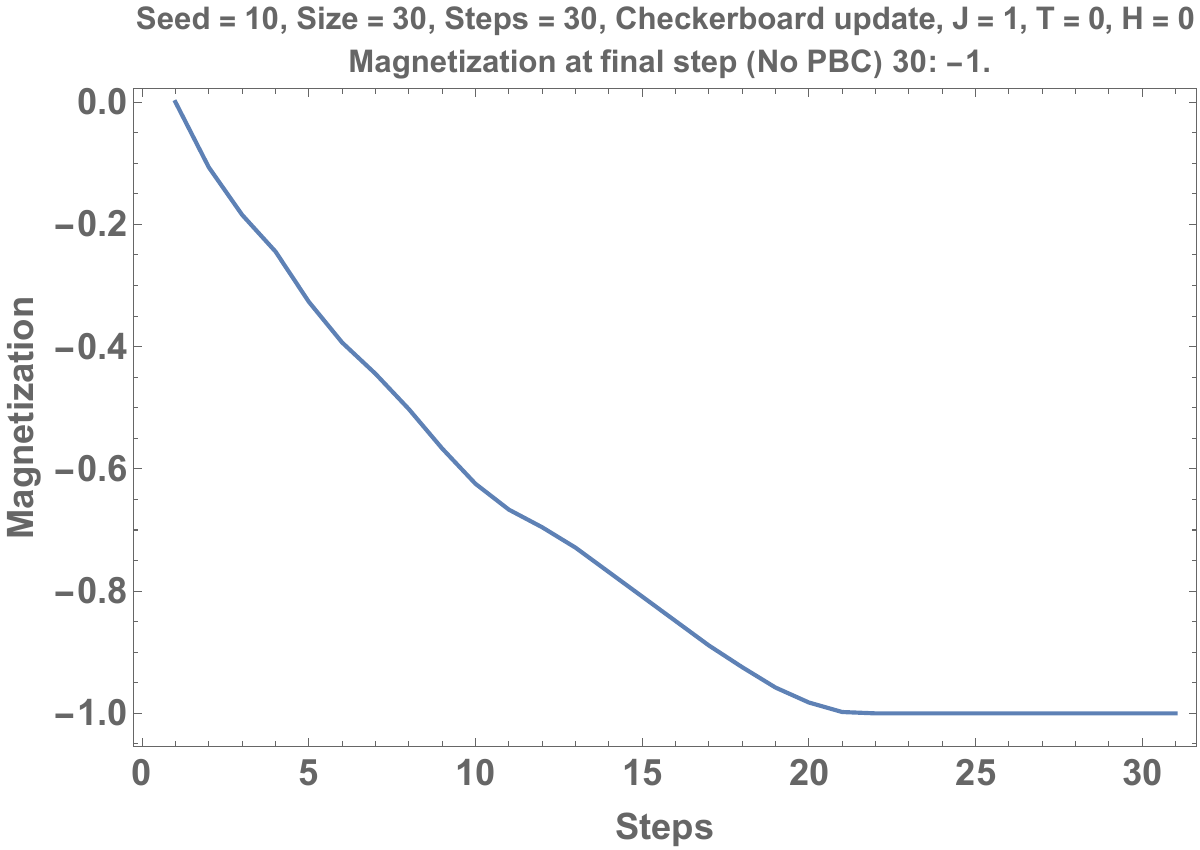}} \hspace{0.002\textwidth}
\subfigure[]{\includegraphics[width=0.32\textwidth]{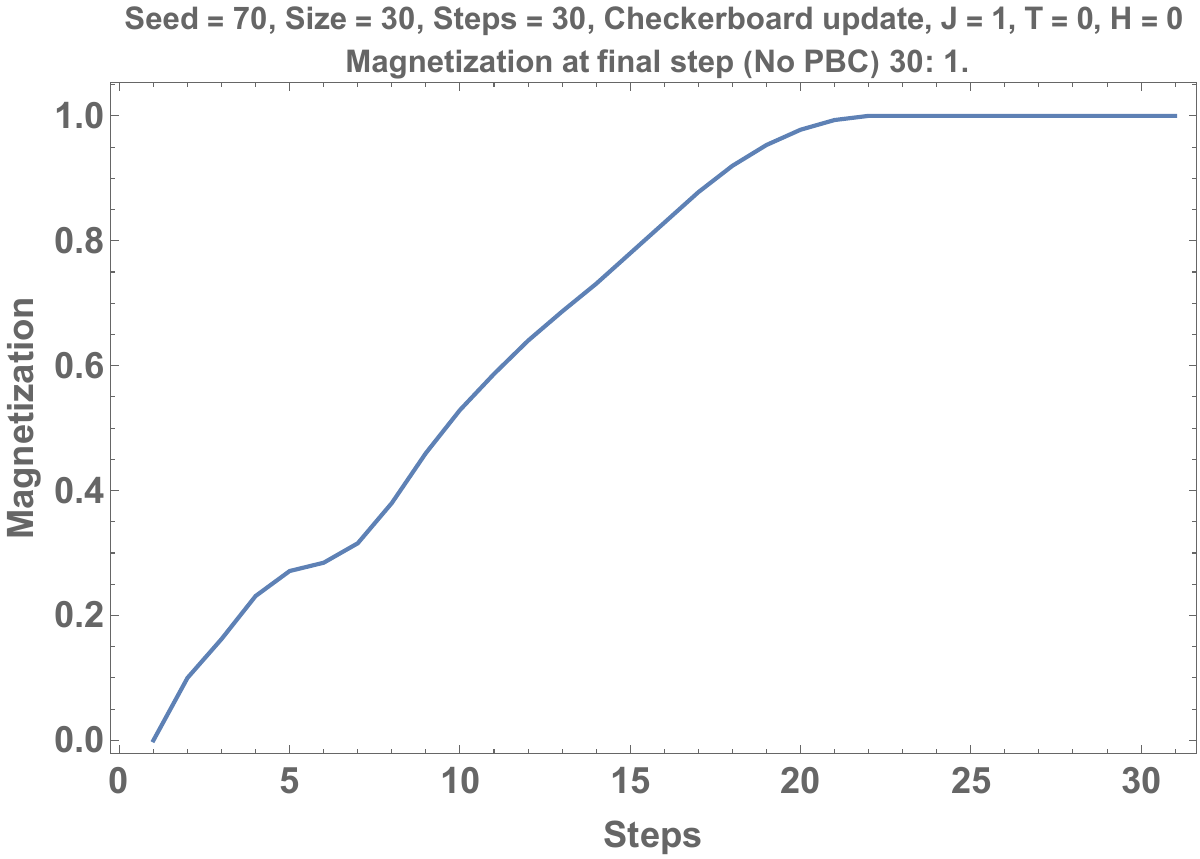}} \hspace{0.002\textwidth}
\subfigure[]{\includegraphics[width=0.32\textwidth]{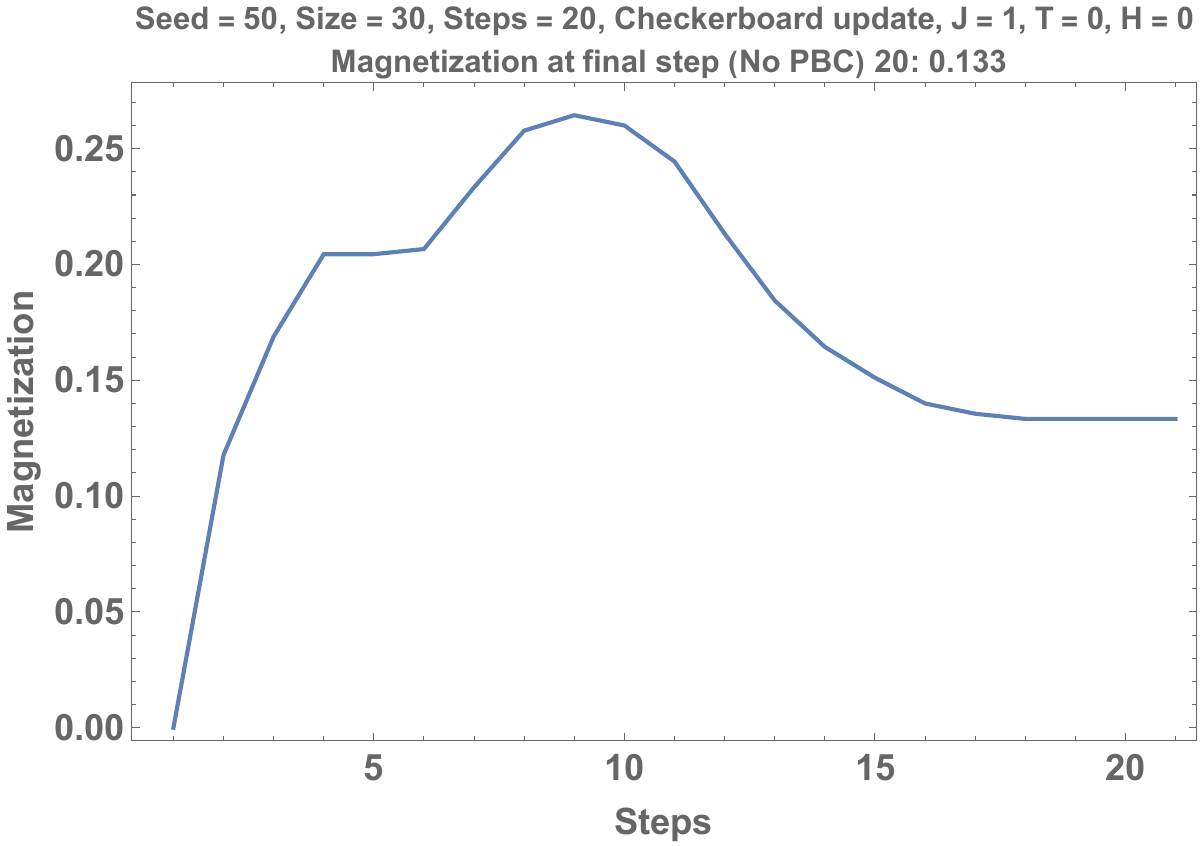}}
\\[0.03\textwidth]
\subfigure[]{\includegraphics[width=0.32\textwidth]{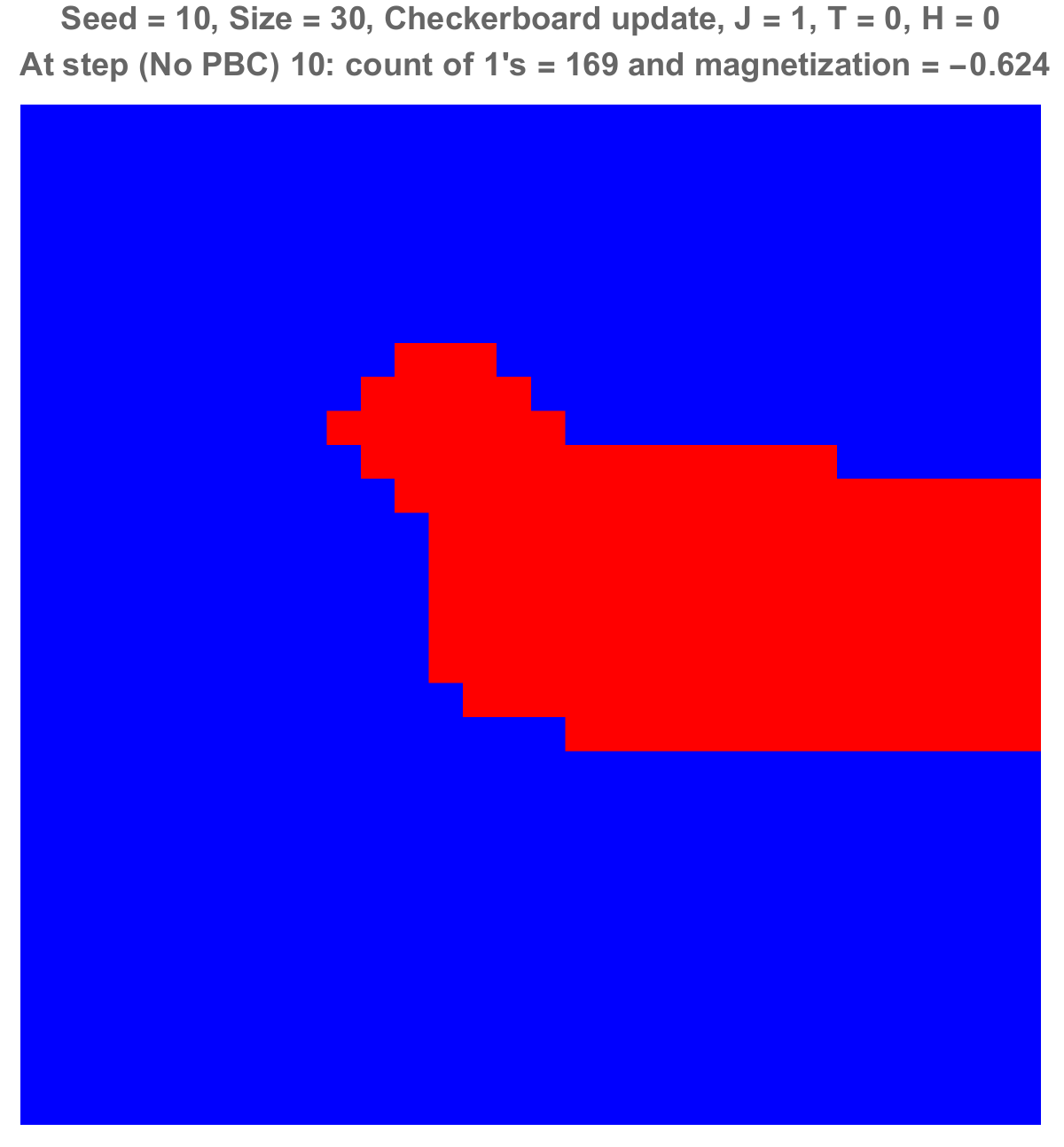}} \hspace{0.002\textwidth}
\subfigure[]{\includegraphics[width=0.32\textwidth]{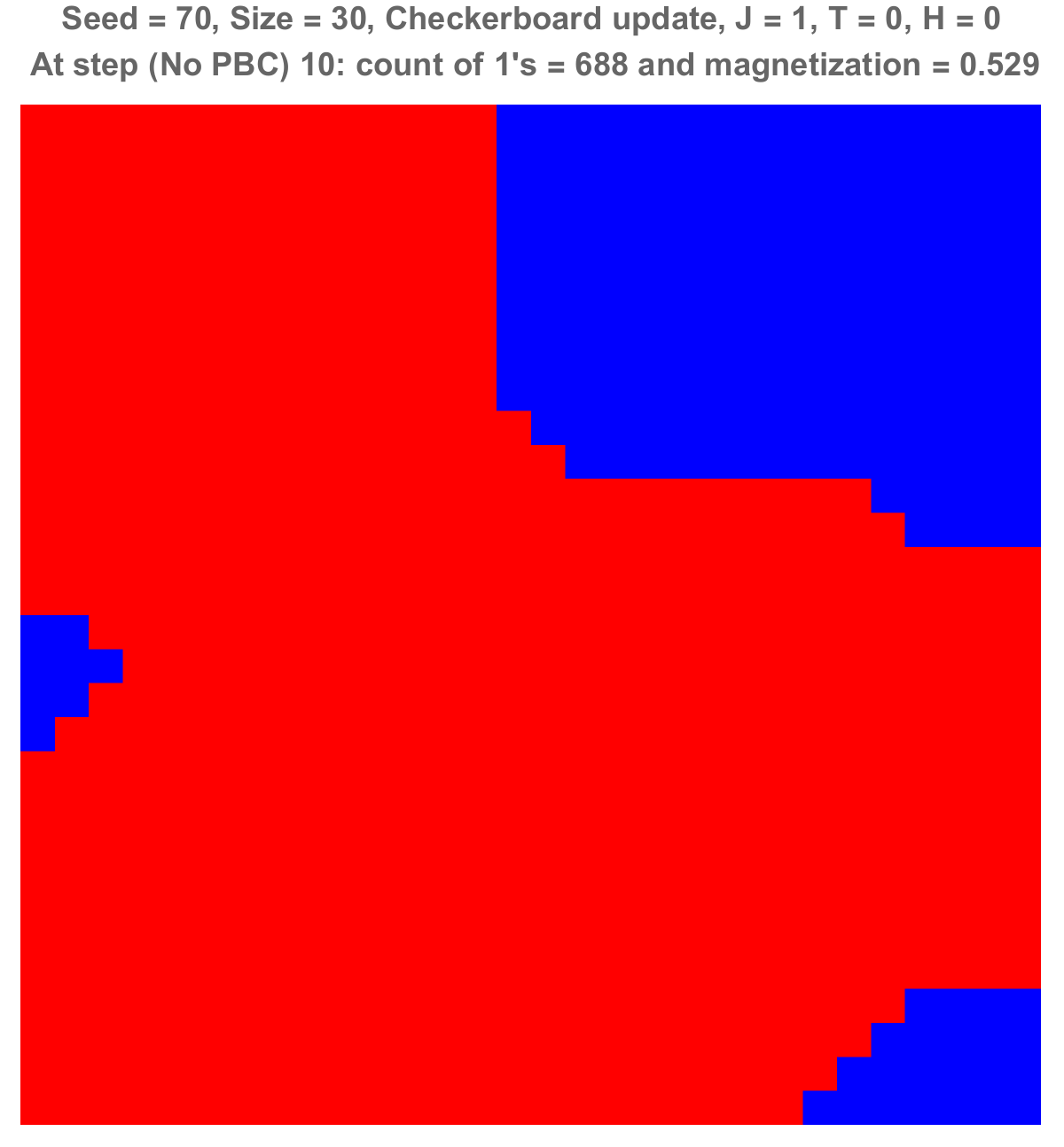}} \hspace{0.002\textwidth}
\subfigure[]{\includegraphics[width=0.32\textwidth]{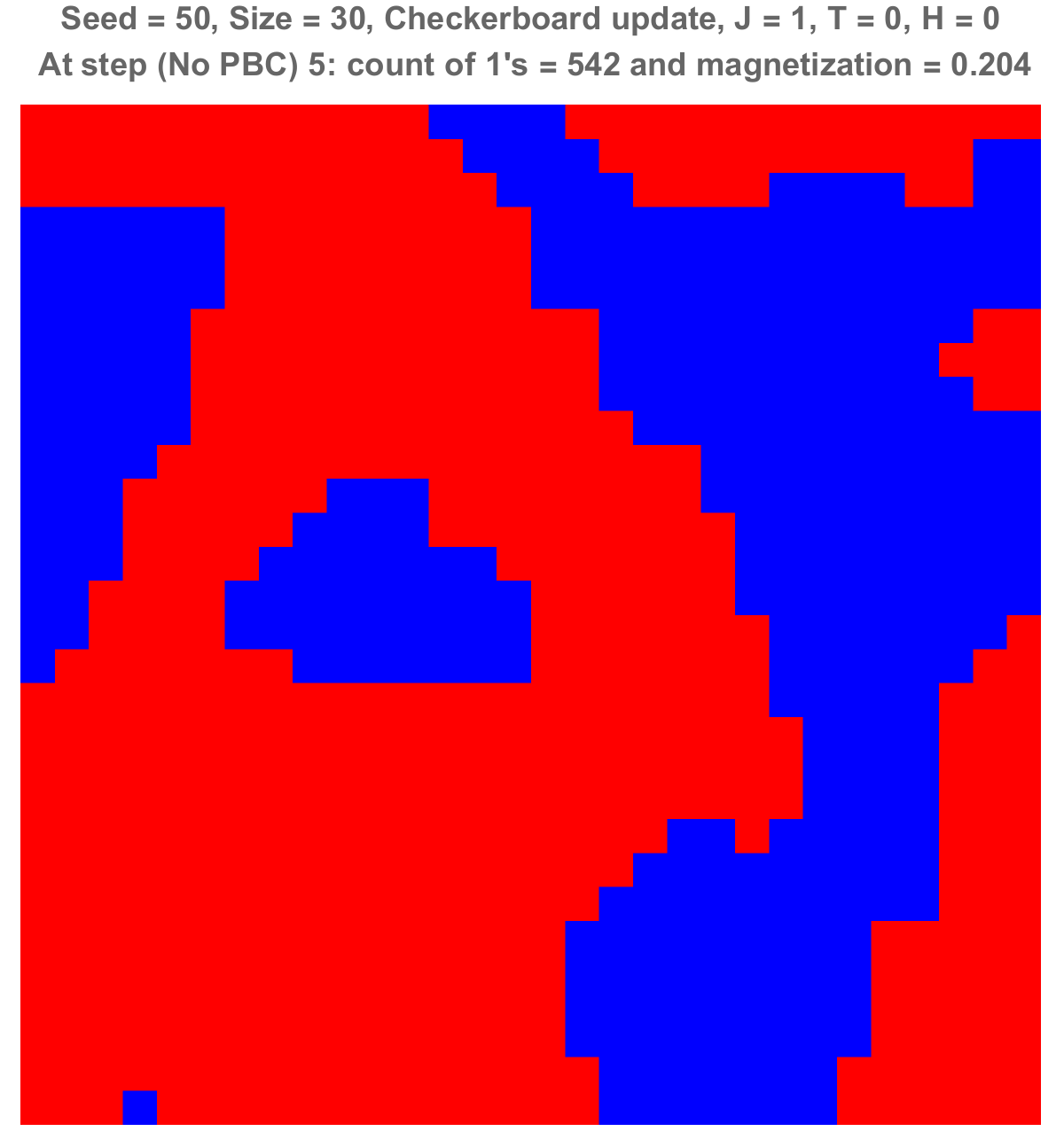}}
\\[0.03\textwidth]
\subfigure[]{\includegraphics[width=0.32\textwidth]{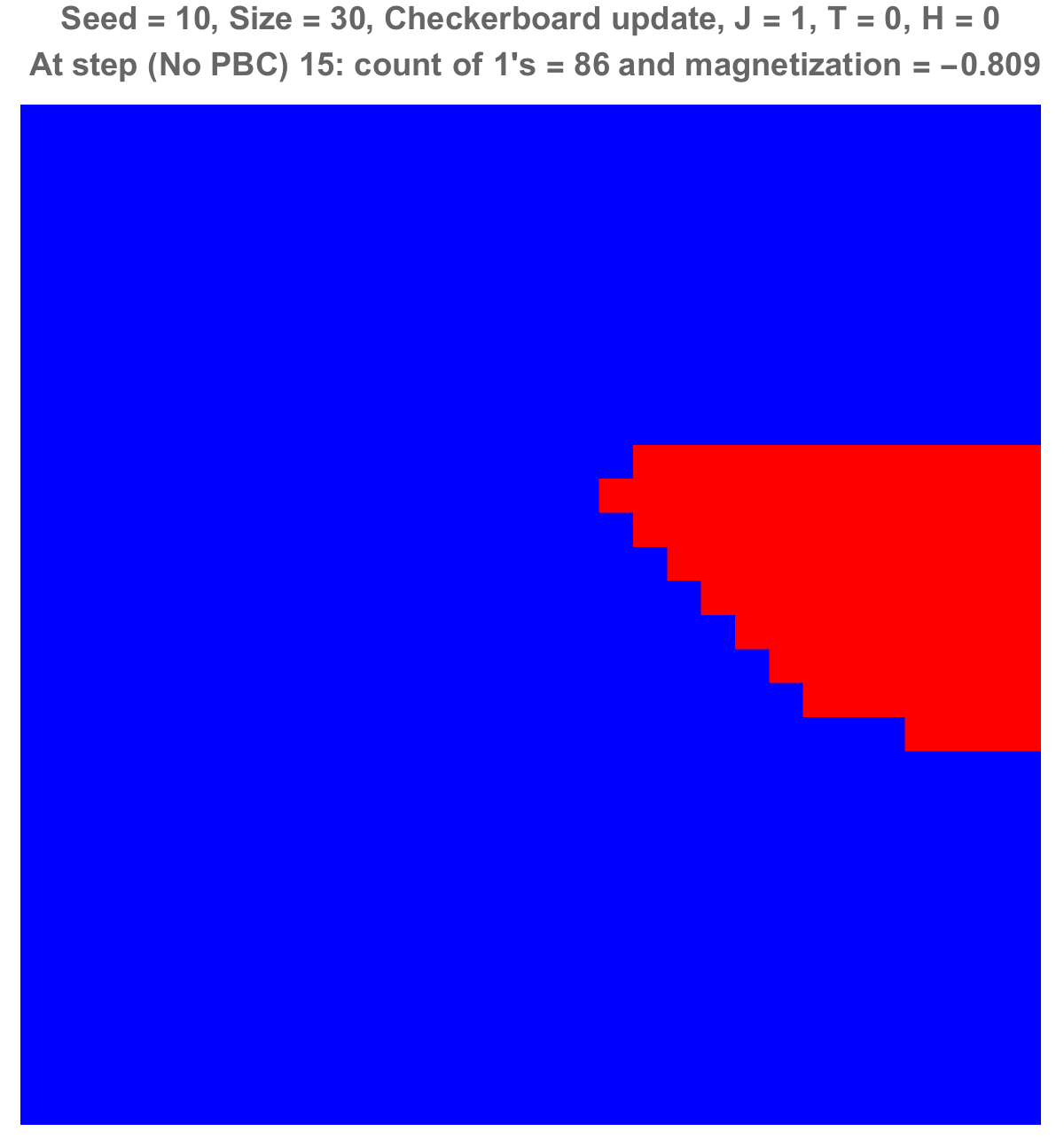}} \hspace{0.002\textwidth}
\subfigure[]{\includegraphics[width=0.32\textwidth]{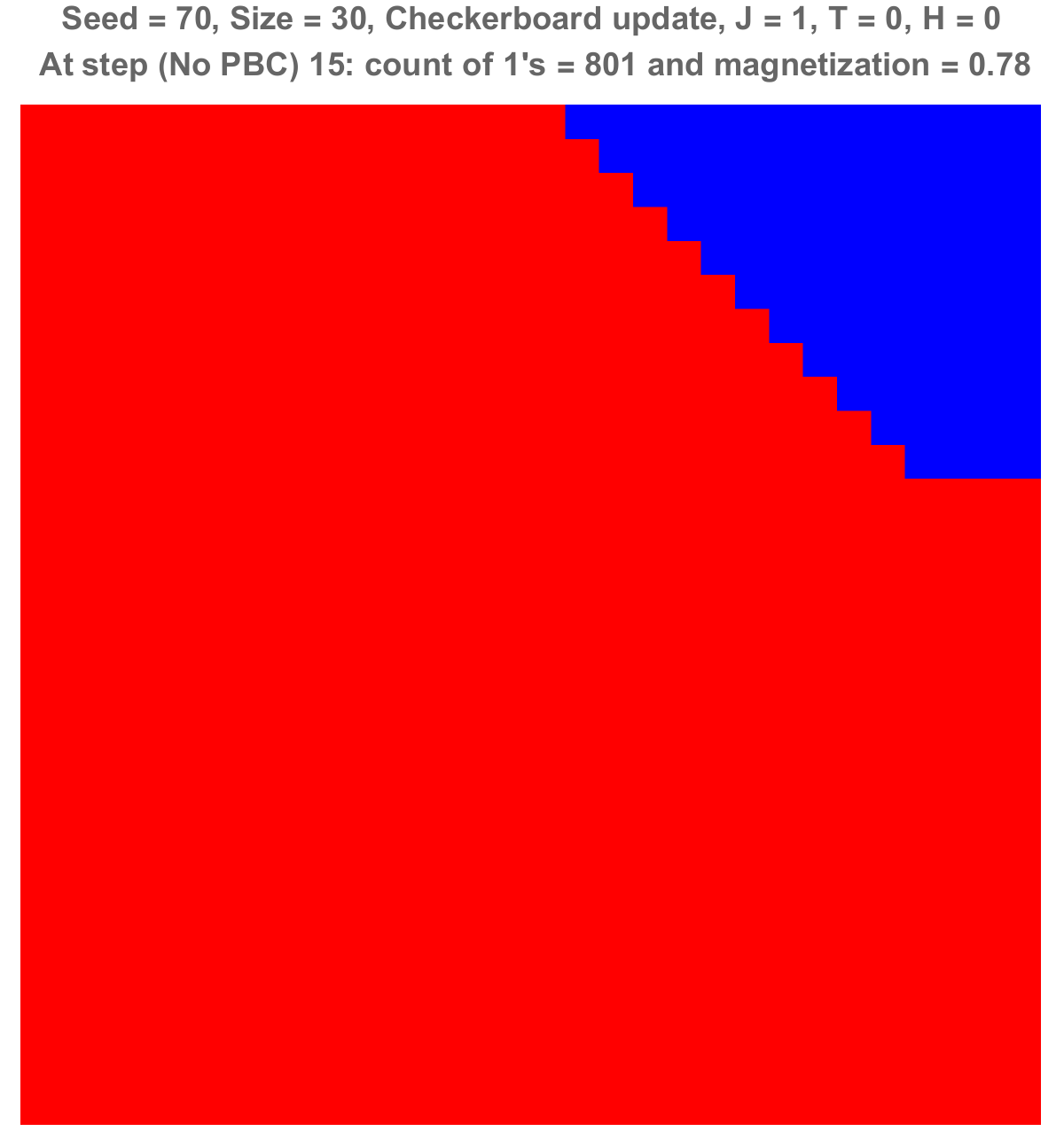}} \hspace{0.002\textwidth}
\subfigure[]{\includegraphics[width=0.32\textwidth]{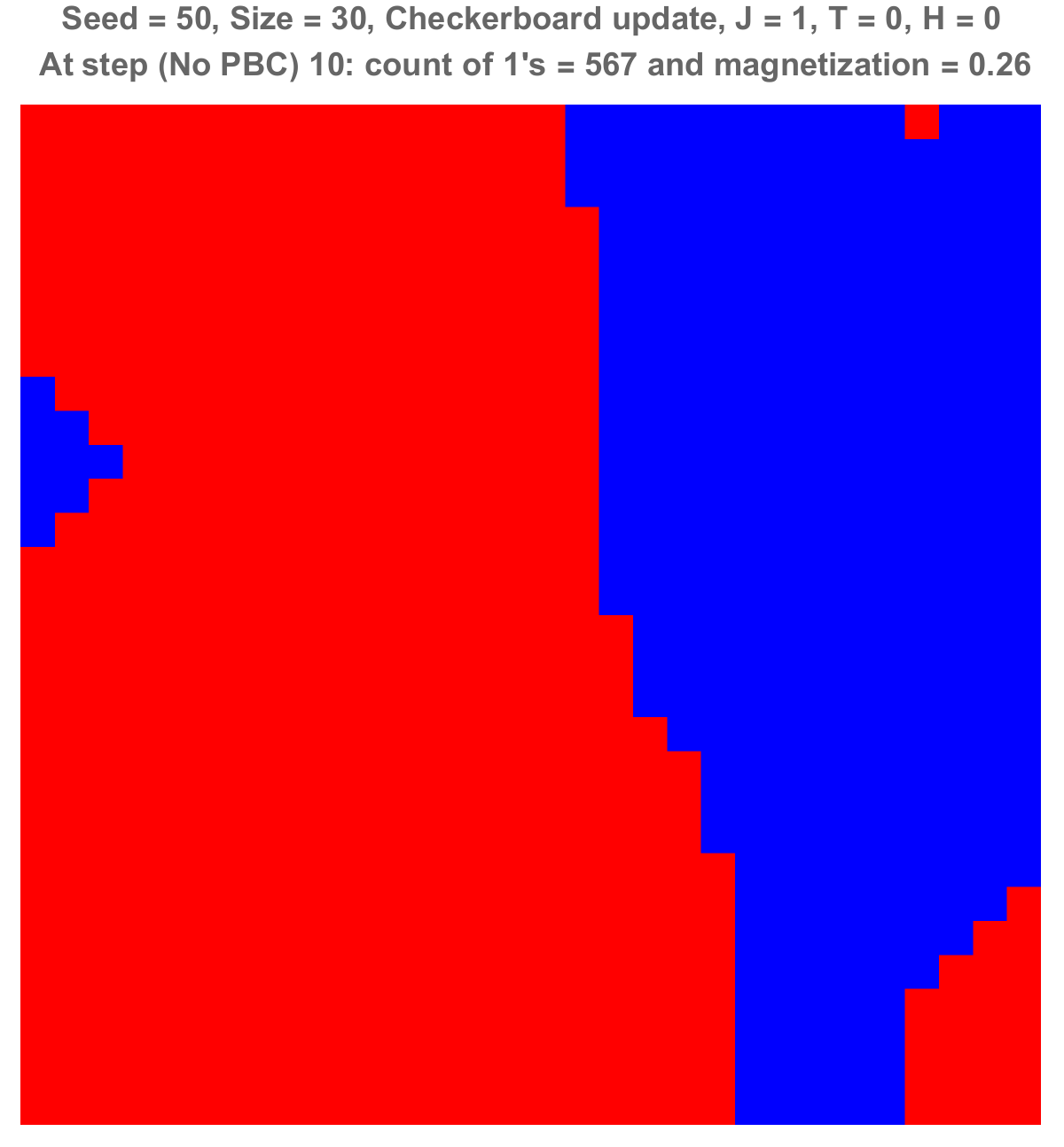}}
\\[0.03\textwidth] 
\end{figure}

\newpage 

\noindent\captionof{figure}{Results of a two step simultaneous update denoted checkerboard update. All sites of each sub-lattice are updated simultaneously one after the other sequentially. Three simulations (sub-cases a, b, c ) are performed with identical initial conditions (Seed = 10, 50, 70) as in Fig. (\ref{r1}) but using checkerboard update instead of random update. The checkerboard update leads to very different results from Fig. (\ref{r1}) with respectively a full symmetry breaking unchanged along $-1$ but now after about only 15 Monte Carlo steps instead of 180, two coexisting domains of different sizes (magnetization 0.253) instead of a full symmetry breaking along $+1$ after about only 15 Monte Carlo steps instead of 90, and two coexisting domains of different sizes (magnetization 0.142) instead of a full symmetry breaking along $-1$ (magnetization -1) after about 20 Monte Carlo steps instead of about 700. Sub-cases (d, e, f) exhibits the same simulations as in sub-cases (a, b, c) but without Periodic Boundary Conditions (PBC). The associated results are slightly different with respectively still a full symmetry breaking along $-1$ but with about 20 Monte Carlo steps instead of 15, a full symmetry breaking along $+1$ instead of two coexisting domains with similar numbers of Monte Carlo steps, still two coexisting domains of different sizes with magnetization 0.133 instead of magnetization 0.142.}
\label{c1}

\section{The requirement of social structures to drive collective symmetry breaking}

The main conclusion which arises from the series of Figs. (\ref{r1}, \ref{r2}, \ref{r3}, \ref{s1}, \ref{si1}, \ref{c1}) is that the initial microscopic distribution of spins within the same overall numbers $N_+$ and $N_-$, matters to determine the final macroscopic equilibrium state of the  sample. Furthermore, the actual algorithm used to monitor the microscopic update is also instrumental in the making of the final macroscopic state. Moreover, the size of the sample, via the inclusion or not of boundary conditions also affect the final state.

In this respect, it is essential to stress that all these features are deemed artifacts in statistical physics. Any solid simulation must make sure to avoid those effects in order to reach the absolute equilibrium state of an infinite system. This true state must be independent of both the algorithm and the initial state being used. In statistical physics all of the above features are deemed technical limitations. Their related effects must be discarded and avoided. 

In contrast,  these features are bounded to the nature of a social system. That is my major statement in connection to the phenomenon of spontaneous symmetry breaking in social systems in contrast to physical systems. I claim that contrary to physics where all these features must be made irrelevant, in social systems these  features are relevant and the most meaningful ones in the making of actual symmetry breaking of social groups and communities.

Although they are also technical, they must be build in the reality of each social community to activate the process of dynamical update of choices and opinions. Once installed they operate to drive the actual emergence of a collective opinion or direction of each social community. In addition, they are likely to be different from a community to another and in turn trigger different final collective states.

In case, agents are left by themselves to coordinate locally their respective choice, that coordination will be likely done via trail and error, a process which requires time and gets exponentially large with the number $N$ of agents. Furthermore, without any coordinating framework, agents try to optimize their respective utility along a path similar to simultaneously update without any order. Above related simulations have shown that  simultaneous update traps the system very quickly in a disorder state with most agents unsatisfied with repeated shifts between the two individual states. This fact explains why every social community needs an effective framework for monitoring the update of individual choices in order to exist to reach with a coherent aggregation beyond small numbers of people.

\section{Statistical fluctuations preserve the random symmetry breaking}

All above simulations were performed starting from a symmetrical initial state with a perfect balance between the two competing choices among the 900 agents ($N_+ = N_- =  450$). However, for most situations the initial values of $N_+$ and $N_-$ deviate a bit from $N/2$ due to statistical fluctuations. The amplitude of the deviation is a function of the size of the sample.

\begin{figure}[!htb]
\centering
\subfigure[]{\includegraphics[width=0.32\textwidth]{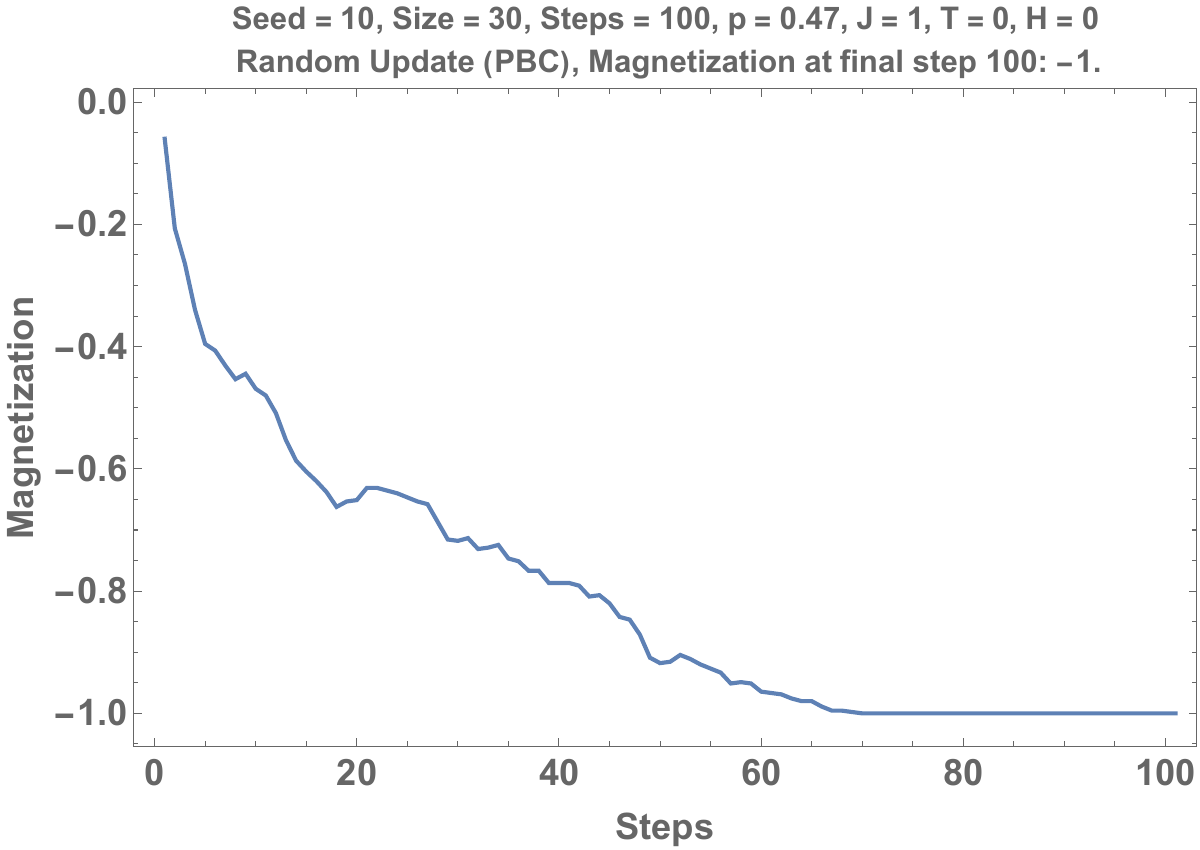}} \hspace{0.002\textwidth}
\subfigure[]{\includegraphics[width=0.32\textwidth]{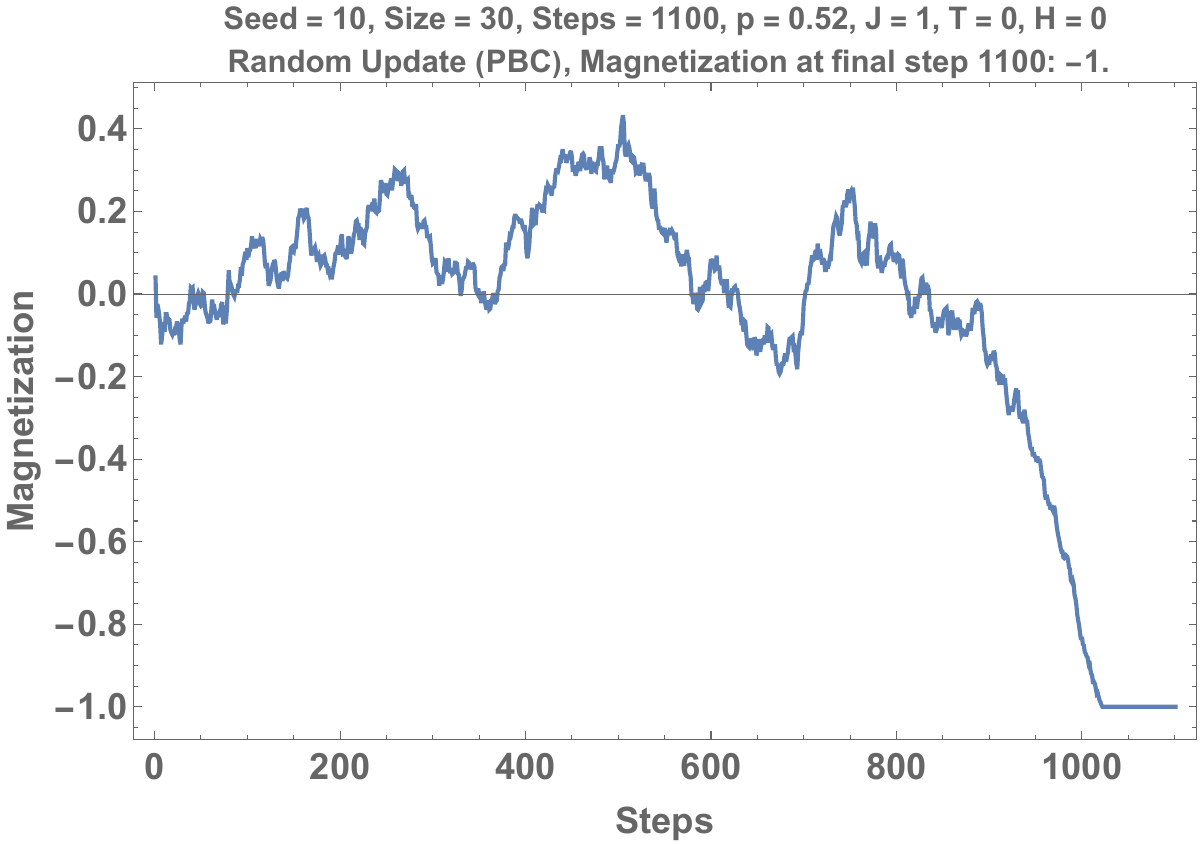}} \hspace{0.002\textwidth}
\subfigure[]{\includegraphics[width=0.32\textwidth]{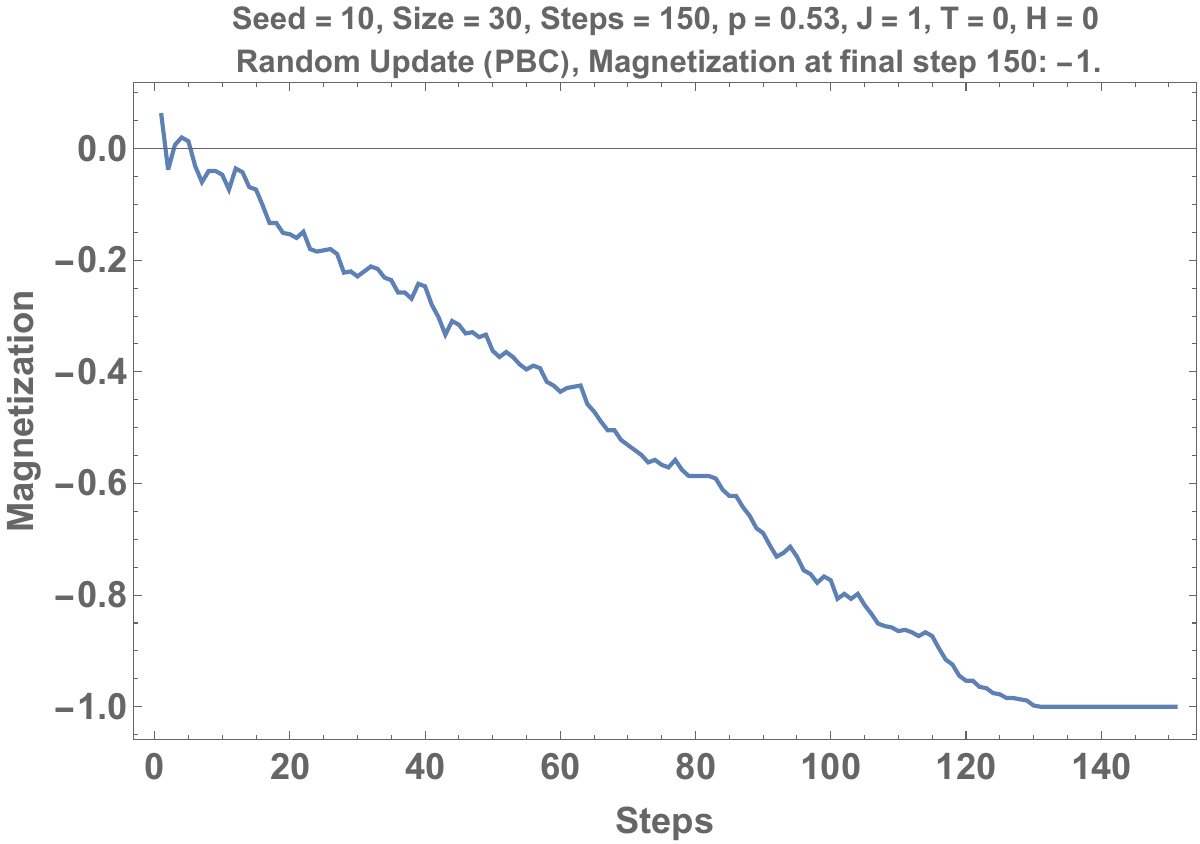}}
\\[0.03\textwidth]
\subfigure[]{\includegraphics[width=0.32\textwidth]{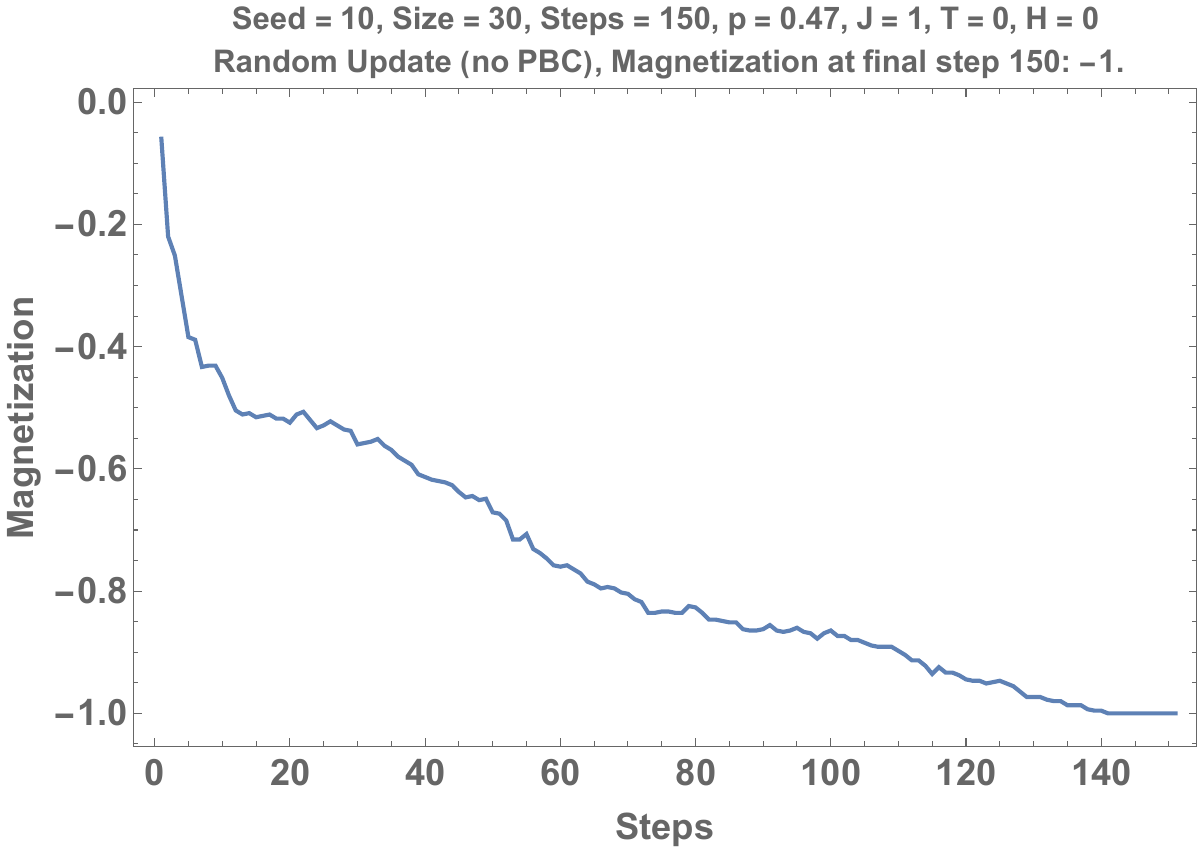}} \hspace{0.002\textwidth}
\subfigure[]{\includegraphics[width=0.32\textwidth]{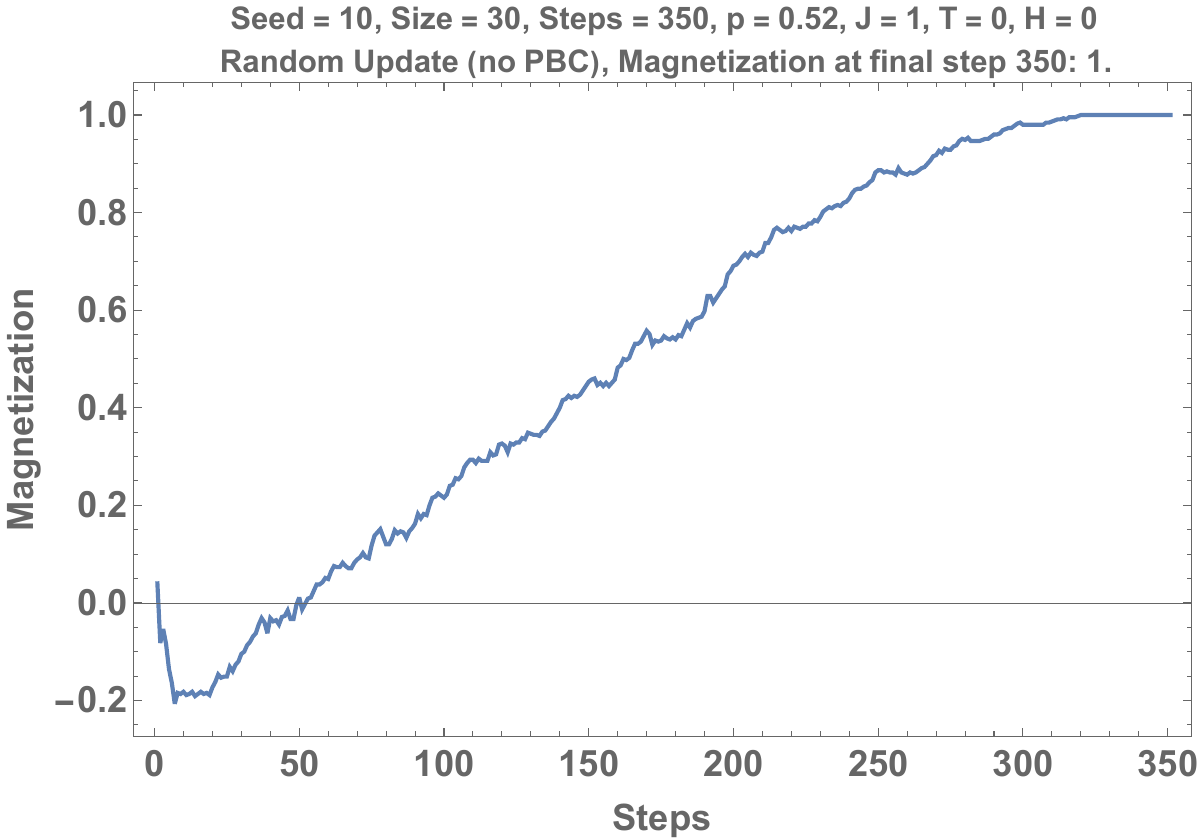}} \hspace{0.002\textwidth}
\subfigure[]{\includegraphics[width=0.32\textwidth]{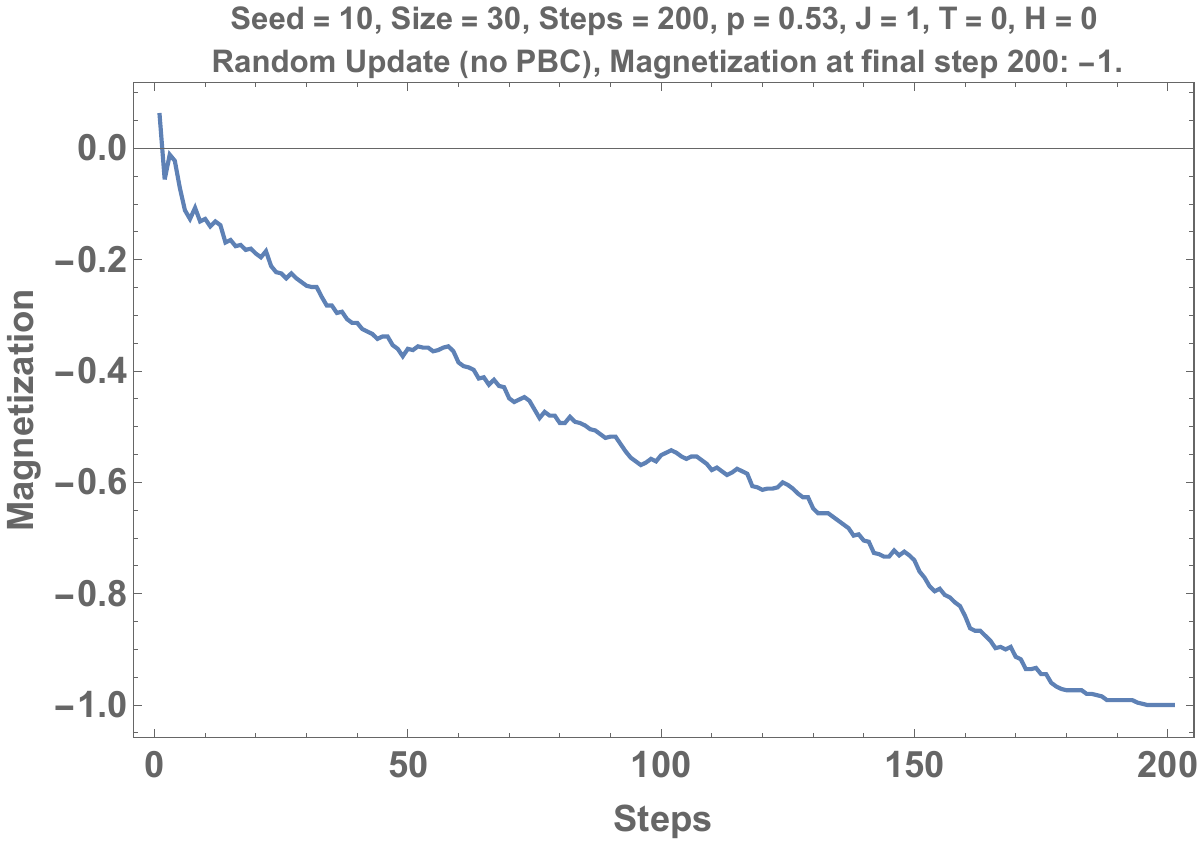}} 
\\[0.03\textwidth]
\subfigure[]{\includegraphics[width=0.32\textwidth]{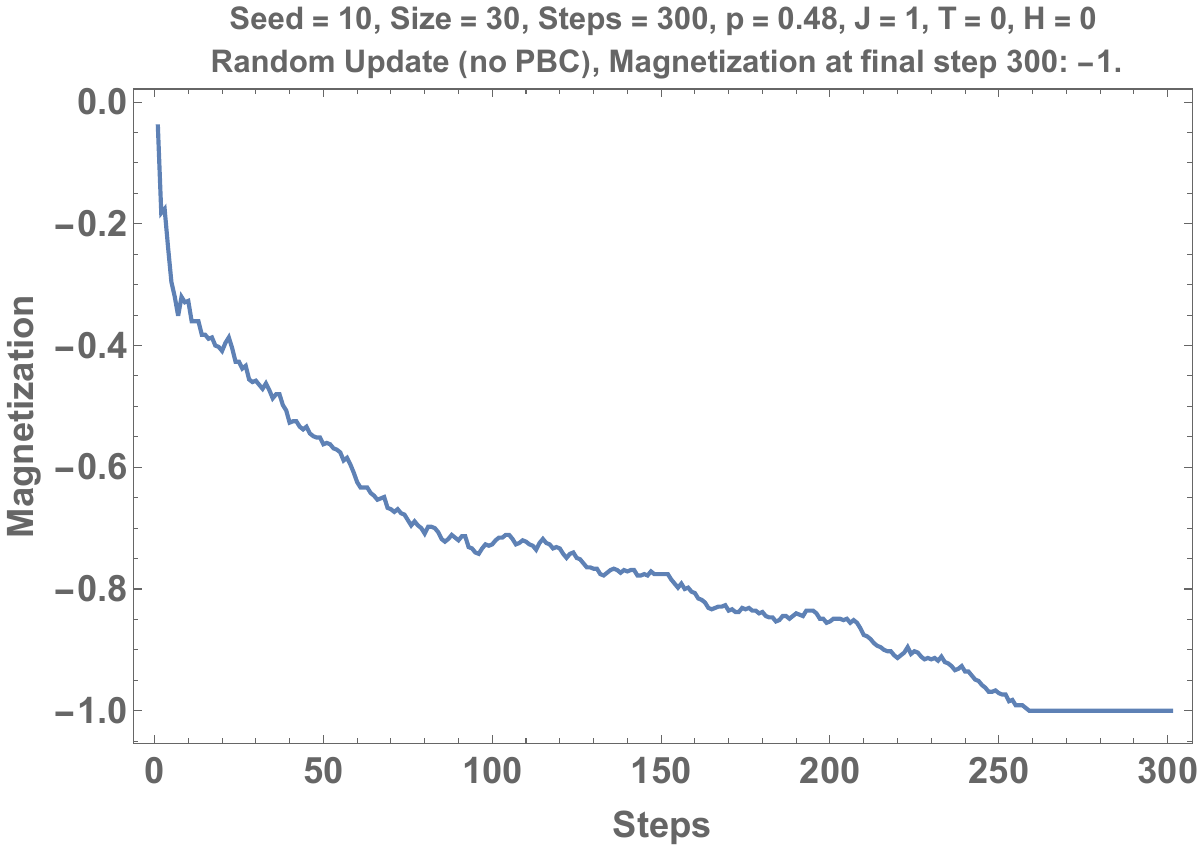}} \hspace{0.002\textwidth}
\subfigure[]{\includegraphics[width=0.32\textwidth]{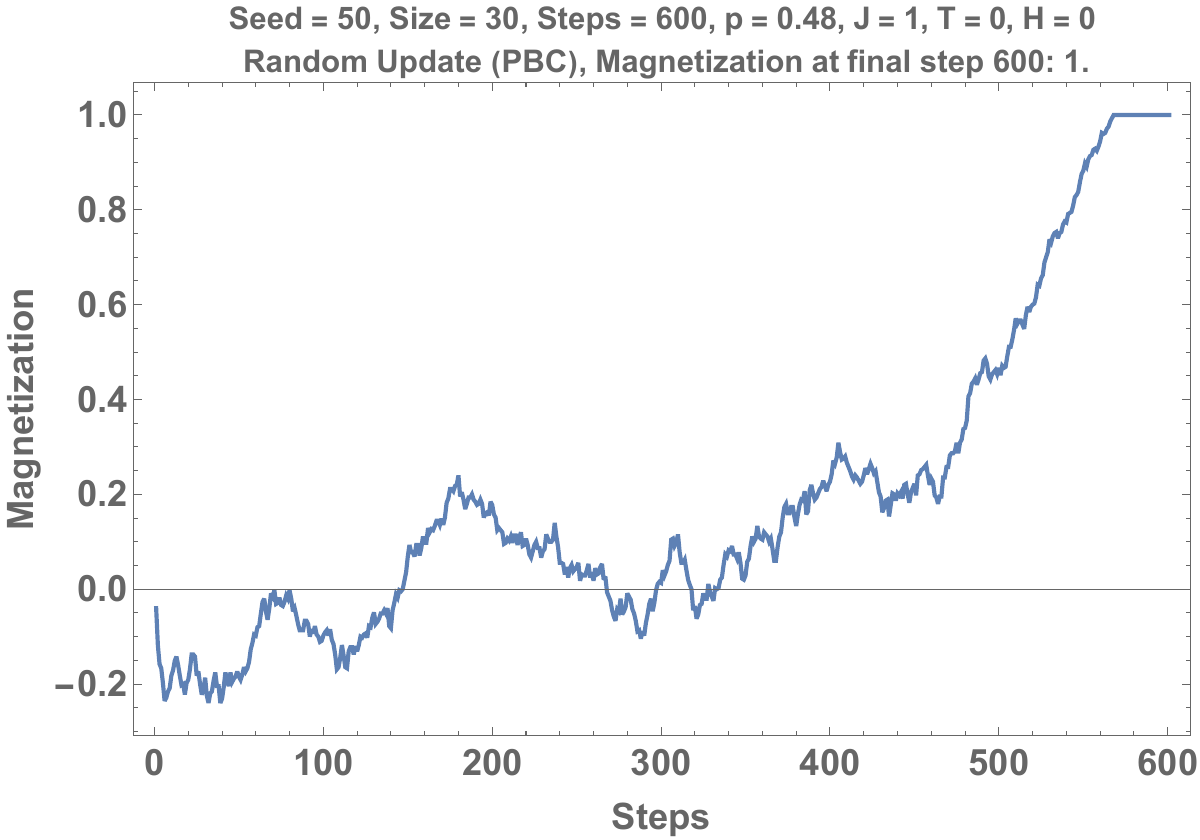}} \hspace{0.002\textwidth}
\subfigure[]{\includegraphics[width=0.32\textwidth]{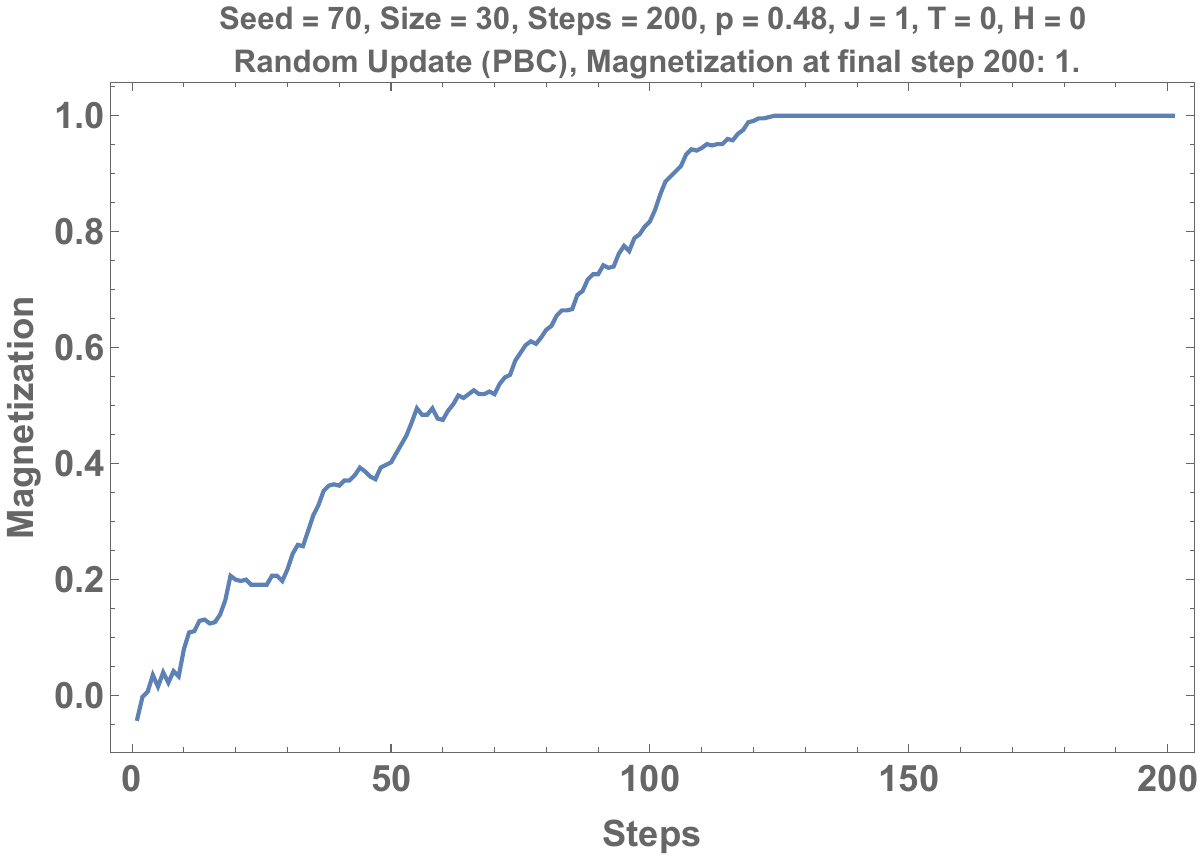}}
\\[0.03\textwidth]
\caption{Results of Monte Carlo simulations for initial respective conditions $p=0.47 (a), 0.52 (b), 0.53 (c)$ with Periodic Boundary Conditions (PBC). Sub-cases (d, e, f) show the results of the same Monte Carlo simulations but with no Periodic Boundary Conditions (no PBC). 
Except for sub-case (e), the dynamics always ends up broken along $-1$. PBC accelerates the process with less Monte Carlo steps than with no PBC.
Sub-cases (g, h, i) show the outcomes for  $p=0.48$ using different initial distributions of spins and no PBC for (g) and PBC for (h, l). Associated numbers of Monte Carlo steps differ.}
\label{rr1}
\end{figure}

But situations with a perfect equality between the numbers of agents sharing the two available choices are very rare. Therefore, it is of importance to check the robustness of the results against statistical fluctuations. To this end, I run additional simulations with small differences between $N_+$ and $N_-$ of order $\sqrt N$. Associated results are shown in Figure (\ref{rr1}) for a sample of size $30 \times 30 =900$ yielding a range $ \pm 0.03$ for the amplitude of statistical fluctuations. Random update is used.

Sub-cases (a, b, c) show the results of Monte Carlo simulations for initial respective conditions $p=0.47, 0.52, 0.53$ with Periodic Boundary Conditions (PBC). Sub-cases (d, e, f) show the results of the same Monte Carlo simulations but with no Periodic Boundary Conditions (no PBC). 

Except for sub-case (e), the dynamics always ends up broken along $-1$. PBC accelerates the process with less Monte Carlo steps than with no PBC.

Sub-cases (g, h, i) show the outcomes for  $p=0.48$ using different initial distributions of spins and no PBC for (g) and PBC for (h, l). Associated numbers of Monte Carlo steps differ.

I also run additional simulations with a larger sample $40 \times 40$ having an associated range of amplitude $\pm 0.025$ for statistical fluctuations. Figure (\ref{mn}) shows the results with Initial conditions $p=0.47$ (a, b, c, d) and $0.53$ (e, f, g, h). PBC are applied in (a, b, e, f) and not in (c, d, g, h). Domains coexistence is found in (a, b, d, e, g). More numerous Monte Carlo steps are needed than for the sample $30 \times 30$.

\begin{figure}
\centering
\subfigure[]{\includegraphics[width=0.45\textwidth]{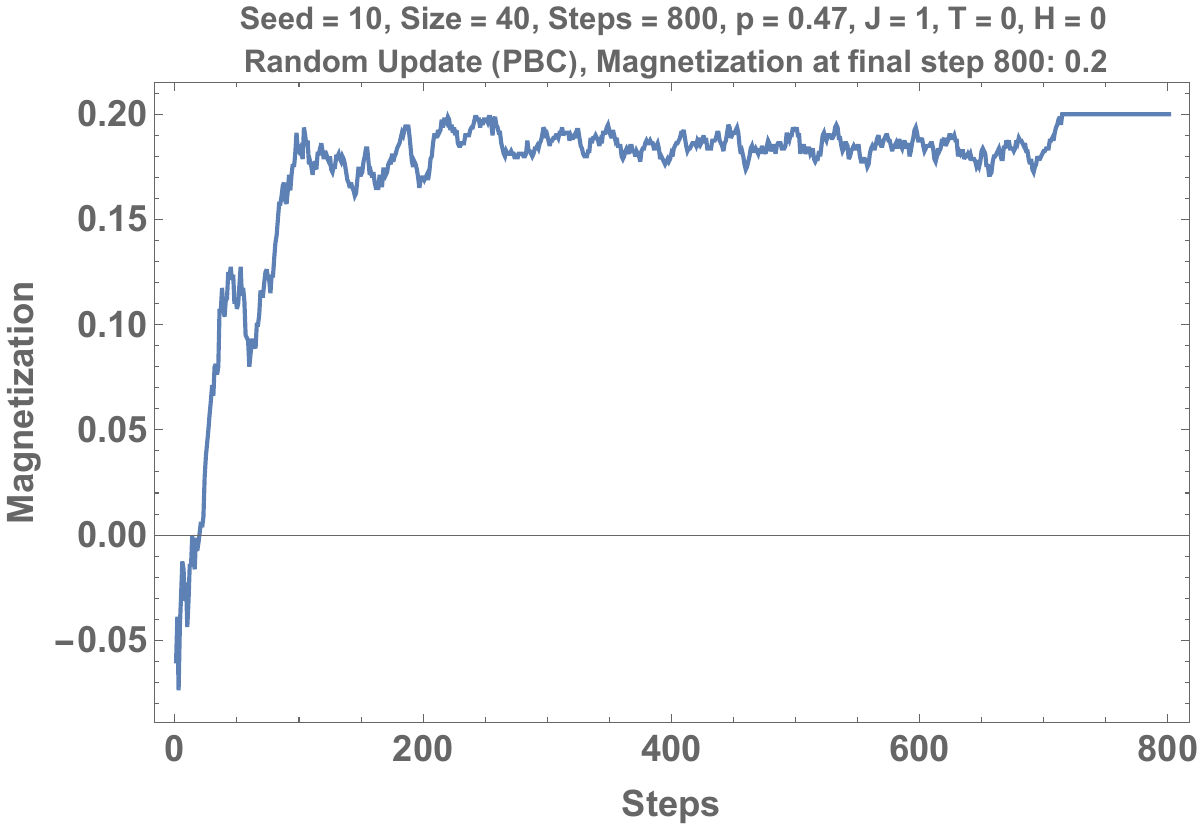}} \hspace{0.002\textwidth}
\subfigure[]{\includegraphics[width=0.45\textwidth]{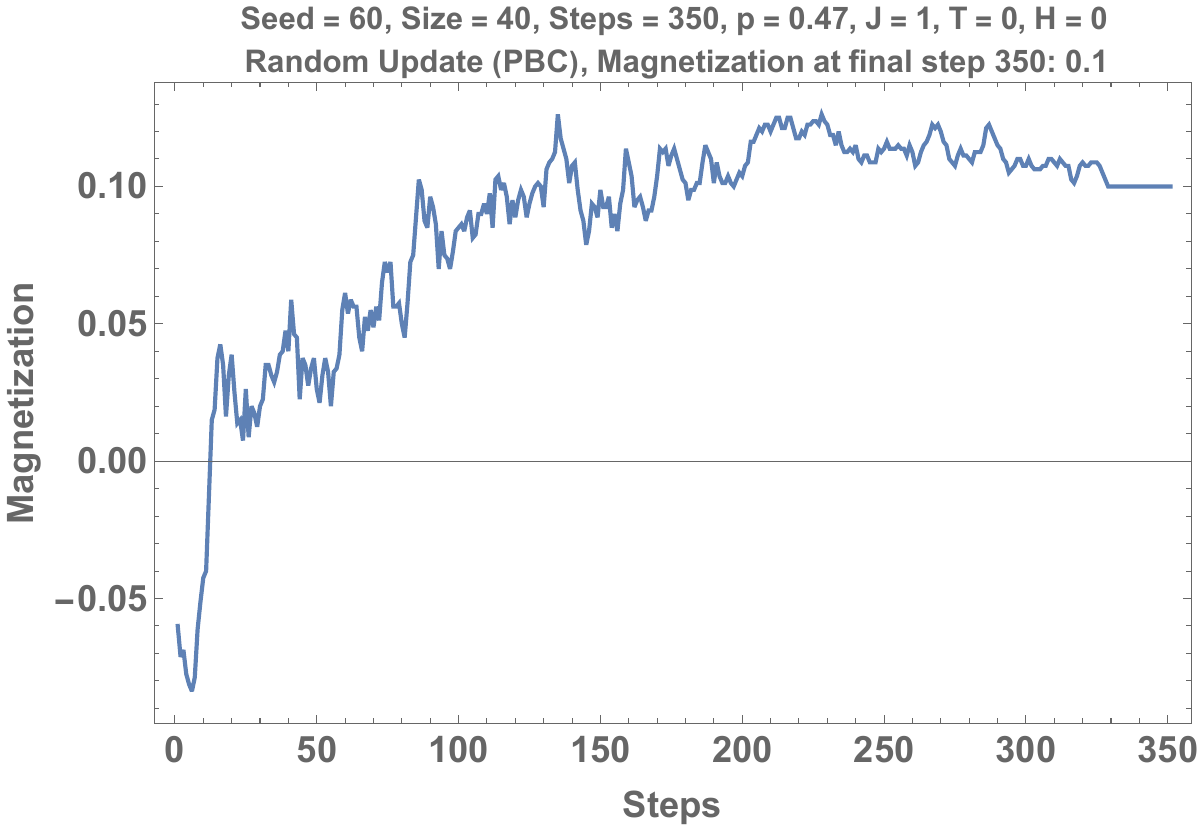}}
\\[0.03\textwidth]
\subfigure[]{\includegraphics[width=0.45\textwidth]{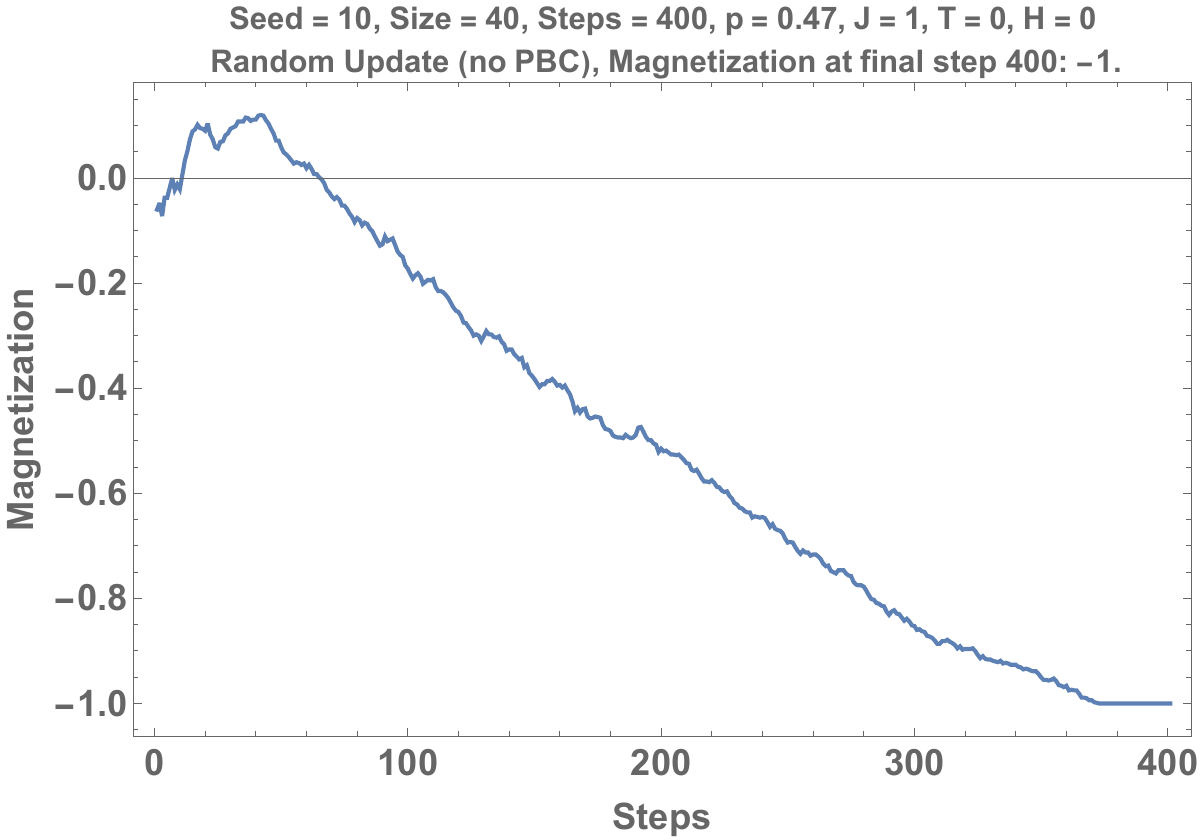}} \hspace{0.002\textwidth}
\subfigure[]{\includegraphics[width=0.45\textwidth]{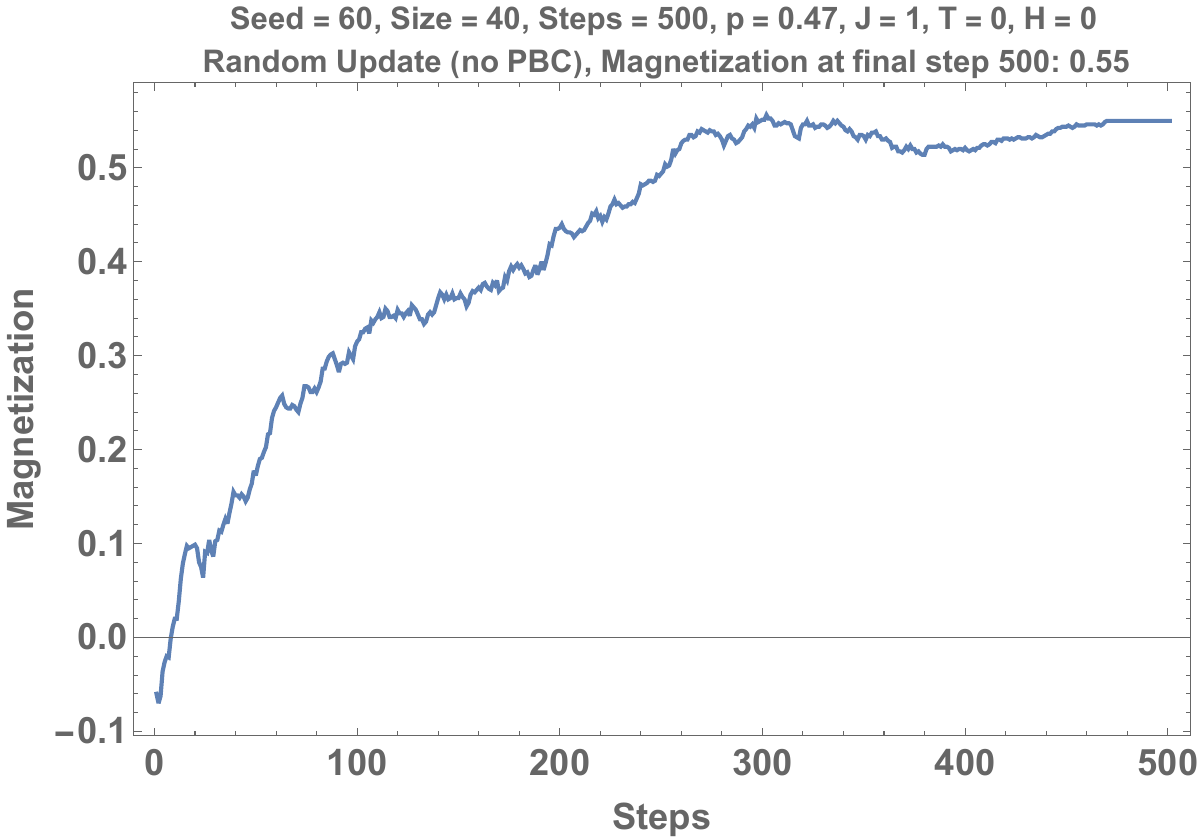}}
\\[0.03\textwidth]
\subfigure[]{\includegraphics[width=0.45\textwidth]{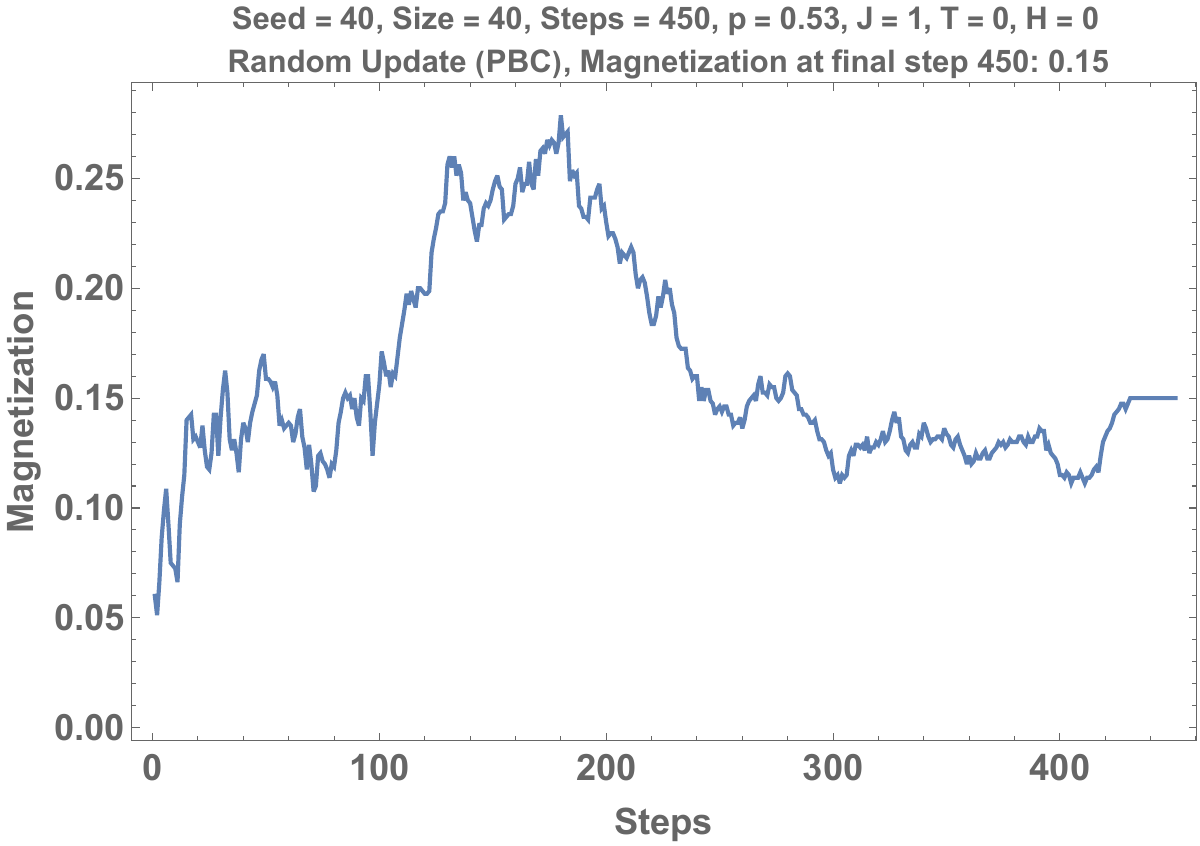}} \hspace{0.002\textwidth}
\subfigure[]{\includegraphics[width=0.45\textwidth]{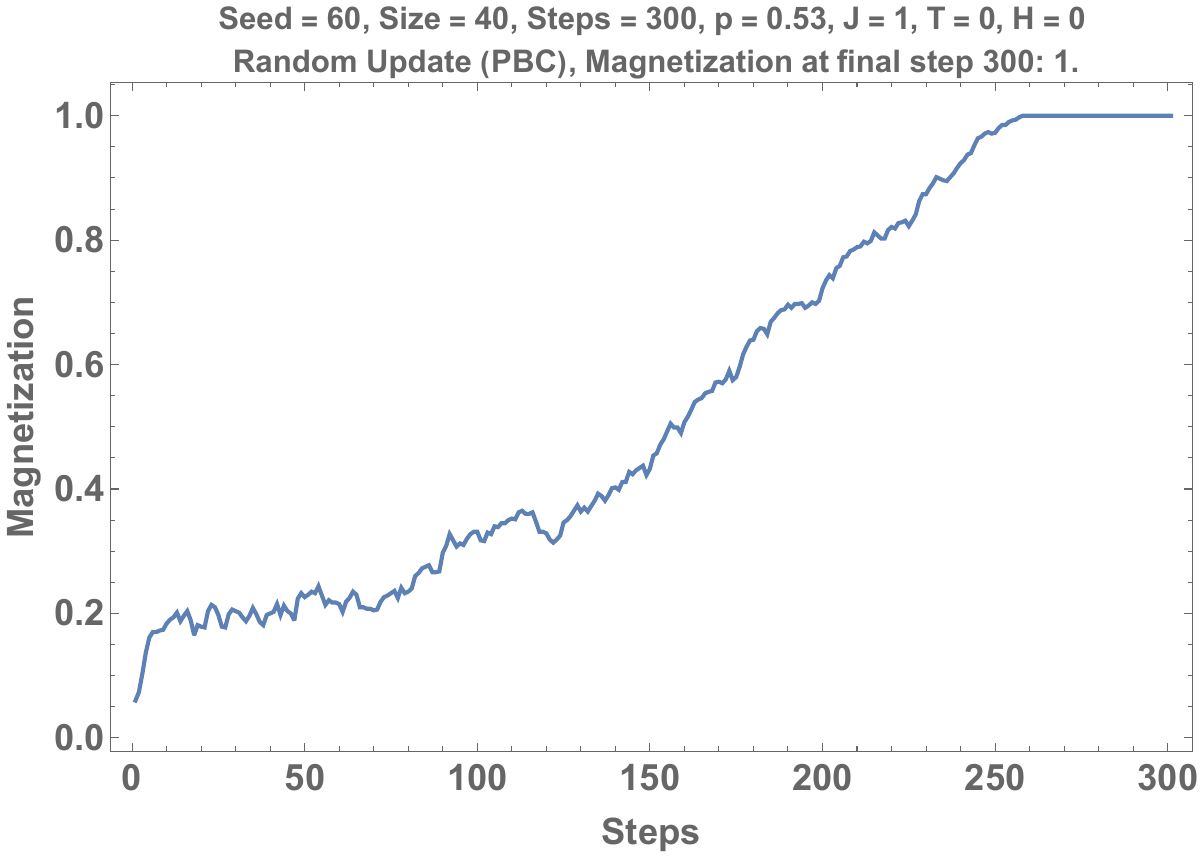}}
\\[0.03\textwidth]
\subfigure[]{\includegraphics[width=0.45\textwidth]{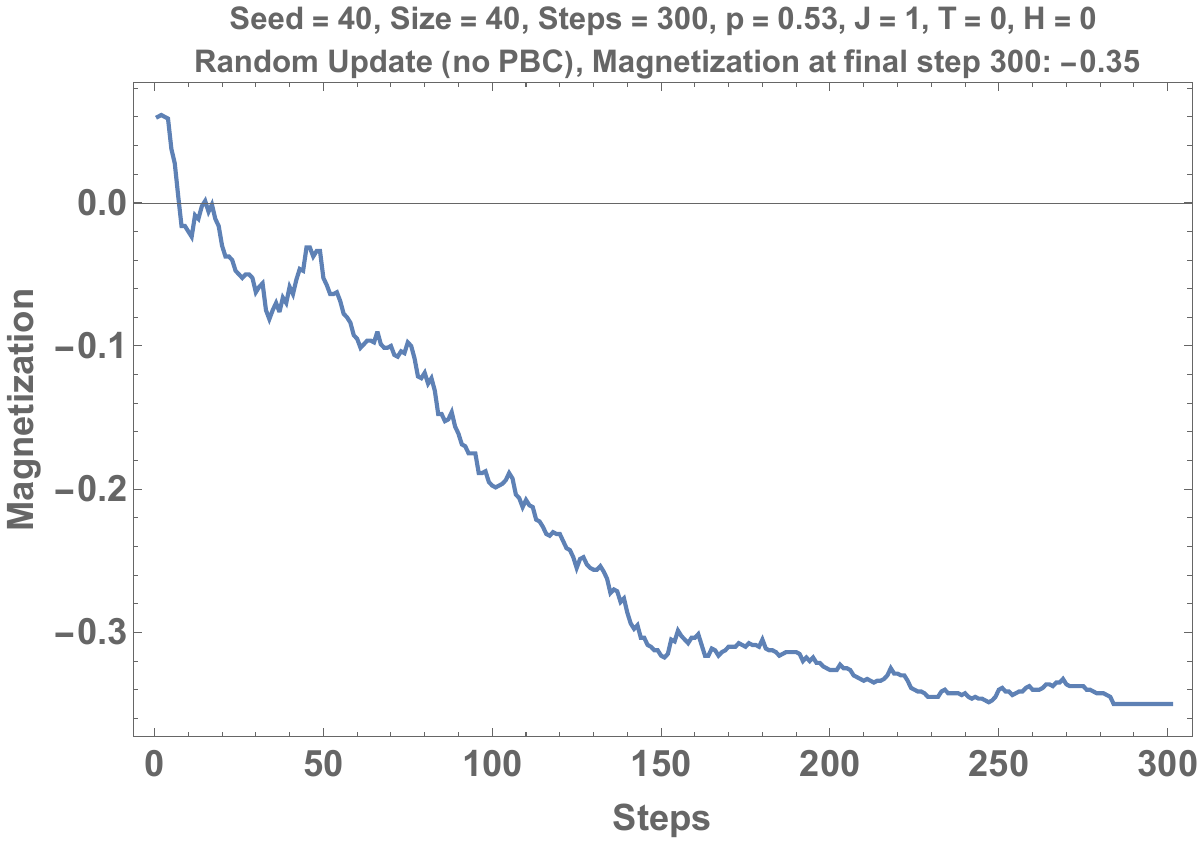}} \hspace{0.002\textwidth}
\subfigure[]{\includegraphics[width=0.45\textwidth]{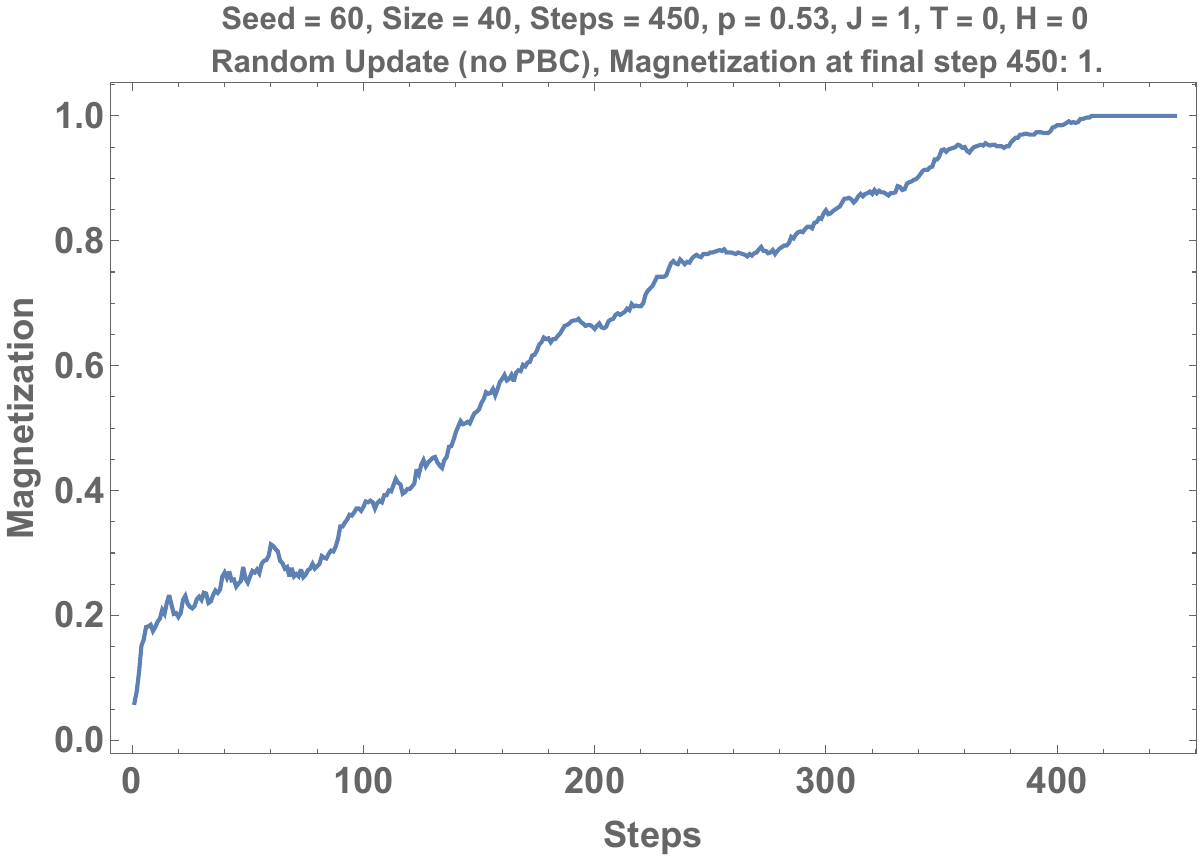}}
\\[0.03\textwidth]
\end{figure}

\newpage 

\noindent\captionof{figure}{Results for a $40 \times 40$ sample with Initial conditions $p=0.47$ (a, b, c, d) and $0.53$ (e, f, g, h). PBC are applied in (a, b, e, f) and not in (c, d, g, h). Domains coexistence is found in (a, b, d, e, g). Much more Monte Carlo steps are needed than for the sample $30 \times 30$.}
\label{mn}

\section{The social consequences of social collective symmetry breaking}

Elaborating about the details of the mentioned effective frameworks, which are required to monitor the update of opinions in a social network, is out the scope of this work. At this stage of the study it is sufficient to assert the existence of such  frameworks noticing that are likely different from one social community to another. They can also share share similar parts. Their existence is sufficient to draw the social consequences of their activation.

\subsection{Echo chambers are the outcome of spontaneous symmetry breaking}

The various results of above simulations demonstrate that when a social network is facing a debate about an issue for which there is no objective difference between the two competing options, the pair interactions among its agents lead to a spontaneous symmetry breaking along one of them. The winning option being selected randomly. 

I claim that it is the process of completing a spontaneous symmetry breaking, which in turn generates quite automatically the appearance of echo chambers. 
Echo chambers are then an unexpected, unplanned but natural outcome of the maximization of agent utilises via pair interactions. They arise step by step following the update dynamics of agent choices with unanimity of choice when full symmetry breaking is achieved.

My suggestion is that it is the interacting structure linking agents by interacting pairs, which triggers  the building of unanimity among related agents. Once the symmetry breaking is fully implemented, the related driving structure turn the associated network into an echo chamber. 

Therefore, the qualification of echo chamber does not apply at the beginning of the opinion dynamics within the network. Its reality emerges only at the end of the update dynamics when reaching final outcome. At the start of what will become an echo chamber stand a collection of agents holding heterogeneous opinions. 

The other surprising and a bit disturbing conclusion from the study is the actual final winning choice is indeed selected randomly when a symmetry of choice holds at the individual level.

My statement goes against current explanations of the making of echo chambers, which all claim echo chambers are the result of a preferential attachment dynamics of agents sharing the same opinion. Here people with different choices start interacting in their network ending all along the same choice. The symmetry breaking being random the actual choice labelling the network could have been the opposite one.

\subsection{Polarization is a byproduct of the random selection of the final equilibrium state}

Since the final state labelling a given echo chamber is selected randomly by the update process used to reach the full symmetry breaking, different echo chambers end up holding different opinions or choices. Echo chambers are thus opposed in their respective unanimities in a quenched opposition. This fact fuels the emergence of possible hate between members of opposed echo chambers. 

Accordingly, a social community divided into several networks become naturally polarized between two opposed subparts. The simulations showed also that sometimes a given network gets polarized with two opposed subparts.

My takeaway from the simulations is counterintuitive and seems contradictory since on the one hand polarization materializes the unconditional support of a specific social choice and the other hand, this very choice has been selected randomly. This  fact sheds a disruptive light on the polarization phenomenon.

\section{Conclusion}

I would like to highlight that figuring out that the equilibrium state associated with Eq. (\ref{U}) is having all agents (spins) making the same choice (orientation) is not of a surprise by itself. What comes as a non trivial and unexpected outcome is the fact that a collection of agents (spins) connected only by few pair interactions manage to come up with a collective coordination aligning all agents while each agent (spin) is connected only to a few other ones. The question is not the final outcome but how can it be reach?

In the geometrical grid used for the simulations the question is how the existence of simultaneous short range order interactions manages to create a long range order. That puzzle has been solved long time ago in statistical physics and is denoted spontaneous symmetry breaking. 

In this work I have recovered the  phenomenon of spontaneous symmetry breaking in the case of a collection of agents having the choice between two opinions and interacting by pairs simultaneously, which also not new. 

The novelty of the work is the exploration of the effect of both initial conditions and update algorithm on the final equilibrium state and the number of steps required to reach it. Especially, since in physics what matters is the equilibrium state, which must be independent of the actual dynamics used to reach it. When running Monte Carlo simulations for which the implementation of a dynamics of repeated updates is a prerequisite, the final state must be independent of the associated dynamics used. With respect to social systems I claim that the opposite holds true. The algorithm used to monitor the individual updates is an instrumental feature of each social network.

I have thus shown that different initial conditions and update scheme produce different equilibrium states. These equilibrium states display either a full spontaneous symmetry breaking or a fragmented spontaneous symmetry breaking with the stable coexistence of two opposed domains of different sizes. Four types of update schemes, namely the random, sequential, simultaneous and checkerboard updates, have exhibited different outcomes.

On this basis the various simulations shed a new light on the making of echo chambers. In my process, echo chambers appear quite naturally at the end of a dynamics of local updates within an existing social network where agents shared initially different opinions. It is the existence of a social network bounding a collection of interacting agents, which leads step by step to their alignment on the same choice. 

This conclusion refutes the usual and well-accepted definition of an echo chamber viewed as a network built via a preferential attachment of people sharing the same opinion. Here the network exists prior to the symmetry breaking achievement.

In addition, the simulations have shown that the selection of the actual choice adopted by all members of a social network is made randomly. Indeed, the opinion shared by all members of a so-called echo chamber could has been the opposite one.

Furthermore, it is the random character of the spontaneous symmetry breaking, which prompts the occurrence of polarization within communities made of several social networks. Polarization turns out to be a side effect of the random selection of the respective unanimous choices made in the various echo chambers present in a given social community. In particular, the simulations have indicated that the details of the update processing are instrumental to monitor the dynamics of individual updates. 

Accordingly, the design of dedicated algorithms aimed at hindering or accelerating the completion of the symmetry breaking among agents opens the way to a purely technical lever to  tackle polarization independently of the issue and opinions at stake. 

While trying to control the initial distribution of individual choices might be tricky, acting on the update process seems more feasible. For instance, implementing a checkerboard update proved to be very efficient and quick in reaching the full symmetry breaking of a social network. In contrast a simultaneous update traps the network in a highly heterogenous distribution of different choices. 

In this paper all individual opinions are symmetrical, which in turn makes random the selection of the final unique choice. In a forthcoming work I intend to investigate the effect of thwarting the random selection of the spontaneous symmetry breaking by applying external homogeneous symmetry breaking pressure as well as internal heterogenous individual pressures.

%%%%%%%%%%%%%%%%%%%%%%%%%%%%%%

%%%%%%%%%%%%%%%%%%%%%%%%%%%%%%%%%%%%%
\end{document}